\documentclass[11pt]{article}
\pdfoutput=1
\usepackage[a4paper, top=2.5cm, bottom=2.5cm, outer=2.1cm, inner=2.1cm]{geometry}
\usepackage{graphicx}
\usepackage{epsfig}
\usepackage{amsfonts,amssymb,amsmath,mathrsfs}
\usepackage{cite}
\usepackage{float}
\usepackage[textsize=footnotesize,textwidth=1.5cm]{todonotes}
\usepackage[font=footnotesize,labelfont=bf,justification=centerlast,width=.94\textwidth]{caption}
\usepackage{authblk}
\usepackage{comment}
\usepackage{hyperref}
\usepackage[utf8]{inputenc}
\usepackage[toc,page]{appendix}
\usepackage{color,xcolor}
\usepackage{cancel}
\usepackage{tikz}
\usetikzlibrary{automata,positioning,arrows}
\usetikzlibrary{shapes,shadows,arrows}
\usepackage{makecell,multirow}

\def\rg{\ensuremath{\mathscr R}}

\numberwithin{equation}{section}
\def\bal#1\eal{\begin{align}#1\end{align}}

\def\){\right)}
\def\({\left( }
\def\]{\right] }
\def\[{\left[ }

\def\nn{\nonumber}
\def\NO{\nonumber}

\def\a{\alpha}
\def\b{\beta}

\def\d{\delta}
\def\e{\epsilon}
\def\f{\phi}
\def\g{\gamma}

\def\j{\psi}
\def\k{\kappa}
\def\l{\lambda}
\def\m{\mu}
\def\n{\nu}
\def\om{\omega}
\def\p{\pi}

\def\r{\rho}
\def\s{\sigma}
\def\t{\tau}
\def\x{\xi}
\def\z{\zeta}
\def\D{\Delta}
\def\F{\Phi}

\def\J{\Psi}
\def\L{\Lambda}
\def\Om{\Omega}

\def\S{\Sigma}

\def\ve{\varepsilon}

\def\vf{\varphi}

\def\cj{{\cal J}}

\def\cl{{\cal L}}
\def\cm{{\cal M}}

\def\co{{\cal O}}
\def\cp{{\cal P}}
\def\cq{{\cal Q}}

\def\cs{{\cal S}}
\def\ct{{\cal T}}
\def\cu{{\cal U}}
\def\cv{{\cal V}}
\def\cw{{\cal W}}

\renewcommand{\bar}{\overline}
\renewcommand{\tilde}{\widetilde}
\renewcommand{\hat}{\widehat}
\renewcommand{\leq}{\leqslant}
\renewcommand{\geq}{\geqslant}

\newcommand{\cL}{\mathcal{L}}

\newcommand{\cO}{\mathcal{O}}

\newcommand{\be}{\begin{equation}}
\newcommand{\ee}{\end{equation}}
\newcommand{\bea}{\begin{eqnarray}}
\newcommand{\eea}{\end{eqnarray}}
\newcommand{\bb}{\mathbb}
\newcommand{\ba}{\begin{align}}
\newcommand{\ea}{\end{align}}
\newcommand{\bad}{\begin{aligned}}
\newcommand{\ead}{\end{aligned}}

\newcommand{\bsub}{\begin{subequations}}
\newcommand{\esub}{\end{subequations}}
\newcommand{\beqx}{\begin{displaymath}}
\newcommand{\eeqx}{\end{displaymath}}
\newcommand{\bmat}{\left(\begin{array}}
\newcommand{\emat}{\end{array}\right)}

\def\bb#1{\ensuremath{\mathbb{#1}}} 

\def\bo{{\raise-.3ex\hbox{\large$\Box$}}}               
\def\pa{\partial}                                       
\def\face{{\raise.2ex\hbox{$\displaystyle \bigodot$}\mskip-2.2mu \llap {$\ddot
        \smile$}}}                                   
\def\>{\rangle}                                      
\def\<{\langle}                                      

\def\sbtx#1{{}_{\rm #1}}                           
\newcommand{\sub}[1]{\phantom{}_{(#1)}\phantom{}}    
\def\wt#1{\widetilde{#1}}                            
\def\Hat#1{\widehat{#1}}                             
\def\lbar#1{\ensuremath{\overline{#1}}}              
\def\leftrightarrowfill{$\mathsurround=0pt \mathord\leftarrow \mkern-6mu
        \cleaders\hbox{$\mkern-2mu \mathord- \mkern-2mu$}\hfill
        \mkern-6mu \mathord\rightarrow$}        
\def\dvec#1{\vbox{\ialign{##\crcr
        \leftrightarrowfill\crcr\noalign{\kern-1pt\nointerlineskip}
        $\hfil\displaystyle{#1}\hfil$\crcr}}}           

\def\-{\hphantom{-}}

\begin{document}


\title{\vspace{-1.0in}
\begin{minipage}[t]{\textwidth}
 \begin{flushright}
 {\small{KIAS-P20068}}\\\vskip2.0in
 \end{flushright}
\end{minipage}\\ 
\vspace{1.0in}\textbf{AdS$_3$ gravity and the complex SYK models}}

\author[a]{Pankaj Chaturvedi\thanks{\noindent E-mail:~cpankaj@iitk.ac.in}}
\author[b]{Ioannis Papadimitriou\thanks{\noindent E-mail:~ioannis@kias.re.kr }}
\author[a,c]{Wei Song\thanks{\noindent E-mail:~wsong2014@mail.tsinghua.edu.cn}}
\author[a]{Boyang Yu\thanks{\noindent E-mail:~yuby16@mails.tsinghua.edu.cn}}

\affil[a]{
\textit{Yau Mathematical Sciences Center, Tsinghua University, Beijing, 100084, China}}

\affil[b]{
\textit{School of Physics, Korea Institute for Advanced Study, Seoul 02455, Korea}}

\affil[c]{
\textit{Institute for Advanced Study, 1 Einstein Drive, Princeton, NJ 08540, USA}}

\maketitle
\thispagestyle{empty}

\begin{abstract}
We provide a non-conformal generalization of the Compère-Song-Strominger (CSS) boundary conditions for AdS$_3$ gravity that breaks the $\widehat u(1)$ Kac-Moody-Virasoro symmetry to two $u(1)$s. The holographic dual specified by the new boundary conditions can be understood as an irrelevant deformation of a warped conformal field theory (WCFT). Upon consistent reduction to two dimensions, AdS$_3$ gravity results in a deformed Jackiw-Teitelboim dilaton gravity model coupled to a Maxwell field. We show that near extremality the boundary conditions inherited from generalized CSS boundary conditions in three dimensions give rise to an effective action exhibiting the same symmetry breaking pattern as the complex Sachdev-Ye-Kitaev models. Besides the Schwarzian term reflecting the breaking of conformal symmetry, the effective action contains an additional term that captures the breaking of the $\widehat u(1)$ Kac-Moody symmetry.  

\end{abstract}

\newpage
\tableofcontents
\addtocontents{toc}{\protect\setcounter{tocdepth}{2}}
\setcounter{page}{1}

\section{Introduction and summary of results}
\label{sec:intro}

Recent progress on the holographic duality between the Jackiw-Teitelboim (JT) gravity and the Sachdev-Ye-Kitaev (SYK) models \cite{Sachdev:1993PhRvL,Kitaev:2015tk, Maldacena:2016hyu} has provided a paradigm towards understanding quantum gravity in two  \cite{Almheiri:2014cka,Jensen:2016pah, Maldacena:2016upp, Engelsoy:2016xyb, Saad:2019lba} and three dimensions \cite{Cotler:2018zff,Cotler:2020ugk,Maxfield:2020ale,Cotler:2020lxj}. JT gravity is a two dimensional theory of gravity coupled to a dilaton \cite{ Jackiw:1984je,Teitelboim:1983ux}, and describes the near horizon dynamics of near extremal black holes. The solutions of JT gravity have a locally AdS$_2$ geometry and a running dilaton, resulting in a Schwarzian on-shell action -- a feature shared by the SYK models \cite{Sachdev:1993PhRvL,Kitaev:2015tk}. The latter are quantum mechanics models of Majorana fermions with random couplings, featuring an emergent conformal symmetry at low energy. Since JT gravity describes the region slightly away from the AdS$_2$ throat of extremal black holes and the effective low energy description of the dual SYK models is applicable slightly away from the conformal fixed point, this correspondence can be viewed as an example of a nAdS$_2$/nCFT$_1$ holographic duality \cite{Mertens:2017mtv,  Larsen:2018iou, Johnson:2019eik, Iliesiu:2019xuh, Witten:2020wvy, Aniceto:2020saj}.\footnote{See recent reviews \cite{Sarosi:2017ykf,Trunin:2020vwy}.} 

An important ingredient in the JT/SYK or nAdS$_2$/nCFT$_1$ correspondence is the emergence of conformal symmetry, and the pattern of symmetry breaking. In \cite{Chaturvedi:2018uov},  a similar emergent symmetry and breaking pattern have been discussed for the complex SYK models, which involve Dirac fermions instead of Majorana \cite{Davison:2016ngz,Bulycheva:2017uqj,Gu:2019jub,Klebanov:2020kck,Berkooz:2020uly}. The effective action of the complex SYK models takes the general form
\be
I_{\rm eff}^{\rm CSYK}=\frac{NK}{2}\int\limits_{0}^{T^{-1}} d\t\;\big(i{\bar \m}\,\pa_\t \bar f(\t)-\pa_\t\L\big)^2-\frac{N\g}{4\pi^2}\int\limits_{0}^{T^{-1}} d\t\;\big\lbrace\tan\big(\pi T \bar f(\t)\big),\t\big\rbrace,\label{CSYKeff}
\ee
where, $T$  is the temperature, $K$ is the compressibility and $\g$ is the specific heat. The effective action \eqref{CSYKeff} includes a Schwarzian term as well as a $u(1)$ term, reflecting the breaking of the so-called warped conformal symmetry \cite{Detournay:2012pc}. The observation of \cite{Chaturvedi:2018uov} is that the complex SYK models can be understood as a nearly warped CFT (WCFT), featuring a Virasoro-Kac-Moody algebra. 

Holography with (unbroken) warped conformal symmetry has been discussed for the warped AdS$_3$ black holes \cite{Anninos:2008fx, Compere:2008cv}, BTZ black holes with CSS boundary conditions \cite{Compere:2013bya}, and the horizon throat of extremal black holes in higher dimensions \cite{Song:2011sr}. Warped conformal symmetry arises also as a hidden symmetry of non-extremal Kerr black holes \cite{Aggarwal:2019iay}. While AdS$_2$ is a universal factor of the near horizon throat of extremal black holes, its $sl(2,\bb R)$ isometry is usually accompanied by an additional $u(1)$ symmetry. For example, the extremal Reissner-Nordstr\"om black hole has a non-zero charge, while extremal Kerr black holes require a non-vanishing angular momentum. With appropriate boundary conditions, warped conformal symmetry may arise as the asymptotic symmetry in such cases. In other words, the $sl(2,\bb R)\times u(1)$ symmetry may be enhanced to a Virasoro-Kac-Moody symmetry.

Slightly away from the horizon throat, warped conformal symmetry is broken and one expects the near horizon dynamics to be captured by JT gravity coupled to a $u(1)$ Maxwell field. Even though such a field does typically arise from the Kaluza-Klein (KK) reduction of higher dimensional gravity  \cite{Cvetic:2016eiv,Nayak:2018qej,Kolekar:2018sba,Castro:2018ffi,Moitra:2019bub,Lozano:2020txg}, the symmetries preserved at or near the horizon depend on the boundary conditions imposed. In particular, it was shown in \cite{Cvetic:2016eiv} that Dirichlet boundary conditions on the $u(1)$ gauge field can be straightforwardly applied both at and near the extremal horizon and in either case result in a global $u(1)$ symmetry. However, Neumann boundary conditions on the gauge field have so far been applied only strictly at the extremal horizon (see e.g. \cite{Sen:2008vm,Grumiller:2014oha}), resulting in an unbroken $\Hat u(1)$ Kac-Moody-Virasoro symmetry. This boundary condition is imposed implicitly whenever the 2D Maxwell field is integrated out \cite{Ghosh:2019rcj}. It is also inherited from CSS boundary conditions for AdS$_3$ gravity \cite{Cvetic:2016eiv}. In order to capture the dynamics of near extremal black holes with an explicitly and spontaneously broken $\Hat u(1)$ Kac-Moody-Virasoro symmetry, one must extend the Neumann boundary conditions for the Maxwell field away from the extremal horizon, i.e. in the presence of a non-constant dilaton. 

In this paper we achieve this goal by first constructing a non-conformal generalization of the CSS boundary conditions for AdS$_3$ gravity. This an interesting problem in its own right, since it amounts to identifying a suitable irrelevant deformation of the holographic dual WCFT. Moreover, it allows us to construct the relevant boundary condition in 2D non-perturbatively, at an arbitrary distance from the extremal horizon. This is because, in contrast to the standard JT dilaton gravity, the 2D Einstein-Maxwell-Dilaton (EMD) model resulting from a consistent KK reduction of AdS$_3$ gravity describes the dynamics of both extremal and non-extremal black holes. Other approaches to a holographic description of the low energy dynamics of the complex SYK models include \cite{Davison:2016ngz,Gaikwad:2018dfc,Moitra:2018jqs,Afshar:2019axx,Godet:2020xpk}.  

In order to obtain a non-conformal generalization of the CSS boundary conditions, we describe a framework within which consistent boundary conditions for AdS$_3$ gravity can be systematically studied and classified. Other approaches to this problem include \cite{Perez:2016vqo,Grumiller:2016pqb} for AdS$_3$ gravity and \cite{Goel:2020yxl} for AdS$_2$.
We argue that in the metric formulation of AdS$_3$ gravity admissible boundary conditions correspond to canonical transformations on the Fefferman-Graham reduced phase space, in direct analogy with general mixed boundary conditions for scalar fields in AdS \cite{Papadimitriou:2007sj}. We identify two distinct classes of such canonical transformations, depending on whether they are generated by a non-trivial boundary term, or they simply correspond to a rotation -- or `spectral flow' -- among the canonical variables. 

An important difference compared to scalars in AdS is that generic canonical transformations for AdS$_3$ gravity correspond to nonlocal deformations of the dual CFT, due to the differential constraints that the canonical variables satisfy (conservation and trace conditions of the stress tensor). Although such deformations are occasionally controllable and can lead to very interesting physics, here we focus exclusively on local deformations of the dual theory. This is achieved by first solving the constraints in order to obtain a reduced phase space  parameterized by unconstrained canonical variables, and then considering canonical transformations within this unconstrained phase space.       

Our first result is the construction of a deformed WCFT (DWCFT) phase space parameterized by an arbitrary function that breaks conformal invariance through a non-zero trace anomaly. The DWCFT phase space is an exact solution of the conservation and trace conditions of the stress tensor and reduces to the WCFT phase space as the deformation parameter approaches zero. Starting with the Dirichlet variational principle on the DWCFT phase space, we identify a canonical transformation that results in a non-conformal generalization of the CSS boundary conditions. At generic values of the deformation parameter this boundary condition preserves two $u(1)$s, which are enhanced to a $\Hat u(1)$ Kac-Moody-Virasoro symmetry as the deformation is turned off.   
         
\begin{figure}[ht!] 
\captionsetup{justification=raggedright,
singlelinecheck=false
}
\centering
\includegraphics[scale=.5]{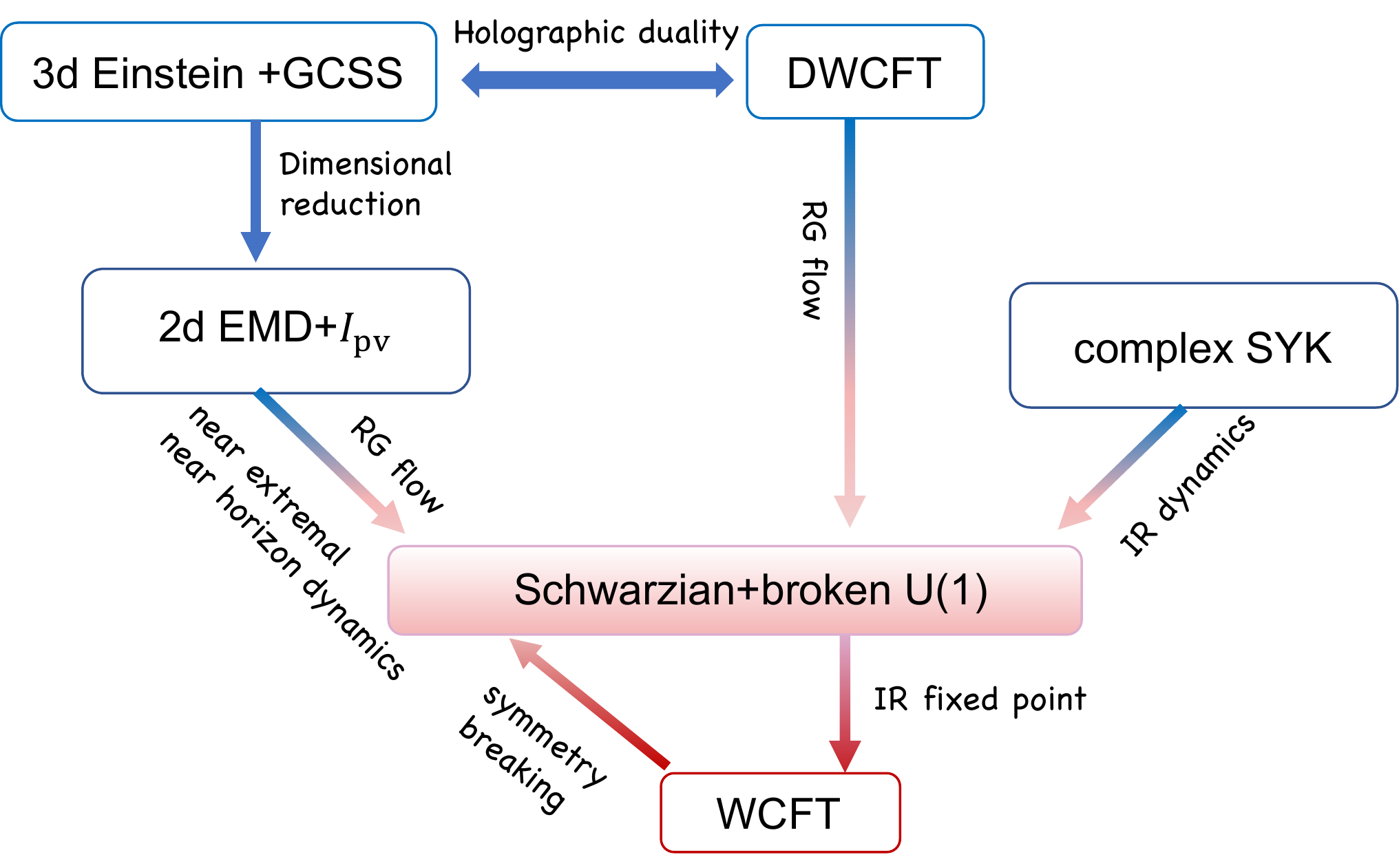} 
\caption{This diagram summarizes the web of relations we have identified between 3D and 2D theories of gravity with specific boundary conditions, and their holographic duals. In particular, these relations connect AdS$_3$ gravity with generalized CSS boundary conditions, warped CFTs, 2D Einstein-Maxwell-dilaton gravities, and complex SYK models.}
\label{relations} 
\end{figure}

A key property of the DWCFT phase space of AdS$_3$ gravity is that it is the minimal extension of the WCFT phase space that incorporates the entire solution space of the EMD model obtained from a consistent KK reduction to 2D \cite{Cvetic:2016eiv}. In particular, besides the extremal black holes that uplift to the WCFT phase space, the DWCFT phase space allows us to uplift near extremal and non-extremal black holes with a running dilaton profile. Moreover, the boundary condition inherited from the generalized CSS boundary conditions of the DWCFT phase space provides a consistent Neumann boundary condition for the 2D gauge field away from the extremal horizon. Evaluating the renormalized on-shell action to linear order around an extremal black hole we obtain an effective action of the form \eqref{CSYKeff}, exhibiting the same symmetry breaking pattern as the complex SYK models. The relations among AdS$_3$ gravity with generalized CSS boundary conditions, 2D EMD gravity, DWCFT and the CSYK models is pictorially summarized in fig.~\ref{relations}. 

The paper is organized as follows. In section \ref{sec:phase-space} we develop a systematic framework for classifying consistent boundary conditions for AdS$_3$ gravity. We then review the Compère-Song-Strominger (CSS) boundary conditions \cite{Compere:2013bya} and construct a non-conformal generalization that admits both extremal and non-extremal black hole solutions. In section \ref{sec:Dimreduce} we review the consistent Kaluza-Klein reduction of pure AdS$_3$ gravity to two dimensions and show that the full phase space of dilaton gravity in two dimensions can be embedded into a deformed version of the WCFT phase space of AdS$_3$ gravity. Incorporating the boundary term that imposes generalized CSS boundary conditions, in section \ref{sec:eff-action} we evaluate the renormalized on-shell action on arbitrary dilaton gravity solutions and we demonstrate that near extremality it reproduces the low energy effective action of the complex SYK models in \eqref{CSYKeff}. Moreover, we show that the thermodynamics of near extremal black holes with generalized CSS boundary conditions coincides with that of the complex SYK models. We conclude in section \ref{sec:discussion} with a brief discussion. A number of technical results are presented in two appendices.

\section{3D gravity and generalized CSS boundary conditions}
\label{sec:phase-space}

In this section we discus aspects of three dimensional (3D) Einstein gravity with a negative cosmological constant. In particular, we describe a framework within which one can classify all possible solutions and boundary conditions. We focus especially on the subspace of solutions corresponding to the BTZ black hole with Compère, Song and Strominger (CSS) boundary conditions, first discussed in \cite{Compere:2013bya}. In addition, we discuss a more general class of solutions which can be understood as a certain deformation of the BTZ black hole with CSS boundary conditions.
\subsection{The space of solutions of AdS$_3$ gravity}\label{subsec:21}
The action of Einstein gravity with a negative cosmological constant in three dimensions is 
\be\label{3Daction}
I_{\rm 3D}=\frac{1}{2\k^2_{\rm 3D}}\bigg[\int_{\cm} d^3x\sqrt{-g}\Big(R^{(3)}+{2\over \ell^2}\Big)+2\int_{\pa\cm} d^2x\sqrt{-h}\;K^{(3)}\bigg],
\ee
where $\k^2_{\rm 3D}=8\p G_{\rm 3D}$ is the 3D gravitational constant and $\ell$ is the AdS$_{\rm 3}$ radius. Moreover, the Gibbons-Hawking term (involving the induced metric, $h_{ij}$, on the regulated boundary and the trace, $K^{(3)}=h_{ij}K^{(3)ij}$, of its extrinsic curvature) ensures that the Dirichlet problem on a finite volume manifold, $\cm$, is well posed. As we will review below, additional boundary terms are required to render the variational principle well posed on infinite volume manifolds, and for imposing boundary conditions other than Dirichlet \cite{Papadimitriou:2005ii,Papadimitriou:2010as}. 

All solutions of 3D Einstein gravity with a negative cosmological constant can be conveniently written in the Fefferman-Graham (FG) gauge, 
\be\label{metric_fg}
ds^2=d\r^2+h_{ij}(\r,x)dx^idx^j,\qquad i=+,-,
\ee
where $\r$ denotes the radial coordinate, and $x^{\pm}$ represent the boundary coordinates. In the absence of matter fields, the FG expansion in three dimensions terminates, namely \cite{Skenderis:1999nb}
\be\label{induced-metric}
h_{ij}=e^{2\r/\ell}\big(g\sub{0}_{ij}(x)+e^{-2\r/\ell}g\sub{2}_{ij}(x)+e^{-4\r/\ell}g\sub{4}_{ij}(x)\big),
\ee
where the boundary metric, $g\sub{0}$, is arbitrary, $g\sub{4}=\frac14 g\sub{2}g_{(0)}^{-1}g\sub{2}$, and $g\sub{2}$ satisfies the constraints
\be\label{reduced-constraints}
g_{(0)}^{ij}g\sub{2}_{ij}=-\frac{\ell^2}{2}R[g\sub{0}],\qquad D_{(0)}^i\big(g\sub{2}_{ij}-g_{(0)}^{kl}g\sub{2}_{kl}g\sub{0}_{ij}\big)=0,
\ee
with $D\sub{0}_i$ denoting a covariant derivative with respect to $g\sub{0}_{ij}$. Introducing the tensor  
\be\label{Tij}
\ct_{ij}=\frac{1}{\k^2_{\rm 3D}  \ell}\big(g\sub{2}_{ij}-g_{(0)}^{kl}g\sub{2}_{kl}g\sub{0}_{ij}\big),
\ee
the constraints \eqref{reduced-constraints} can be written in the more suggestive form 
\be\label{2D-constraints}
D\sub{0}_i\ct^i_j=0,\qquad \ct^i_i=\frac{c}{24\p}R[g\sub{0}],\qquad c=\frac{12\p \ell}{\k^2_{\rm 3D}},
\ee
where $c$ is the Brown-Henneaux central charge \cite{Brown:1986nw}. This form of the constraints allows us to identify $\ct_{ij}$ with the stress tensor in a dual 2D CFT. Expressing $g\sub{2}$ and $g\sub{4}$ in terms of $\ct_{ij}$, the metric \eqref{metric_fg}-\eqref{induced-metric} takes the form
\bal
\label{3D-metric}
ds^2=&\;d\r^2+e^{2\r/\ell}\bigg[g\sub{0}_{ij}+2e^{-2\r/\ell}\Big(\frac{\k^2_{\rm 3D}\ell}{2}\ct_{ij}-\frac{\ell^2}{4}R[g\sub{0}]g\sub{0}_{ij}\Big)\NO\\
&\hspace{.65cm}+e^{-4\r/\ell}\Big(\frac{\k^2_{\rm 3D}\ell}{2}\ct_{ik}-\frac{\ell^2}{4} R[g\sub{0}]g\sub{0}_{ik}\Big)
\Big(\frac{\k^2_{\rm 3D}\ell}{2}\ct^k_j-\frac{\ell^2}{4} R[g\sub{0}]\d^k_j\Big)\bigg]dx^idx^j.
\eal
Subject to the constraints \eqref{2D-constraints}, this is the most general solution of the 3D Einstein equations with a negative cosmological constant. It is a more general form of the Ba\~nados metric \cite{Banados:1998gg}.

The space of solutions of the form \eqref{3D-metric} is a symplectic manifold parameterized by the variables $(g_{(0)ij},\hat\p^{ij})$, where $\hat\p^{ij}=\frac12\sqrt{-g\sub{0}}\;\ct^{ij}$ is the renormalized canonical momentum. Notice that $\ct^{ij}$, or equivalently $\hat\p^{ij}$, is not a free symplectic variable, due to the constraints \eqref{2D-constraints} discussed earlier. We will refer to the symplectic space of solutions parameterized by these variables as the `reduced phase space'. Its symplectic form is \cite{Papadimitriou:2010as}
\be\label{S-form}
\Om=\int_{\pa\cm} d^2x\;\d\hat\p^{ij}\wedge\d g_{(0)ij},
\ee
which leads to a Poisson bracket between the conjugate variables, namely
\be\label{P-bracket}
\big\{\hat\p^{ij}(x),g_{(0)kl}(x')\big\}=\d^{(i}_{k}\d^{j)}_{l}\d^{(2)}(x-x').
\ee
This acts on arbitrary functionals $A[\hat\p,g\sub{0}]$ and $B[\hat\p,g\sub{0}]$ on the reduced phase space as
\be\label{P-bracket1}
\big\{A[\hat\p,g\sub{0}],B[\hat\p,g\sub{0}]\big\}=\int d^2x\Big(\frac{\d A}{\d\hat\p^{ij}}\frac{\d B}{\d g_{(0)ij}}-\frac{\d B}{\d\hat\p^{ij}}\frac{\d A}{\d g_{(0)ij}}\Big).
\ee

\subsubsection{Reduced phase space symmetries}

The FG gauge \eqref{metric_fg}-\eqref{induced-metric} breaks the 
Diff$_{\rm 3D}$ symmetry of 3D gravity to the subgroup Diff$_{\rm 2D}\times$Weyl. These residual symmetry transformations correspond to bulk diffeomorphisms that leave the metric \eqref{metric_fg} or \eqref{3D-metric} form-invariant and are known as Penrose-Brown-Henneaux (PBH) diffeomorphisms \cite{penrose_rindler_1984,Brown:1986nw,Imbimbo:1999bj}. Their infinitesimal form can be determined as follows.

An infinitesimal 3D diffeomorphism, $\d x^\m=-\xi^{\m}(\r,t)$, preserves the FG gauge provided
\bal
\label{pbh-diff-cond}
\d_{\xi}g_{\r\r}=&\;\cL_{\xi}g_{\r\r}=\dot{\xi}^{\r}=0,\nn\\
\d_{\xi}g_{\r i}=&\;\cL_{\xi}g_{\r i}=h_{ij}\dot{\xi}^{j}+\pa_{i}\xi^{\r}=0.\eal
These equations imply that the radial and transverse components of the vector field $\xi^{\m}(\r,t)$ take respectively the form
\bal\label{pbh-gen}
\xi^{\r}=&\;\s(x),\nn\\
\xi^{i}=&\;\xi_o^i(x)+\pa_{j}\s(x)\int_{\r}^{\infty}d\r'\;h^{ij}(\r',x)\NO\\
=&\;\xi_o^i(x)+\frac{\ell}{2}\;e^{-2\r/\ell}\Big(g_{(0)}^{ij}-\frac{1}{2}e^{-2\r/\ell}g_{(2)}^{ij}+\co\big(e^{-4\r/\ell}\big)\Big)\pa_{j}\s(x),
\eal
where $\s(x)$ and $\xi^{i}_{o}(x)$ are arbitrary functions of the boundary coordinates. 

Applying the bulk diffeomorphism \eqref{pbh-gen} to the transverse components $g_{ij}$ of the bulk metric, one finds that the symplectic variables parameterizing the reduced phase space transform as 
\bal\label{pbh-trans}
\d_{\xi}g\sub{0}_{ij}=&\;D\sub{0}_{i}\xi_{oj}+D\sub{0}_{j}\xi_{oi}+\frac{2}{\ell}\s(x)g\sub{0}_{ij},\NO\\
\d_{\xi}\ct_{ij}=&\;\ct_{il}D\sub{0}_{j}\xi_{o}^{l}+\ct_{jl}D\sub{0}_{i}\xi_{o}^{l}+\xi_{o}^{l} D\sub{0}_{l}\ct_{ij}+\frac{1}{\k^2_{\rm 3D}}\(D\sub{0}_{i}D\sub{0}_{j}\s-g\sub{0}_{ij}\Box\sub{0}\s\).
\eal
The action of these diffeomorphisms on the boundary metric $g\sub{0}$ makes it manifest that the residual local symmetry on the reduced phase space is Diff$_{\rm 2D}\times$Weyl. Moreover, the symplectic variable $\ct_{ij}$ possesses an anomalous transformation under local 2D Weyl transformations, in agreement with its identification with the stress tensor of a dual CFT$_2$ with the Brown-Henneaux central charge.

\subsection{Unconstrained phase space and the variational principle} 

A well posed variational principle on infinite volume manifolds requires further boundary terms, besides the Gibbons-Hawking term. For asymptotically locally AdS manifolds, the additional terms ensure that the variational problem is formulated in terms of boundary conformal equivalence classes and coincide with the boundary counterterms that render the on-shell action finite \cite{Papadimitriou:2005ii,Papadimitriou:2010as}. For AdS$_3$ gravity, the boundary counterterms take the form \cite{Henningson:1998gx}
\be\label{3DactionCT}
I^{\rm ct}_{\rm 3D}= -\frac{1}{\k^2_{\rm 3D}}\int_{\pa\cm_{\r_c}} d^2x\sqrt{-h}\;\Big(\frac{1}{\ell}-\frac{\ell}{4}\log(e^{-2\r_c/\ell})R[h]\Big),
\ee
where $\r_c$ is the radial cutoff, and ensure that the renormalized action  
\be\label{3DactionRen}
I^{\rm ren}_{\rm 3D} \equiv \lim_{\r_c\to\infty}\big(I_{\rm 3D}^{\rm reg}+I^{\rm ct}_{\rm 3D}\big),\qquad I_{\rm 3D}^{\rm reg}\equiv \left.I_{\rm 3D}\right|_{\r_c},
\ee
is finite and admits a well defined variational principle as $\r_c\to\infty$, namely
\be\label{var-3Daction}
\d I^{\rm ren}_{\rm 3D}=\int_{\cm}d^3x\;\sqrt{-g}\,(e.o.m.)\,\d g_{\m\n}-\frac 12\int_{\pa\cm} d^2x\;\sqrt{-g\sub{0}}\;\ct_{ij}\;\d g_{(0)}^{ij}.
\ee

The variational principle \eqref{var-3Daction} is well posed provided a Dirichlet boundary condition is imposed on the boundary metric $g\sub{0}$. However, alternative boundary conditions are possible for AdS$_3$ gravity and they generically correspond to keeping fixed a combination of the symplectic variables $(g_{(0)ij},\hat\p^{ij})$. Generalized boundary conditions are in one to one correspondence with additional {\em finite} boundary terms that depend on the reduced phase space variables $(g_{(0)ij},\hat\p^{ij})$ and are added to the renormalized Dirichlet action \eqref{3DactionRen} \cite{Papadimitriou:2007sj}. 

The constraints \eqref{2D-constraints} for AdS$_3$ gravity, however, imply that the form of the finite boundary terms that lead to a well defined variational problem, other than the Dirichlet one, is highly restricted. For example, the naive Neumann problem corresponding to adding a term of the form 
\be
\int_{\pa\cm} d^2x\,\Hat\p^{ij}g_{(0)ij},
\ee 
to \eqref{3DactionCT} is not well posed, since $\Hat\p^{ij}$ is a constrained variable. To remedy this, we will solve the constraints \eqref{2D-constraints} explicitly in terms of unconstrained symplectic variables, before considering alternative boundary conditions. Solving the constraints explicitly typically requires breaking general covariance, and hence the unconstrained phase space only preserves a subspace of the Diff$_{\rm 2D}\times$Weyl symmetry of the reduced phase space parameterized by $(g_{(0)ij},\hat\p^{ij})$. This in turn implies that the asymptotic symmetry algebra preserved by alternative boundary conditions is necessarily a subset of the maximal asymptotic symmetry preserved by Dirichlet boundary conditions.

In order to solve the constraints \eqref{2D-constraints} and obtain the unconstrained space of solutions of AdS$_{3}$ gravity, it is convenient to introduce light-cone coordinates on the boundary, $x^{\pm} = t \pm \f$, with $\f \sim \f + 2\pi$. Without loss of generality, the boundary metric can then be parameterized as  
\be\label{BD-metric}
ds^2_{(0)}=g\sub{0}_{ij}dx^idx^j=-e^{2\Phi(x^{+},x^-)}\big(dx^{+}+\m^+(x^{+},x^-)dx^{-}\big)\big(dx^{-}+\m^-(x^{+},x^-)dx^{+}\big),
\ee
where $\Phi(x^{+},x^-)$, $\m^-(x^{+},x^-)$ and $\m^+(x^{+},x^-)$ are arbitrary functions. In terms of these variables, the variation \eqref{var-3Daction} becomes (ignoring the bulk term and using the trace constraint in \eqref{2D-constraints})
\bal\label{var-3Daction3}
\d I^{\rm ren}_{\rm 3D}=&\;-\frac12\int_{\pa\cm} d^2x\Big[\frac{2}{|1-{{\m^+}}{{\m^-}}|}\Big((\ct_{++}-{{\m^-}}\ct_{+-})\d{{\m^+}}+(\ct_{--}-{{\m^+}}\ct_{+-})\d{{\m^-}}\Big)\NO\\
&\;\hskip7.0cm-\frac{c}{24\p}e^{\F} R[g\sub{0}]\d\big(e^{\F}|1-{{\m^+}}{{\m^-}}|\big)\Big].
\eal

The unconstrained phase space is obtained by determining the most general form of the stress tensor that solves the constraints \eqref{2D-constraints} on the metric \eqref{BD-metric}. Such a general analysis is not required for the purposes of the present paper, however. Instead, we will focus on two special classes of solutions of the constraints \eqref{2D-constraints} that are relevant for describing the phase space of 2D dilaton gravity in the subsequent sections. 

\subsubsection{Generalized Warped CFT solution space}\label{subsec:RF-GWCFT-sol}

A class of metrics \eqref{BD-metric} that allows us to solve the constraints \eqref{2D-constraints} is specified by setting  
\be\label{GWCFT-BC}
\Phi=\Phi(x^{+}),\qquad \m^-=\m^-(x^{+}),\qquad  \m^+=0,
\ee
where $\Phi(x^{+})$, $\m^-(x^{+})$ are arbitrary functions of $x^+$. An important property of this specific choice is that the corresponding boundary metric \eqref{BD-metric} is Ricci flat,\footnote{The Ricci flatness condition is more generally satisfied for $\Phi(x^+,x^-)=\Phi(x^{+})+ax^-$, where $a$ is an arbitrary constant that we set to zero in this paper. Clearly, $\Phi(x^+,x^-)=\Phi(x^{-})+ax^+$, along with $\m^+=\m^+(x^-)$ and $\m^-=0$, is an equivalent solution obtained by interchanging $x^+$ and $x^-$.} i.e. $R[g\sub{0}] = 0$, and so $\d\Phi$ drops out of the variational principle \eqref{var-3Daction3}. In particular, the conformal anomaly is numerically zero and so the variational problem is defined on conformal equivalence classes \cite{Papadimitriou:2005ii,Papadimitriou:2010as}. A second noteworthy property of the choice \eqref{GWCFT-BC} is that it manifestly violates Lorentz invariance, as well as parity. In contrast to the flat Minkowski metric, therefore, it is not compatible with Brown-Henneaux boundary conditions. As we will discuss shortly, appropriate boundary conditions in this case include the CSS boundary conditions \cite{Compere:2013bya}.  

Writing the constraints \eqref{2D-constraints} in terms of the metric \eqref{BD-metric} and imposing the conditions \eqref{GWCFT-BC} leads to a set of three equations for the components of the stress tensor, namely
\be\label{GWCFT-constraints}
\ct_{+-}=\m^-\ct_{--},\qquad (\pa_{+}-\m^-\pa_{-})\ct_{--}=0,\qquad \pa_{-}\big(\ct_{++}-(\mu^-)^2\ct_{--}\big)=0.
\ee
Introducing the new coordinate \be z^-\equiv x^{-}+\int^{x^+} dx^+ \m^-(x^{+}),\,\label{zminus}\ee
the general solution of these equations can be parameterized as
\bal\label{GWCFT-T}
\ct_{--}=&\; \cl_-(z^-),\nn\\
\ct_{+-}=&\; \m^{-}(x^{+})\cl_-(z^-),\nn\\
\ct_{++}=&\; {{\wt\cl}}_+(x^{+})+(\m^-(x^{+}))^{2}\cl_-(z^-),
\eal 
where $\cl_-$ and ${{\wt\cl_+}}$ are arbitrary functions of their arguments. 

In the special case $\Phi=0$ and $\cl_-=\text{constant}$, the space of solutions parameterized by these variables coincides with that discussed in the context of CSS boundary conditions for AdS$_3$ gravity \cite{Compere:2013bya} and warped CFT (WCFT) \cite{Detournay:2012pc}. The more general case described here, which we will refer to as the generalized WCFT (GWCFT) space of solutions, has been discussed in connection with the $J\lbar T$ deformation of 2D CFTs in \cite{Bzowski:2018pcy}. However, neither the WCFT nor the GWCFT classes of solutions constitute a well defined phase space with non-degenerate symplectic form.

Inserting the GWCFT solution \eqref{GWCFT-BC}-\eqref{GWCFT-T} in the variational principle \eqref{var-3Daction3} leads to 
\bal\label{var-GWCFT}
\d I^{\rm ren}_{\rm 3D}=&\; -\int_{\pa\cm} d^2x\;\big(\ct_{--}\d\m^-+(\ct_{++}-(\m^-)^2\ct_{--})\d\m^+\big)\NO\\
=&\;-\int_{\pa\cm} d^2x\;\big(\cl_-\d\m^-+{{\wt\cl_+}}\d\m^+\big),
\eal
where we have allowed for infinitesimal deformations of ${{\m^+}}$ away from the GWCFT value ${{\m^+}}=0$. Crucially, the variational principle is now formulated in terms of the unconstrained symplectic conjugate variables $({{\m^-}},\cl_-)$ and $({{\m^+}},{{\wt\cl_+}})$. As we will discuss in the next subsection, this will allow us to consider boundary conditions other than Dirichlet. It is clear from \eqref{var-GWCFT}, however, that in order to obtain a well defied phase space with a non-degenerate symplectic form, it is necessary to allow for a finite, non-zero ${{\m^+}}$, which is the task we turn to next.

\subsubsection{Deformed Warped CFT phase space}

As we have seen, the GWCFT space of solutions does not admit a non-degenerate symplectic form. We will see in the next section that it is also insufficient for describing the most general solutions of the 2D dilaton gravity obtained by a circle reduction of AdS$_3$ gravity. Both these issues can be addressed by a more general space of solutions, corresponding to turning on a finite function ${{\m^+}}(x^+)$, so that the boundary metric \eqref{BD-metric} takes the form    
\be
\label{BD-metricGEN}
ds^2_{(0)}=-e^{2\Phi(x^{+})}(dx^{+}+\m^+(x^{+})dx^{-})(dx^{-}+\m^-(x^{+})dx^{+}).
\ee
When ${{\m^+}}(x^+)\neq 0$, this metric is not Ricci flat and corresponds to a deformation of the GWCFT boundary metric. Moreover, the metric \eqref{BD-metricGEN} may restore parity invariance and the local parameter ${{\m^+}}(x^+)$ can interpolate between a WCFT and a regular CFT$_2$. As we will see later on, the uplift of the 2D dilaton gravity solutions provides an explicit realization of such an interpolating flow. 

\paragraph{Ricci scalar}
The Ricci scalar of the metric \eqref{BD-metricGEN} is given by
\be\label{Ricci}
R[g\sub{0}]=-\frac{4e^{-2\F}}{1-{{\m^+}}{{\m^-}}}\pa_+\mathscr F,
\ee
where
\be\label{anomaly}
\mathscr F(x^+)\equiv \frac{e^{-2\F}}{1-{{\m^+}}{{\m^-}}}\pa_{+}\big(e^{2\F}{{\m^+}}\big).
\ee

\paragraph{Stress tensor} 
Similar to the discussion for the WCFT phase space, the constraints \eqref{2D-constraints} on the metric \eqref{BD-metricGEN} can be solved using the following nonlocal coordinates,
\be\label{zs}
z^-\equiv x^-+\int^{x^+} dx^+{{\m^-}}(x^+), \qquad z^+\equiv x^-+\int^{x^+} dx^+(\m^+(x^+))^{-1}.
\ee
The most general stress tensor that solves \eqref{2D-constraints} can then be written as 
\bal\label{DWCFT-stress-tensor}
&\ct_{++}=\;- {{\bar\co}_{+}}-(\m^-)^2{{\bar\co}_{-}}+\frac{c{{\m^-}}}{24\p}\Big(\frac{3+{{\m^+}}{{\m^-}}}{1-{{\m^+}}{{\m^-}}}\Big)\pa_+\mathscr F ,\NO\\
&\ct_{--}=\;-{{\bar\co}_{-}}-(\m^+)^2{{\bar\co}_{+}}+\frac{c{{\m^+}}}{24\p}\Big(\frac{3+{{\m^+}}{{\m^-}}}{1-{{\m^+}}{{\m^-}}}\Big)\pa_+\mathscr F,
\NO\\
&\ct_{-+}=\;-{{\m^-}}{{\bar\co}_{-}}-{{\m^+}}{{\bar\co}_{+}}+\frac{c}{24\p}\Big(\frac{1+3{{\m^+}}{{\m^-}}}{1-{{\m^+}}{{\m^-}}}\Big)\pa_+\mathscr F,
\eal
where 
\bal 
\label{Jvariables}
{{\bar\co}_{-}}\equiv \; -\cl_-(z^-)+\frac{c}{48\p}\mathscr{F}^2,\qquad 
{{\bar\co}_{+}}\equiv \;-(\m^+)^{-2}\Big(\cl_+(z^+)-\frac{c}{48\p}\big(\mathscr F^2-2{{\m^+}}\pa_+\mathscr F\big)\Big)\, , 
\eal
and $\cl_-$ and $\cl_+$ are arbitrary functions of their arguments, generalizing the definitions of \eqref{GWCFT-T}.
The physical meaning and significance of the variables $\cl_\pm,\,\bar\co_\pm$ will be clear in the next subsection. 

Note that in the limit $\m^+\to0$, $z^+$ effectively only depends on $x^+$, 
$\mathscr F\to0$, and \eqref{DWCFT-stress-tensor} reduces to \eqref{GWCFT-T} provided that 
\be\label{mp0}
{{\bar\co}_{+}} =-(\m^+)^{-2}\Big(\cl_+(z^+)-\frac{c}{48\p}\big(\mathscr F^2-2{{\m^+}}\pa_+\mathscr F\big)\Big)  =-\wt\cl_+(x^+)+\co(\m^+),
\ee 
As we discuss in section 3, this condition is the near extremality condition for black holes and is also required for a running dilaton solution to flow to a constant dilaton solution in the two dimensional Einstein-Maxwell-Dilaton gravity. We refer to the phase space parameterized by the boundary metric \eqref{BD-metricGEN} and the stress tensor \eqref{DWCFT-stress-tensor} as the Deformed Warped CFT (DWCFT) phase space. As we now discuss, it contains both the GWCFT space of solutions and 2D dilaton gravity as subspaces. Moreover, it admits a non-degenerate symplectic form and a well defined variational principle.

\subsubsection{Variational principle on the DWCFT phase space}

The DWCFT phase space allows us to formulate the variational principle in terms of unconstrained variables. From the solution \eqref{DWCFT-stress-tensor}, we note that  the following relations hold: \bal\label{T-relations}
&\ct_{--}-{{\m^+}}\ct_{+-}=\;-(1-{{\m^-}}{{\m^+}}){{\bar\co}_{-}}+\frac{c}{12\p}{{\m^+}}\pa_+\mathscr F,\NO\\
&\ct_{++}-{{\m^-}}\ct_{+-}=\;-(1-{{\m^-}}{{\m^+}}){{\bar\co}_{+}}+\frac{c}{12\p}{{\m^-}}\pa_+\mathscr F.
\eal
Inserting these, together with the metric \eqref{BD-metricGEN}, in \eqref{var-3Daction3} leads to the variational principle 
\be\label{var-DWCFT}
\d I^{\rm ren}_{\rm 3D}=-\int_{\pa\cm} d^2x\,\big(\cl_-\d{{\m^-}}+(\m^+)^{-2}\cl_+\d\m^+\big)+\d I\sbtx{local},
\ee
where $I\sbtx{local}$ is the local expression  
\be\label{Ilocal}
I\sbtx{local}
=\frac{c}{24\p}\int d^2x\,\frac{1-{{\m^+}}{{\m^-}}}{2{{\m^+}}}\mathscr F^2.
\ee
This local part of the renormalized on-shell action is directly related with the conformal anomaly of the 2D CFT at the boundary of AdS$_3$. Moreover, it is precisely the term that gives rise to the Schwarzian effective action upon a circle reduction \cite{Cvetic:2016eiv}. As we will see later on, this term is also the origin of the Schwarzian effective action in the complex SYK model.

Another important observation regarding $I\sbtx{local}$ is that it drops out of the symplectic form. This can be evaluated either by inserting the DWCFT solution \eqref{BD-metricGEN} and \eqref{DWCFT-stress-tensor} in \eqref{S-form}, or through a second variation of $\d I^{\rm ren}_{\rm 3D}$ in \eqref{var-DWCFT}. In the latter approach, $\d I^{\rm ren}_{\rm 3D}$ can be interpreted as the canonical one-form or symplectic potential on the DWCFT symplectic space of solutions and so, its phase space exterior derivative is the symplectic  form
\be\label{SymplecticForm-DWCFT}
\Om=-\int_{\pa\cm} d^2x\,\big(\d\cl_-\wedge\d{{\m^-}}+(\m^+)^{-2}\d\cl_+\wedge\d{{\m^+}}\big).
\ee
Since $I\sbtx{local}$ is local, $\d I\sbtx{local}$ is an exact phase space one-form and drops out of  the symplectic form.

Although $I\sbtx{local}$ does not contribute to the symplectic form, it affects the variational principle. A general variation of $I\sbtx{local}$ takes the from
\be
\d I\sbtx{local}=\frac{c}{48\p}\int d^2x\, \big(\mathscr F^2\d{{\m^-}}+(\m^+)^{-2}(\mathscr F^2-2{{\m^+}}\pa_+\mathscr F)\d{{\m^+}}-4\pa_+\mathscr F\d\F\big).
\ee
This expression is consistent with the general result that in the presence of a conformal anomaly (corresponding to ${{\m^+}}\neq 0$ or $\mathscr F\neq 0$ here), the variational problem can only be defined by keeping the conformal representative of the boundary metric fixed \cite{Papadimitriou:2005ii}. In particular, whenever ${{\m^+}}\neq 0$, we must impose the condition $\d\F=0$ in order to obtain a well posed variational problem. 

Except for the term proportional to $\d\F$,  the variation of $I\sbtx{local}$ can be absorbed by a redefinition of the canonical variables. In particular, the variational principle \eqref{var-DWCFT} takes the form
\be\label{var-DWCFT-new}
\d I^{\rm ren}_{\rm 3D}=\int_{\pa\cm} d^2x\,\Big({{\bar\co}_{-}}\d{{\m^-}}+{{\bar\co}_{+}}\d{{\m^+}}-\frac{c}{12\p}\pa_+\mathscr F\d\F\Big),
\ee
where ${{\bar\co}_{-}}$ and ${{\bar\co}_{+}}$ are the variables introduced in \eqref{Jvariables}. 
In terms of these variables, the symplectic form \eqref{SymplecticForm-DWCFT} becomes 
\be\label{SymplecticForm-DWCFT-new}
\Om=\int_{\pa\cm} d^2x\,\big(\d{{\bar\co}_{-}}\wedge\d{{\m^-}}+\d{{\bar\co}_{+}}\wedge\d{{\m^+}}\big),
\ee
since the difference between the original and new canonical variables is due to $I\sbtx{local}$ and so does not contribute. We have therefore arrived at a variational principle in terms of the unconstrained symplectic variables $({{\m^-}},{{\bar\co}_{-}})$ and $({{\m^+}},{{\bar\co}_{+}})$ that parameterize a well defined phase space with a non-degenerate symplectic form.  This enables us to impose alternative boundary conditions for this class of solutions, as we discuss in the next subsection.

\subsubsection{Spectral flow}

The variational principle can be modified through a change of variables that preserves the symplectic form \eqref{SymplecticForm-DWCFT-new}. Generic canonical transformations correspond to the addition of a further finite boundary term to the renormalized action (i.e. to a shift of the symplectic potential one-form by an exact form) and can be used in order to classify the admissible boundary conditions. However, there exists a class of canonical transformations that preserve not only the symplectic form, but also the variational principle, i.e the symplectic potential. In other words, there exist canonical transformations that whose corresponding boundary term is identically zero.  

We start with the canonical variables 
\be
{\bar\cj}^-\equiv{{\m^-}},\qquad {{\bar\co}_{-}},\qquad {\bar\cj}^+\equiv{{\m^+}},\qquad {{\bar\co}_{+}}, 
\ee
with the variational principle \eqref{var-DWCFT-new}.
A canonical transformation $\{\bar\cj^i,{{\bar\co}_{i}}\}\to\{\cj^i,{\co}_{i}\},$ 
preserves this symplectic potential provided the two sets of variables satisfy the relations 
\bal\label{flow-generic}
&\co_j\frac{\pa\cj^j}{\pa{\bar{\cj}}^i}=\bar{\co}_{i},
\quad{{\co}_{j}}\frac{\pa\cj^j}{\pa {{\bar\co}_{i}}}=0,
\eal
where the indices $i,\,j$ take the values $+,\,-$, and a summation is assumed whenever the same lower and upper indices appear. Such canonical transformations, therefore, correspond to a spectral flow.    

In the present context, we wish to treat $\m^+$ as a parameter and so we consider canonical transformations that preserve $\m^+$, i.e. 
\be 
\cj^+=\m^+.\label{tildetau}
\ee
In this case, the relations \eqref{flow-generic} determine that $\cj^-$ is independent of ${{\bar\co}_{-}}$ and ${{\bar\co}_{+}}$, and hence
$\cj^-=\cj^-(\mu^-,\mu^+)$. We will further take the  following ansatz for the function $\cj^-({{\m^-}},{{\m^+}})$
\be\label{RG-function}
\cj^-({{\m^-}},{{\m^+}})=\rg(\l){{\m^-}},\qquad \rg(0)=1,
\ee
where $\rg(\l)$ is an arbitrary function satisfying $\rg(0)=1$ and $\l\equiv{{\m^+}}{{\m^-}}$ is a dimensionless composite scalar coupling. Plugging \eqref{tildetau} and \eqref{RG-function} into \eqref{flow-generic}, the new variables are then determined in terms of the arbitrary function $\rg(\l)$ and the original variables by
\be\label{flow}\boxed{
\cj^+={{\m^+}},\qquad\cj^-=\rg{{\m^-}},\qquad {{\co}_{-}}=\frac{{{\bar\co}_{-}}}{(\l\rg)'},\qquad 
{\co}_{+} ={{\bar\co}_{+}}-\frac{\rg'}{(\l\rg)'}{(\m^-)}^2{{\bar\co}_{-}}.}
\ee
where the prime denotes a derivative with respect to $\lambda$. 
The significance of the function $\rg(\l)$ will become clear in the subsequent analysis.

\subsubsection{DWCFT phase space symmetries }

Before discussing admissible boundary conditions, we need to identify the symmetries of the DWCFT phase space. These correspond to the subset of PBH transformations \eqref{pbh-trans} preserving the form of the DWCFT metric \eqref{BD-metricGEN} and stress tensor \eqref{DWCFT-stress-tensor}. Under PBH diffeomorphisms with parameters $\x_o^\pm(x^+, x^-)$ and $\s(x^+, x^-)$, the functions parameterizing the metric \eqref{BD-metricGEN} transform as  
\bal\label{PBH-coords}
\d\sbtx{PBH}{\m^-}=&\;(\pa_+-{{\m^-}}\pa_-)(\x_o^-+{{\m^-}}\x_o^+)
,\NO\\
\d\sbtx{PBH}(\m^+)^{-1}=&\;(\pa_+-(\m^+)^{-1}\pa_-)(\x_o^-+(\m^+)^{-1}\x_o^+)
,\NO\\
\d\sbtx{PBH}\F=&\;
\frac{\s}{\ell}+\frac{1}{2} \big((\pa_++{{\m^-}} \pa_-)\x_o^++(\pa_-+{{\m^+}} \pa_+)\x_o^-+2 \Phi ' \x_o^+\big).
\eal

Inserting these transformations in the variational principle \eqref{var-DWCFT-new} and using the invariance of $I^{\rm ren}_{\rm 3D}$ under 2D diffeomorphisms one may derive the two Ward identities (cf. (6.9) in \cite{Cvetic:2016eiv}) 
\be\label{WIds}
(\pa_+-{{\m^-}}\pa_-)\cl_-=0,\qquad (\pa_+-(\m^+)^{-1}\pa_-)\cl_+=0,
\ee
require that $\cl_-$ and $\cl_+$ are functions respectively of the variables $z^-$ and $z^+$, defined in \eqref{zs}. Of course, this is expected and is in agreement with the solution \eqref{DWCFT-stress-tensor} of the constraints \eqref{2D-constraints}. 

The transformations \eqref{PBH-coords} preserve the form of the metric \eqref{BD-metricGEN} provided 
\be\label{conditions}
\pa_-\d\sbtx{PBH}{{\m^-}}=\pa_-\d\sbtx{PBH}{{\m^+}}=\pa_-\d\sbtx{PBH}\F=0.
\ee
The first two of these conditions determine that the parameters $\x_o^\pm(x^+, x^-)$ must satisfy
\be
\x_o^-+{{\m^-}}\x_o^+=\vf(x^+)+{{\m^-}}{{\ve}}(x^+)+{\e^-}(z^-),\qquad \x_o^-+(\m^+)^{-1}\x_o^+=(\m^+)^{-1}{{\ve}}(x^+)+{\e^+}(z^+),
\ee
where $\vf,\,\ve$ and ${\e^\pm}$ are arbitrary functions of their arguments. Combining these with the third condition in \eqref{conditions}, we determine that the DWCFT phase space symmetries are parameterized by five arbitrary functions, ${{\ve}}(x^+),\, \vf(x^+)$, ${\e^-}(z^-)$, ${\e^+}(z^+)$ and $\om(x^+)$, and correspond to PBH transformations with parameters 
\bal\label{resPBH}
\x_o^+(x^+, x^-)=&\;\ve-{\m^+\over 1-\lambda}(\vf+{\e^-}-{\e^+}),\NO\\
\x_o^-(x^+, x^-)=&\;{1\over 1-\lambda}({\vf+\e^--\lambda \e^+}),\NO\\
\frac{2}{\ell}\s(x^+, x^-)=&\;({\e^-}-{\e^+})\mathscr F-\big((\e^-)'+(\e^+)'\big)+\om.
\eal

Inserting these parameters in the transformations \eqref{PBH-coords} we determine that the residual DWCFT symmetries act on the functions parameterizing the metric \eqref{BD-metricGEN} as
\bal\label{resPBH-coords}
&\d\sbtx{DWCFT}{{\m^-}}=\pa_+(\vf+{{\m^-}}{{\ve}}),\qquad
\d\sbtx{DWCFT}{{\m^+}}={{\ve}}\pa_+{\m^+}-\pa_+{\ve}\,{{\m^+}},
\NO\\
&2\d\sbtx{DWCFT}\F=-\vf\mathscr F+e^{-2\F}\pa_+(e^{2\F}{{\ve}})+\om.
\eal
Notice that ${\e^-}$ and ${\e^+}$ drop out of these transformations and so the metric continues to depend only on $x^+$, as required. Similarly, from the general PBH transformation of the stress tensor in \eqref{pbh-trans}, we find that the functions $\cl_-$ and $\cl_+$ parameterizing the stress tensor \eqref{DWCFT-stress-tensor} transform as
\bal\label{resPBH-G}
\d\sbtx{DWCFT}\cl_-=&\;(\vf+{{\m^-}}{{\ve}}+{\e^-})\cl_-'+2(\e^-)'\cl_--\frac{c}{24\p}(\e^-)''',\NO\\
\d\sbtx{DWCFT}\cl_+=&\;\big((\m^+)^{-1}{{\ve}}+{\e^+}\big)\cl'_++2(\e^+)'\cl_+-\frac{c}{24\p}(\e^+)'''.
\eal

One may wonder why these transformations do not depend only on $z^\pm$.
This is because they are to be understood as functional variations of $\cl_\pm$ as functions of $x^\pm$, through $z^\pm(x^-,x^+)$, i.e. 
\be 
\d\sbtx{DWCFT}\cl_\pm=\d_{z}\cl_\pm+\d z^i \pa_{z^i}\cl_\pm, 
\ee
where $\d\sbtx{DWCFT}$ denotes the functional variation in terms of the DWCFT phase space coordinates $x^\pm$, and $\d_{z}$ denotes the functional variation as a function of $z^\pm$. Due to the field dependence of the composite coordinates $z^\pm(x^-,x^+)$ introduced in \eqref{zs}, the functional variation of $z^\pm$ in the $x^\pm$ coordinate system is non-zero, e.g.
\be
\d z^-=\int d x^+\d\sbtx{DWCFT}{{\m^-}}=\vf+{{\m^-}}{{\ve}},
\ee
which leads to the term $(\vf+{{\m^-}}{{\ve}})\cl_-'$ in the transformation of $\cl_-$ in \eqref{resPBH-G}. 
The same applies to the term $(\m^+)^{-1} {{\ve}}\cl_+'$ in the transformation of $\cl_+$.
Hence, the variation of $\cl_\pm$ as functions of $z^\pm$ are
\bal\label{resPBH-Gz}
\d_{z}\cl_-=&\;\e^- \cl'_-+2(\e^-)'\cl_--\frac{c}{24\p}(\e^-)''',\NO\\
\d_{z}\cl_+=&\;\e^+\cl'_++2(\e^+)'\cl_+-\frac{c}{24\p}(\e^+)'''.
\eal

On the other hand, the non-local coordinates $z^\pm$ also contain explicit dependence on the coordinates $x^\pm$. The DWCFT transformations \eqref{resPBH} correspond to a total variation of $z^\pm$, e.g. 
\be
\d^{\rm tot} z^-=-(\x_o^-+{{\m^-}}\x_o^+)+\int d x^+\d\sbtx{DWCFT}{{\m^-}}=-{\e^-}(z^-),
\ee
and, similarly, $\d^{\rm tot} z^+=-{\e^+}(z^+)$. Therefore, in the $z^\pm$ coordinate system, the transformations of $\cL_\pm$ become the standard ones for a holomorphic stress tensor in the variables $z^-$ and $z^+$, except that these variables depend on, respectively, ${{\m^-}}$ and ${{\m^+}}$. 
The resulting symmetry algebra is similar to the non-local Virasoro in the context of $T\bar T$ \cite{Guica:2019nzm} and $J\bar T$ \cite{Bzowski:2018pcy} deformations. We would like to further explore the relation in future work. 

The transformations \eqref{resPBH-coords} and \eqref{resPBH-G} determine those of the spectrally flowed variables \eqref{flow} under the residual DWCFT phase space symmetries, which are given in appendix \ref{sec:transformations}.

\subsection{Generalized CSS boundary conditions and symmetry algebra}

In the previous subsection we constructed a non-degenerate symplectic space of solutions of the constraints \eqref{2D-constraints}, and obtained a well posed variational principle in terms of unconstrained symplectic data. Moreover, we saw that there exists a class of canonical transformations that preserve the variational principle. We now discuss canonical transformations that modify the variational principle, allowing for more general boundary conditions. We will see that the boundary conditions relevant for describing the CSYK model are a generalization of the CSS boundary conditions \cite{Compere:2013bya}.  

Starting with the spectrally flowed variables \eqref{flow}, the variational principle takes the form 
\be\label{var-DWCFT-flow}
\d I^{\rm ren}_{\rm 3D}=\int_{\pa\cm} d^2x\,\Big({{\co}_{-}}\d\cj^-+{\co}_{+}\d\cj^+-\frac{c}{12\p}\pa_+\mathscr F\d\F\Big).
\ee
This can be modified by adding an arbitrary boundary term 
\be\label{canonical}
I^{\rm ren}_{\rm 3D}\to I'^{\rm ren}_{\rm 3D}=I^{\rm ren}_{\rm 3D}+\int_{\pa\cm}d^2x\; \mathscr K(\cj^i,{{\co}_{i}}),
\ee
for some function $\mathscr K$. More generally, $\mathscr K$ may contain a finite number of derivatives with respect to the boundary coordinates, $x^\pm$, and can depend on $\F$ as well. The argument given earlier for $I\sbtx{local}$ in \eqref{Ilocal} implies that the boundary term $\mathscr K$ does not affect the symplectic form \eqref{SymplecticForm-DWCFT}, and so it corresponds to the generating function of a canonical transformation within the DWCFT symplectic space, parameterized by the variables $\{\cj^i, \, {\co}_{i}\}, \, i=-,+$ \cite{Papadimitriou:2010as}. This is analogous to mixed boundary conditions for scalars with an AdS mass in the window where both modes are normalizable \cite{Papadimitriou:2007sj}. However, a generic $\mathscr K$ preserves a subset of the DWCFT phase space symmetries, if any at all. 

Under the canonical transformation generated by $\mathscr K$, the variational principle is modified as
\bal\label{var-DWCFT-new-tranformed}
\d I'^{\rm ren}_{\rm 3D}=&\;\int_{\pa\cm} d^2x\,({\cp_-}\d{\cq^-}+{\cp_+}\d{\cq^+})
=\int_{\pa\cm} d^2x\,\Big[\Big(\frac{\pa \mathscr K}{\pa \cj^i}+{{\co}_{i}}\Big)\d\cj^i+\frac{\pa \mathscr K}{\pa {{\co}_{i}}}\d{{\co}_{i}}\Big],
\eal
where the new symplectic variables, $({\cq^i},{\cp_i})$, are related to the original ones, $(\cj^i,{{\co}_{i}})$, as
\be\label{canonical-variables}
{\cp_j}\frac{\pa{\cq^j}}{\pa \cj^i}=\frac{\pa \mathscr K}{\pa \cj^i}+{{\co}_{i}},\qquad
{\cp_j}\frac{\pa{\cq^j}}{\pa {{\co}_{i}}}=\frac{\pa \mathscr K}{\pa {{\co}_{i}}}.
\ee
The spectral flow equations \eqref{flow-generic} are a special case of these relations, obtained by setting $\mathscr K=0$.

Given arbitrary functions  ${\cq^i}(\cj^j,{\co}_{j})$ and ${\cp^i}(\cj^j,{\co}_{j})$, 
these relations uniquely determine the boundary term, $\mathscr K(\cj^i,{{\co}_{i}})$, provided it exists. Existence of $\mathscr K$ is equivalent to preserving the symplectic form and so, for any choice of new variables that preserve the symplectic form, there exists a unique boundary term $\mathscr K$. However, a given boundary term, $\mathscr K(\cj^i,{{\co}_{i}})$, does not uniquely determine the new symplectic variables ${\cq^i}(\cj^j,{{\co}_{j}})$ and ${\cp_i}(\cj^j,{{\co}_{j}})$ due to the existence of canonical transformations that preserve the variational principle. Nevertheless, additional conditions may be imposed by the requirement that the deformed theory preserves certain symmetries or that it has a well defined holographic dual. In the case of scalar fields, for example, the form of the new canonical variables is dictated by the requirement that the canonical transformation corresponds to a multitrace deformation of the dual field theory in the large-$N$ limit \cite{Papadimitriou:2007sj}. 

\subsubsection{Generalized CSS boundary conditions}

The CSS boundary conditions for AdS$_3$ gravity, first introduced in \cite{Compere:2013bya}, correspond to modifying the variational principle \eqref{var-GWCFT} at ${{\m^+}}=0$ by adding the boundary term\footnote{As we show here, in order to obtain the CSS asymptotic symmetry algebra at ${{\m^+}}=0$, the boundary term must be specified to linear order in ${{\m^+}}$. However, the boundary term is \cite{Compere:2013bya} is specified explicitly only to zeroth order in ${{\m^+}}$.} 
\be\label{Ipv-CSS}
\int_{\pa\cm} d^2x\,\cl_-{{\m^-}},
\ee
which amounts to a Legendre transform in one symplectic pair. The resulting variational principle, 
\be\label{var-GWCFT-CSS}
\d I^{\rm CSS}_{\rm 3D}=\int_{\pa\cm} d^2x\,\big({{\m^-}}\d\cl_--{{\wt\cl_+}}\d\m^{+}\big),
\ee
requires that $\cl_-$ and ${{\m^+}}$ be kept fixed. As we will briefly review momentarily, the asymptotic symmetry generators obtained from these boundary conditions form a single right-moving (along $x^{+}$-coordinate) non-compact $u(1)$ Virasoro-Kac-Moody algebra.  

In this subsection, we would like to generalize the CSS variational principle to the DWCFT phase space such that \eqref{var-GWCFT-CSS} is recovered in the limit ${{\m^+}}\to 0$. Such a generalization, however, is not unique. In fact, turning on ${{\m^+}}$ amounts to an irrelevant deformation of the ${{\m^+}}=0$ theory \cite{Castro:2014ima,Cvetic:2016eiv} and, therefore, the generalization of the variational principle \eqref{var-GWCFT-CSS} to non-zero ${{\m^+}}$ is not unique. Different possibilities correspond to different ultraviolet completions. 

A natural generalization of \eqref{Ipv-CSS} to finite ${{\m^+}}$ is the Lorentz and parity violating boundary term
\be\label{Ipv}\boxed{
I_{\rm pv}=-\int_{\pa\cm} d^2x\,\cj^-{{\co}_{-}},}
\ee
where $(\cj^-,{{\co}_{-}})$ are the spectrally flowed symplectic variables in \eqref{flow}, parameterized by the arbitrary function $\rg(\l)$. Depending on the choice of this function, the resulting total action
\be\label{Iren-GCSS}
I_{\rm 3D}^{\rm GCSS}\equiv I^{\rm ren}_{\rm 3D}+I_{\rm pv},
\ee
is invariant under a subset of the local symmetries of the DWCFT phase space. By identifying the local symmetries the boundary term \eqref{Ipv} preserves, we will determine the conditions that the function $\rg(\l)$ must satisfy so that \eqref{Iren-GCSS} coincides with the effective action of the CSYK model.

From \eqref{var-DWCFT-flow} it follows that the boundary term \eqref{Ipv} leads to the variational principle 
\be\label{varDWCFT}
\d I_{\rm 3D}^{\rm GCSS}=\int_{\pa\cm}d^2x\;\Big(-\cj^-\d{{\co}_{-}}+{\co}_{+}\d\cj^+-\frac{c}{12\p}\pa_+\mathscr F\d\F\Big).
\ee
The boundary conditions compatible with this variational principle are
\be\label{DWCFT-BC}
\d{{\co}_{-}}=0,\qquad \d\cj^+=0,\qquad \d\F=0,
\ee
which provide a generalization of the CSS boundary conditions \cite{Compere:2013bya} to finite ${{\m^+}}$.

\subsubsection{Conserved charges and asymptotic symmetry algebra}

A well defined variational principle on a non-degenerate symplectic space defines a symplectic potential, $\J({\cp}_a,\d{\cq}^a)$, which depends on the generalized momenta, ${\cp}_a$, and the variations of the generalized coordinates, $\d{\cq}^a$. The index $a$ runs over the number of symplectic pairs; in our case $a=i=\pm$. The symplectic potential is the integrand of the variational principle and contains all information necessary for identifying the asymptotic symmetries and  associated conserved charges. For Dirichlet boundary conditions, it has been shown \cite{Papadimitriou:2005ii} that this symplectic potential coincides with that used in the covariant phase space approach \cite{Wald:1999wa}. However, in contrast to the covariant phase space approach, the holographic construction of the conserved charges correctly accounts for the Casimir contribution and, as we will now discuss, it is generalizable to boundary conditions other than Dirichlet.        

To determine the asymptotic symmetries and conserved charges, one starts by identifying the local symmetries that leave the action invariant up to anomalies, i.e. up to terms that depend locally on the generalized coordinates, ${\cq}^a$, but not on the momenta, ${\cp}_a$. Let us parameterize the $I$th local symmetry by $y^{i_1i_2\cdots i_I}_I$, where the indices $i_1i_2\cdots i_I$ reflect the tensor structure of the symmetry parameter, such as scalar for Weyl or U(1) transformations, and vector for diffeomorphisms. When evaluated on symmetry variations of the coordinates ${\cq}^a$, the symplectic potential takes the form   
\be\label{localsym}
\J({\cp}_a,\d_{y}{\cq}^a)=\pa_i{\mathscr V}^i(y)+y^{i_1i_2\cdots i_I}_I{\mathscr W}_{i_1i_2\cdots i_I}^I,
\ee
for some quantities ${\mathscr V}^i$ and ${\mathscr W}_{i_1i_2\cdots i_I}^I$ that depend on the canonical variables. This expression involves a sum over the index $I$, which counts the number of independent local symmetry parameters. If the
local parameters $y^{i_1i_2\cdots i_I}_I$ are arbitrary functions of all spacetime coordinates, invariance of the action (up to possible anomalies) leads to the Ward identities 
\be
{\mathscr W}_{i_1i_2\cdots i_I}^I=0.
\ee
As in some of the examples we will consider, the parameters $y^{i_1i_2\cdots i_I}_I$ may in general be arbitrary functions of a subset of spacetime coordinates, resulting in a partially integrated Ward identity. 

The local symmetries allow us to identify not only the Ward identities, but also the  conserved charges associated with global symmetries. Global symmetries correspond to the subset of local symmetries that leave the generalized coordinates invariant \footnote{We use this notion of global symmetry throughout this paper. For example, the global symmetry algebra for Brown-Henneaux boundary conditions is two copies of the Virasoro algebra. This is not to be confused with the sl(2,\bb R) subalgebra, which is globally well defined on the sphere.}, that is
\be
\d_{y_I}{\cq}^a=0.\label{global}
\ee 
For each global symmetry satisfying the condition \eqref{global} we have 
\be\label{psigeneral}
0=\J({\cp}_a,\d_{ y}{\cq}^a)=\pa_i\cv^i(y)+ y^{i_1i_2\cdots i_I}_I\cw_{i_1i_2\cdots i_I}^I=\pa_i\cv^i(y),
\ee
which implies that the charges 
\be
Q[y_I]\equiv\int_\S d\s_i\cv^i(y_I),
\ee
where $\S$ is a boundary Cauchy surface and $d\s_i$ is the associated volume form, are conserved. A summary of the various types of symmetries used in our analysis can be found in the following table:
\vskip-.05cm
\begin{table}[htbp]
\captionsetup{justification=raggedright,
singlelinecheck=false
}
\centering
\begin{tabular}{|c|c|}
\hline
\makecell{{\bf phase space}\\ symmetries} & \makecell[l]{- preserve the {\em form} of the symplectic variables\\
- independent of boundary conditions}\\ \hline
\makecell{{\bf local} symmetries\\ of the theory} & \makecell[l]{- depend on the {\em form} of the boundary conditions,\\ 
\hskip.25cm i.e. on the variables kept fixed \\
- render the symplectic potential a total derivative\\
\hskip.25cm i.e. they leave the action invariant\\
- local symmetry parameters $y^{i_1i_2\cdots i_I}_I$ are arbitrary\\ 
\hskip.25cm functions of (at least some of the) spacetime coordinates}\\ \hline
\makecell{{\bf global} symmetries\\  of the theory} & \makecell[l]{- preserve the {\em value} of the variables kept fixed by\\ \hskip.25cm the boundary conditions\\
- render the symplectic potential identically zero}\\ \hline
\end{tabular}
\caption{Glossary of symmetries discussed in this paper. The term ``theory'' refers to the phase space together with a specification of the variables kept fixed in the variational principle, i.e. the holographic sources. Global symmetries are a subset of both phase space and local symmetries. There is no generic relation between local and phase space symmetries, except that, for the phase spaces and boundary conditions we consider, local symmetries are always a subset of PBH transformations \eqref{pbh-gen}, i.e. of the phase space symmetry of the Fefferman-Graham phase space.}
\label{table:symmetries}
\end{table}

Instead of starting with the local symmetries and determining the global ones as a subset, it is often convenient to first identify the global symmetries and then obtain the conserved charges via the Noether procedure. The last step in this approach requires generalizing the global symmetry parameters to local ones that leave the action invariant, i.e. determining the local symmetries. As a result, the two approaches are completely equivalent. The only difference is a practical one; namely the Noether procedure typically does not require a general analysis of the local symmetries, but rather only a specific class of local symmetries need be identified.

We will now apply this general procedure to the generalized CSS variational principle \eqref{varDWCFT}, which corresponds to the symplectic potential 
\be\label{CSS-symplectic-potential}\boxed{
\J_{\rm GCSS}({\cp}_a,\d{\cq}^a)= -\cj^-\d{{\co}_{-}}+{\co}_{+}\d\cj^+-\frac{c}{12\p}\pa_+\mathscr F\d\F.}
\ee
In the following, we consider first the case ${\m^+}=0$ and later generalize the analysis to non-zero ${{\m^+}}$.

\paragraph{Symmetry algebra at ${{\m^+}}=0$} As we discussed briefly earlier,  the DWCFT phase space admits a smooth ${{\m^+}}\to0$ limit provided the extremality condition \eqref{mp0} holds, 
which implies that the spectrally flowed current $ {{\co}_{+}}$ in \eqref{flow} satisfies 
\be\label{limit-0}
{\co}_{+} =-{{\wt\cl}}_+(x^+)-\rg'(0){{\co}_{-}} (\cj^-)^2+\co(\m^+),
\ee 
and hence, unless $\rg'(0)=0$, there is operator mixing even at ${{\m^+}}=0$. 

In order to determine the conserved charges and the corresponding symmetry algebra at $\m^+=0$, we will follow the first approach discussed above, first identifying the local symmetries and then determining the subset corresponding to the global ones. It turns out, however, that the DWCFT phase space symmetries \eqref{resPBH} are too restrictive to be identified with the correct local symmetries. In fact, we will show that the $\m^+\to0$ limit of the DWCFT phase space symmetries coincides with the global symmetries of the theory. The local symmetries can instead be identified with a suitable subset of the $\m^+\to 0$ limit of generic PBH transformations. 

As $\m^+\to 0$, the PBH transformations \eqref{PBH-coords} and \eqref{PBH-flow} in appendix \ref{sec:transformations} imply that 
\bal\label{CSSPBH}
\d_{\rm PBH}\cj^+=&\;\pa_-\x_o^++\co(\m^+),\NO\\
\d_{\rm PBH}{{\co}_{-}}=&\;\x^+_o\pa_{+}\co_{-}+\x^-_o\pa_-\co_{-}+2\big(\pa_{-}\x^-_o+(1-\rg'(0))\m^-\pa_{-}\x^+_o\big)\co_{-}-\frac{c} {12\pi\ell}\pa^2_{-}\s+\co(\m^+),\NO\\
\d\sbtx{PBH}\F=&\;
\frac{\s}{\ell}+\frac{1}{2} \big((\pa_++{{\m^-}} \pa_-)\x_o^++\pa_-\x_o^-+2 \Phi ' \x_o^+\big)+\co(\m^+).
\eal
Inserting these in the symplectic potential \eqref{CSS-symplectic-potential} we find that it can be written as 
\bal\label{sym-var-mu=0}
\J_{\rm GCSS}({\cp}_a,\d_{\rm PBH}{\cq}^a)
=&-\pa_-\Big[\big({{\wt\cl}}_+(x^+)+(2-\rg'(0))(\m^-)^2{{\co}_{-}} \big)\x_o^++2\x_o^-\m^-\co_--\frac{c} {12\pi\ell}\m^-\pa_{-}\s\Big]\NO\\
&\hskip-.2cm+\m^-\big(\x^-_o+(1-\rg'(0))\m^-\x_o^+\big)\pa_-\co_--\x_o^+\m^-(\pa_+-\m^-\pa_-)\co_-+\co(\m^+).
\eal
Except for the first term in the second line, this is of the form \eqref{psigeneral} that the symplectic potential must take when evaluated on local symmetries. In particular, the first line is a total derivative that determines the conserved charges, while the last term in the second line is the first Ward identity in \eqref{WIds}, since $\co_-=-\cl_-+\co(\m^+)$. The first term in the second line is the only deviation from \eqref{psigeneral} and therefore determines the subset of PBH transformations that correspond to local symmetries. 

Since the variational principle requires that $\co_-(z^-)$ is kept fixed, the local symmetries in the limit $\m^+\to 0$ depend on the choice of the function $\co_-(z^-)$ that is set to. Following \cite{Compere:2013bya}, we demand that $\co_-(z^-)=-\D/2\p+\co(\m^+)$, i.e. it approaches a constant as $\m^+\to 0$, so that $\pa_-\co_-=\co(\m^+)$. With this choice, \eqref{sym-var-mu=0} implies that arbitrary PBH transformations are local symmetries as $\m^+\to 0$.   
   
Having identified the local symmetries, the next step is to determine the global symmetries that satisfy the generalized CSS boundary conditions \eqref{DWCFT-BC}. From the local symmetry transformations of the sources in \eqref{CSSPBH}, of $\F$ in \eqref{PBH-coords}, and the choice $\co_-(z^-)=-\D/2\p+\co(\m^+)$ follows that, in the limit $\m^+\to 0$, the global symmetries correspond to a subset of PBH transformations satisfying
\be
\pa_-\x_o^+=0,\qquad  \D\pa_{-}\x^-_o-\frac{c} {24}\pa^3_{-}\x_o^-=0,\qquad \frac{\s}{\ell}=-\frac{1}{2} \big(\pa_+\x_o^++\pa_-\x_o^-+2 \F' \x_o^+\big).
\ee
The general solution of this set of equations takes the form 
\be\label{CSS-global-symmetries}
\x_o^+=v(x^+),\qquad \x_o^-=a_0(x^+)+a_+(x^+)e^{\sqrt{\frac{24\D}{c}} z^-}+a_-(x^+)e^{-\sqrt{\frac{24\D}{c}} z^-},
\ee 
where $v(x^+)$, $a_0(x^+)$ and $a_\pm(x^+)$ are arbitrary functions. Notice that we have chosen to parameterize the exponentials in terms of $z^-$ instead of $x^-$. This is because only the subset of these symmetries with constant $a_\pm$ preserves the DWCFT phase space and, therefore, corresponds to the global symmetries at $\m^+= 0$. However, we find it useful to keep the $x^+$ dependence of $a_\pm$ in the analysis of the global symmetries, setting them to constants at the very end. As it will become clear shortly, closure of the algebra requires that $a_\pm$ are either arbitrary functions of $x^+$, or identically zero.    

In fact, the global symmetries at $\m^+=0$ can be identified with the $\m^+\to0$ limit of the DWCFT phase space symmetries \eqref{resPBH}. In particular, as ${{\m^+}}\to 0$, $z^+$ effectively depends on $x^+$, and hence $\e^+(z^+)$ reduces to a function of $x^+$, and ${\e^+(z^+)}'=\m^+\pa_+\e^+(z^+)$. The transformation of the current ${\co}_{+}$ in \eqref{resPBH-flow}, therefore, remains well defined in this limit, provided that $\m^+\e^+(z^+)$ is finite. On the other hand, the finite part of $\m^+\e^+(z^+)$ as a function of $x^+$ is indistinguishable from $\ve(x^+)$. It follows that the $\m^+\to0$ limit of the DWCFT phase space symmetries \eqref{resPBH} takes the form
\bal\label{m0lim-sym}
\xi_o^+=&\;\ve(x^+)+\mu^+\e^+(z^+)\to\wt\ve(x^+),\NO
\\
\xi_o^-=&\;\vf(x^+)-\l\e^+(z^+)+\e^-(z^-)\to\wt\varphi(x^+)+\e^-(z^-),
\NO\\
\frac{\s}{\ell}\to &\;-\frac12(\e^{-})'(z^-)+\frac12\om(x^+).
\eal
Comparing with \eqref{CSS-global-symmetries}, we identify 
\be
\wt\ve=v,\qquad \wt\vf=a_0,\qquad \e^-=a_+e^{\sqrt{\frac{24\D}{c}} z^-}+a_-e^{-\sqrt{\frac{24\D}{c}} z^-},\qquad \om=-e^{-2\F}\pa_+\big(e^{2\F}v\big),
\ee 
provided $a_\pm$ are constants. Hence, the global symmetries at $\m^+=0$,  corresponding to \eqref{CSS-global-symmetries} with constant $a^\pm$, can be identified with the $\m^+\to0$ limit of the local DWCFT symmetries \eqref{resPBH}.

The Killing vectors associated with the symmetries \eqref{CSS-global-symmetries} take the form 
\be\label{Killing-fixed-point}
\z_{\rm Witt}[v]=v(x^+)\pa_+,\qquad  \z^0_{\Hat{sl}(2)}[a_0]=a_0(x^+)\pa_-,\qquad \z^\pm_{\Hat{sl}(2)}[a_\pm]=a_\pm(x^+) e^{\pm\sqrt{\frac{24\D}{c}} z^-}\pa_-,
\ee
and under the Lie bracket they satisfy the algebra 
\bal
&\big[\z_{\rm Witt}[v_1],\z_{\rm Witt}[v_2]\big]=\z_{\rm Witt}[v_1\pa_+v_2-v_2 \pa_+v_1],\NO\\
&\big[\z_{\rm Witt}[v],\z_{\Hat{sl}(2)}[a_0]\big]=\z_{\Hat{sl}(2)}[v\pa_+a_0],\qquad \big[\z_{\rm Witt}[v],\z^\pm_{\Hat{sl}(2)}[a_\pm]\big]=\pm\sqrt{\frac{24\D}{c}}\z^\pm_{\Hat{sl}(2)}[v a_\pm],\\
&\big[\z^+_{\Hat{sl}(2)}[a_+],\z^-_{\Hat{sl}(2)}[a_-]\big]=2\sqrt{\frac{24\D}{c}}\z^0_{\Hat{sl}(2)}[a_+a_-],\qquad \big[\z_{\Hat{sl}(2)}[a_0],\z^\pm_{\Hat{sl}(2)}[a_\pm]\big]=\pm\sqrt{\frac{24\D}{c}}\z^\pm_{\Hat{sl}(2)}[a_0 a_\pm].\NO
\eal
In particular, $v(x^+)$ generates one copy of the Witt algebra, while $a_0(x^+)$, $a_\pm(x^+)$ generate an $\Hat{sl}(2,\bb R)$ Kac-Moody algebra at level zero \cite{Avery:2013dja}. However, only the subalgebra Witt$\oplus \Hat u(1)$ corresponds to the Killing symmetry of the theory, as well as the DWCFT phase space symmetries \eqref{resPBH} as $\m^+\to 0$. In fact, as we show next, the conserved charges associated with the two off diagonal $\Hat{sl}(2,\bb R)$ generators vanish identically within the DWCFT phase space and therefore the physical symmetry algebra is the subalgebra Witt$\oplus \Hat u(1)$ \cite{Compere:2013bya}. 

The conserved charges associated with the global symmetries can be read off from the total derivative terms in the symplectic potential \eqref{sym-var-mu=0}, namely
\bal\label{conserved-charges}
Q_{\text{Vir}}[v]=&\; \oint dx^+\big(\wt{\cl}_+^\F-(2-\mathscr R'(0))(\mu^{-})^2\cl_-\big)v,\NO\\
Q_{\Hat{sl}(2)}[a_0]=&\; -2\oint dx^+\mu^-\cl_-a_0,\NO\\
Q_{\Hat{sl}(2)}[a_\pm]=&\; -\oint dx^+\mu^-\Big(2\cl_--\frac{c}{24\pi}\pa_-^2\Big)a_\pm e^{\pm\sqrt{\frac{24\D}{c}} z^-}=0,
\eal
where we have shifted $\wt{\cl}_+$ as
\be
\wt{\cl}_+^\F\equiv \wt{\cl}_+-\frac{c}{12\p}(\F''-\F'^2),
\ee
so that the Virasoro charges, besides being conserved, transform correctly in the presence of a fixed but arbitrary conformal factor $\F(x^+)$.\footnote{This shift can also be deduced by a suitable change of coordinates that eliminates the conformal factor \cite{Cvetic:2016eiv}.} As advertised, the charges associated with the off diagonal $\Hat{sl}(2,\bb R)$ generators are identically zero, while the Cartan subalgebra of $\Hat{sl}(2,\bb R)$ can be identified with the $\Hat u(1)$ Kac-Moody algebra.

The algebra of conserved charges follows from the transformation of the Kac-Moody and Virasoro currents, respectively $\cj^-$ and $ {{\co}_{+}}$. These are obtained by inserting the symmetries \eqref{CSS-global-symmetries}  in the transformations \eqref{PBH-flow}, setting $a_\pm=0$, and taking the limit ${{\m^+}}\to 0$. This determines that
\bal
\d\cj^-=&\;\pa_+(\wt\vf+{{\m^-}}\wt\ve),\\
\d{\co}_{+}=&\;\wt\ve\pa_{+}\Big(\co_{+}+\frac{c}{12\p}(\F''-\F'^2)\Big)+2\pa_{+}\wt\ve\Big(\co_{+}+\frac{c}{12\p}(\F''-\F'^2)\Big)+\frac{c}{24\pi}\pa_+^3\wt\ve -2\rg'(0)\m^-\pa_{+}\wt\vf\,\co_{-},\NO
\eal
and hence, since the conformal factor $\F$ is invariant under the global symmetries,
\bal
&\d\big(\wt{\cl}_+^\F-(2-\mathscr R'(0))(\m^{-})^2\cl_-\big)=-\d\Big(\co_++\frac{c}{12\p}(\F''-\F'^2)-2(1-\mathscr R'(0))(\cj^{-})^2\co_-\Big)\NO\\
&\hspace{2.cm}=\wt\ve\pa_{+}\big(\wt{\cl}_+^\F-(2-\mathscr R'(0))(\m^{-})^2\cl_-\big)+2\pa_{+}\wt\ve\big(\wt{\cl}_+^\F-(2-\mathscr R'(0))(\m^{-})^2\cl_-\big)-\frac{c}{24\pi}\pa_+^3\wt\ve\NO\\
&\hspace{2.45cm}-2(2-\mathscr R'(0))\m^-\pa_+\wt\vf\,\cl_-.
\eal

Inserting these transformations in the charges \eqref{conserved-charges} determines their algebra. In particular,  
\bal
\d_{\wt\vf} Q_{\rm Vir}[\wt\ve]=&\;\big\{Q_{\hat{u}(1)}[\wt\vf],Q_{\rm Vir}[\wt\ve]\big\}= -{\Delta\over \pi}\oint \big(2-\rg'(0)\big)\m^- \wt\ve\, \pa_+\wt\vf,\NO\\
\d_{\wt\ve} Q_{\hat u(1)}[\wt\vf] =&\;\big\{Q_{\rm Vir}[\wt\ve],Q_{\hat{u}(1)}[\wt\vf]\big\}= -{\Delta\over \pi}\oint \pa_+(\m^- \wt\ve) \,\wt\vf.
\eal
However, closure of the algebra requires that
\be
\d_{\wt\vf} Q_{\rm Vir}[\wt\ve]=-\d_{\wt\ve} Q_{\hat u(1)}[\wt\vf],
\ee
and so we deduce that the spectral flow function, $\rg(\l)$, must satisfy the additional condition  
\be\label{slope-condition}\boxed{
\rg'(0)=1.
}
\ee

When \eqref{slope-condition} is satisfied, the modes 
\be
L_n=Q_{\rm Vir}[e^{-inx^+}],\qquad J_n=Q_{\Hat u(1)}[e^{-inx^+}],\qquad n\in\bb Z, 
\ee
satisfy the symmetry algebra
\bal
i\{L_m,L_n\}=&\; (m-n)L_{m+n}+\frac{c}{12}m^3\d_{n+m,0}\,,\NO\\
i\{L_m,J_n\}=&\; -nJ_{m+n}\,,\NO\\
i\{J_m,J_n\}=&\; \frac{k_{\Hat u(1)}}{2} 
 m\,\d_{m+n,0}\,,
\eal
where the Kac-Moody level is given by \cite{Compere:2013bya}
\be\label{level} 
k_{\Hat u(1)}=-4\D.
\ee

\paragraph{Symmetry algebra at finite ${{\m^+}}$} Having determined the symmetry algebra at ${{\m^+}}= 0$, we next consider the case of non-zero ${{\m^+}}$, while maintaining the conditions required for the limit ${{\m^+}}\to 0$ to be well defined. In particular, we demand that the extremality condition \eqref{mp0} holds, $\cl_-=\D/2\p$ is constant, and the spectral flow function satisfies \eqref{slope-condition}. As for the case $\m^+=0$, the DWCFT phase space symmetries do not include the local symmetries that determine the conserved charges. We could again resort to PBH transformations, as we did for $\m^+=0$, but 
it is practically easier to follow Noether's procedure in this case, first determining the global symmetries and then promoting the symmetry parameters to arbitrary spacetime functions in order to read off the conserved charges.

The global symmetries are a subset of the DWCFT phase space symmetries \eqref{resPBH}, and hence, they can be determined from the transformations of the sources $\cj^+$, $\co_-$ in \eqref{resPBH-flow} and of the conformal factor $\F$ in \eqref{resPBH-coords}, which we reproduce here for easy reference:
\bal
\rule{0pt}{.5cm}\d\cj^+=&\;-({\m^+})^2\pa_+((\m^+)^{-1}{{\ve}}),\NO\\
\rule{0pt}{.5cm}\d{{\co}_{-}}=&\;(\vf+{{\m^-}}{{\ve}}+{\e^-})\pa_-{{\co}_{-}}+\Big(2(\e^-)'-\frac{(\l\rg)''}{(\l\rg)'}{{\m^+}}\big(\pa_+(\vf+{{\m^-}}{{\ve}})-\l\pa_+((\m^+)^{-1}\ve)\big)\Big){{\co}_{-}}\NO\\
&+\frac{c}{24\p}\frac{{{\m^+}}\mathscr F(\pa_+\om-\pa_+\mathscr F(\vf+{{\m^-}}{{\ve}}-(\m^+)^{-1}{{\ve}}))}{(\l\rg)'(1-\l)}+\frac{c}{24\p}\frac{1}{(\l\rg)'}\big((\e^-)'''-\mathscr F^2(\e^-)'\big),\NO\\
2\d\F=&\;-\vf\mathscr F+e^{-2\F}\pa_+(e^{2\F}{{\ve}})+\om.
\eal
Setting these variations to zero and using the condition that $\cl_-=\D/2\p$ is constant determines 
\bal\label{finite-mu-global-parameters}
&\vf+\m^-\ve=b_1,\qquad {{\ve}}=b_2{{\m^+}},\qquad \om=(b_1-b_2)\mathscr F,\NO\\
&\e^+(z^+)\;\text{arbitrary},\qquad \e^-(z^-)=a_0+a_+e^{\sqrt{\frac{24\D}{c}} z^-}+a_-e^{-\sqrt{\frac{24\D}{c}} z^-},
\eal
where $b_1$, $b_2$, $a_0$, and $a_\pm$ are arbitrary constants. From \eqref{resPBH} follows that the constants $b_1$ and $b_2$ are not independent, since they can be absorbed into the zero modes of ${\e^-}$ and ${\e^+}$, respectively. Moreover, as we saw above, in order for the algebra to close with constant $a_0$, $a_\pm$, the constants $a_\pm$ must vanish identically. Then we are left with
\be\label{finite-mu-Killing}
\z[\e^-]=\frac{\e^-}{1-\l}(\pa_--{{\m^+}}\pa_+)=a_0\pa_{z^-},\qquad\z[\e^+]=\frac{{\e^+}{{\m^+}}}{1-\l}(\pa_+-{{\m^-}}\pa_-)={\e^+}\pa_{z^+},
\ee
where  ${\e^+}(z^+)$ is an arbitrary function of $z^+$, and $a_0$ is a constant. 

The conserved charges associated with \eqref{finite-mu-Killing} may be determined using Noether's procedure, by evaluating the symplectic potential \eqref{CSS-symplectic-potential} on these global symmetry transformations, but promoting $\e^\pm$ to suitable functions of $x^\pm$. These must be as general as possible, but still leave the action invariant, i.e. they must render the symplectic potential a total derivative. The relevant local symmetry associated with $\e^-$ corresponds to promoting $\e^-$ from the special form in \eqref{finite-mu-global-parameters} to an arbitrary function of $z^-$. The global symmetry associated with $\e^+$ already allows for an arbitrary dependence on $z^+$, and so the corresponding local symmetry requires promoting $\e^+$ to an arbitrary function of both $z^+$ and $z^-$, or equivalently of $x^\pm$. 

In order to evaluate the symplectic potential \eqref{CSS-symplectic-potential} on such transformations, it is necessary to correctly specify the value of the function $\s$ in the PBH transformation of the warp factor $\F$ in \eqref{PBH-coords}, since the DWCFT form of $\s$ in \eqref{resPBH} does not have a well defined generalization to an arbitrary $\e^+(x^+,x^-)$. From \eqref{PBH-coords} we see that maintaining the condition $\d\F=0$ determines  
\be
\frac{\s}{\ell}=-\frac{1}{2} \big((\pa_++{{\m^-}} \pa_-)\x_o^++(\pa_-+{{\m^+}} \pa_+)\x_o^-+2 \Phi ' \x_o^+\big),
\ee
where $\x_o^\pm$ are as in \eqref{resPBH}, but with $\e^+$ promoted to an arbitrary function of $x^\pm$. With this choice for $\s$, the resulting expression for the symplectic potential \eqref{CSS-symplectic-potential} is
\bal\label{sym-var}
\J_{\rm GCSS}({\cp}_a,\d{\cq}^a)=&\;
\Big[-(\m^+)^2\bar\co_++\l^2\pa_\l\Big(\frac{\rg}{(\l\rg)'}\Big)\bar\co_-\Big]\big(\pa_+-(\m^+)^{-1}\pa_-\big)\e^+\NO\\
&+\frac{c}{24\p}\frac{\l\rg}{(1-\l)(\l\rg)'}\Big(\mathscr F+\frac{1+\l}{1-\l}\pa_-\Big)\big(\mathscr F+\pa_+\m^++\m^+\pa_+\big)\big(\pa_+-(\m^+)^{-1}\pa_-\big)\e^+\NO\\
&+\pa_-\Big(\frac{\m^-\rg}{(\l\rg)'}\big(2\e^-\cl_--\frac{c}{24\p}\pa_-^2\e^-\big)\Big),
\eal 
where recall that $\bar\co_\pm$ were defined in \eqref{Jvariables}.

The form \eqref{sym-var} of the symplectic potential confirms that an arbitrary $\e^-(z^-)$ is a local symmetry for any $\m^+$. The conserved charge associated with the corresponding global symmetry in \eqref{finite-mu-global-parameters} is
\be\label{conserved-charge-finite-mu-I} 
Q[a_0]=-\oint d x^{+} \frac{\mathscr{R} \mu^{-}}{(\lambda \mathscr{R})^{\prime}}2 \mathcal{L}_{-} a_0.
\ee
Notice that, in the limit $\m^+\to0$, this coincides with the zero mode of the $\Hat u(1)$ Kac-Moody charge $Q_{\Hat{sl}(2)}[a_0]$ in \eqref{conserved-charges}. Hence, the global $u(1)$ at finite $\m^+$ in enhanced to a Kac-Moody $\Hat u(1)$ as $\m^+\to 0$.  

The conserved charge associated with $\e^+$ is less straightforward to identify. Firstly, the symplectic potential \eqref{sym-var} is not a total derivative for arbitrary $\e^+(x^+,x^-)$ and so, such an $\e^+$ is not in general a local symmetry of the theory. However, to leading order in $\m^+$ as $\m^+\to 0$, the $\e^+$ part of the symplectic potential \eqref{sym-var} takes the form 
\be
\J_{\rm GCSS}({\cp}_a,\d_{\e^+}{\cq}^a)=
-\pa_-\Big(\big(\wt\cl_+(x^+)-(\m^-)^2\rg'(0)\co_-\big)\m^+\e^++\frac{c}{24\p}\m^-e^{-2\F}\pa_+\pa_-\big(e^{2\F}\m^+\e^+\big)\Big).
\ee
Since the relations \eqref{resPBH} imply that $\x_o^\pm$ are related to $\e^+$ 
as 
\be
\x_o^+\sim \m^+\e^+,\qquad \x_o^-\sim -\m^-\m^+\e^+,
\ee
this symplectic potential coincides with the one in \eqref{sym-var-mu=0} that was obtained earlier at $\m^+=0$. The conserved charges associated with this emergent Virasoro symmetry in the limit $\m^+\to0$ are those given in the first line of \eqref{conserved-charges}.

At finite $\m^+$, the $\e^+$ part of the symplectic potential \eqref{sym-var} can be expressed in the form
\bal
\J_{\rm GCSS}({\cp}_a,\d_{\e^+}{\cq}^a)=&\;
\big(\pa_+-(\m^+)^{-1}\pa_-\big)(\e^+\cl_+)+\d_{\e^+}\Big(-\cj^-\co_-+\frac{c}{24\p}\frac{1-\l}{2\m^+}\mathscr F^2\Big)\NO\\
&-\frac{c}{24\p}\pa_+\big(\mathscr F(\pa_--\m^+\pa_+)\e^+\big).
\eal
Although we shall not do so here, the term in the second line could be eliminated by modifying the local transformation of $\s$ given above to 
\be
\frac{\s}{\ell}=-\frac{1}{2} \big((\pa_++{{\m^-}} \pa_-)\x_o^++(\pa_-+{{\m^+}} \pa_+)\x_o^-+2 \Phi ' \x_o^+\big)-\frac{1}{2}(\pa_--\m^+\pa_+)\e^+,
\ee
so that
\be
\d\F=-\frac{1}{2}(\pa_--\m^+\pa_+)\e^+.
\ee
It follows that, for Dirichlet boundary conditions (i.e. in the absence of the boundary term \eqref{Ipv}), there exists a modified effective action that differs from the original only by a local functional of the sources, such that arbitrary $\e^+(x^+,x^-)$ are a local symmetry. The corresponding conserved charges are the standard Virasoro charges 
\be\label{conserved-charge-finite-mu-II} 
Q[\e^{+}]=\oint\big(d x^{-}+(\m^+)^{-1}d x^{+}\big) \mathcal{L}_{+} \e^{+}=\oint d z^{+} \mathcal{L}_{+} \e^{+},
\ee
associated with the global symmetries $\e^+(z^+)$. For GCSS boundary conditions, however, the modified action preserving arbitrary $\e^+(x^+,x^-)$ is non-locally related to the original effective action due to the term $\cj^-\co_-$. Such a non-local change would modify the theory and it is therefore not acceptable. As a result, there are no conserved charges associated with generic $\e^+(z^+)$ for GCSS boundary conditions. 

Nevertheless, there is a (different) conserved charge associated with the {\em zero mode} of $\e^+(z^+)$, $\e_0^+$. This symmetry can be gauged by promoting the global parameter $\e_0^+$ to a function of $x^-$, i.e. $\e_0^+(x^-)$. On such local transformations, the symplectic potential \eqref{sym-var} reduces to
\bal
\J_{\rm GCSS}({\cp}_a,\d_{\e_0^+}{\cq}^a)=&\;
-\frac{1}{\m^+}\pa_-\Big[\Big(\cl_+-\frac{c}{48\p}\big(\mathscr F^2-2{{\m^+}}\pa_+\mathscr F\big)\Big)\e_0^+-\l^2\pa_\l\Big(\frac{\rg}{(\l\rg)'}\Big)\Big(\cl_--\frac{c}{48\p}\mathscr{F}^2\Big)\e_0^+\NO\\
&+\frac{c}{24\p}\frac{\l\rg}{(1-\l)(\l\rg)'}(\mathscr F+\pa_+\m^+)\Big(\mathscr F+\frac{1+\l}{1-\l}\pa_-\Big)\e_0^+\Big]+(\m^+)^{-1}\cl_+'\e_0^+.
\eal 
It follows that, provided $\cl_+$ is a constant (recall that we already assume that $\cl_-$ is a constant so that the limit $\m^+\to 0$ is well defined), any function $\e_0^+(x^-)$ is a local symmetry. The conserved $u(1)$ charge associated with the global symmetry corresponding to the zero mode $\e_0^+$ is 
\bal\label{conserved-charge-finite-mu-III} 
Q[\e_0^+]=&\;\oint \frac{d x^{+}}{\m^+}\Big[\cl_+-\frac{c}{48\p}\big(\mathscr F^2-2{{\m^+}}\pa_+\mathscr F\big)-\l^2\pa_\l\Big(\frac{\rg}{(\l\rg)'}\Big)\Big(\cl_--\frac{c}{48\p}\mathscr{F}^2\Big)\NO\\
&\hskip7.cm+\frac{c}{24\p}\frac{\l\rg}{(1-\l)(\l\rg)'}(\mathscr F+\pa_+\m^+)\mathscr F\Big]\e_0^+.
\eal
It is straightforward to check that this charge reduces to the zero mode of the Virasoro charges in \eqref{conserved-charges} in the limit $\m^+\to 0$ with constant $\Phi$. So the $u(1)$ symmetry at finite $\m^+$ is enhanced to a Virasoro algebra at $\m^+=0$. In combination with the $u(1)$ charge \eqref{conserved-charge-finite-mu-I} associated with the zero mode of $\e^-$, therefore, we have shown that the $u(1)\oplus u(1)$ symmetry at finite $\m^+$ is enhanced to the Virasoro-Kac-Moody symmetry \eqref{conserved-charges} at $\m^+=0$. We summarize the phase space and global symmetries of the various phase spaces we have discussed in table~\ref{table:phase-spaces}. 
\begin{table}[htbp]
\captionsetup{justification=raggedright,
singlelinecheck=false
}
\centering
\begin{tabular}{|c|c|c|c|c|}
\hline
phase space & \makecell{phase space\\ symmetry} & BCs & global symmetry & dual \\ \hline
\multirow{2}{*}{FG} & \multirow{2}{*}{\makecell{PBH:\\ $\x_o^i(x^+,x^-)$, $\s(x^+,x^-)$}} & D & Vir$\,\oplus$Vir & CFT$_2$\\ \cline{3-5}
& & - & - & - \\ \hline
\multirow{2}{*}{DWCFT} & \multirow{2}{*}{\makecell{$\ve(x^+)$, $\vf(x^+)$, $\om(x^+)$\\ $\e^-(z^-)$, $\e^+(z^+)$}} & D & Vir$\,\oplus$Vir & CFT$_2$\\ \cline{3-5}
 & & GCSS  & $u(1)\oplus u(1)$ & non-relativistic QFT$_2$\\ \hline
\multirow{2}{*}{GWCFT} & \multirow{2}{*}{\makecell{$\wt\ve(x^+)$, $\wt\vf(x^+)$, $\om(x^+)$\\ $\e^-(z^-)$}} & D  & Vir$\,\oplus$Vir & CFT$_2$\\ \cline{3-5}
 & & CSS & Vir$\,\oplus\,\Hat u$(1) & WCFT$_2$\\ \hline
\end{tabular}
\caption{Summary of the three phase spaces we study in this paper: GWCFT $\subset$ DWCFT $\subset$ FG. For each case we list the phase space symmetries, which are independent of the choice of boundary conditions, as well as the global symmetry algebra corresponding to either Dirichlet (D) or Compère-Song-Strominger (CSS) boundary conditions.}
\label{table:phase-spaces}
\end{table}

We conclude this section we a couple of remarks. Firstly, we emphasize again that only states with constant $\cl_+$ preserve a second $u(1)$ global symmetry at finite $\m^+$, while those with non-constant $\cl_+$ break it spontaneously. As we will see in the next section, the subsector of the DWCFT phase space with constant $\cl_-$ and $\cl_+$ coincides with the phase space of 2D dilaton gravity upon Kaluza-Klein reduction. In particular, the $s$-modes possess two U(1) symmetries corresponding to the 2D mass and electric charge. The limit $\m^+\to0$ results in 2D solutions with a constant dilaton, which have an enhanced Virasoro-Kac-Moody symmetry.        

Secondly, the symmetry analysis at finite $\m^+$ in this section assumes a generic spectral function $\rg(\l)$, subject only to the two conditions \eqref{RG-function} and \eqref{slope-condition}, which are necessary for the emergence of the Virasoro-Kac-Moody symmetry in the limit $\m_+\to 0$. If these conditions are relaxed, however, there exist special choices of $\rg(\l)$ that lead to enhanced symmetry at finite $\m^+$. For example, when $\cl_-$ and $\cl_+$ are constant and the spectral flow function is of the form
\be\label{special-R}
\rg(\l)=\frac{\rg_o}{\l(1-\l)^{n}},
\ee
where $\rg_o$ and $n\neq 1$ are constants, then the symplectic potential \eqref{CSS-symplectic-potential} reduces to a total derivative, except for the terms involving $\om$. In that case, all DWCFT phase space symmetries are preserved by the boundary term \eqref{Ipv}, except for Weyl transformations, $\om$. If $n=1$, the terms involving $\om$ also become a total derivative, restoring Weyl invariance. However, for any value of $n$, the form \eqref{special-R} of $\rg(\l)$ is incompatible with the ${{\m^+}}\to 0$ limit, and so it is not suitable for our present purposes.

\section{2D dilaton gravity and the DWCFT phase space}
\label{sec:Dimreduce}

In this section, we review the consistent Kaluza-Klein (KK) reduction of AdS$_3$ gravity to two dimensions, as well as the space of solutions of the resulting 2D dilaton gravity \cite{Cvetic:2016eiv}. Moreover, we demonstrate that any 2D solution can be embedded in the DWCFT phase space of 3D gravity. The discussion in this section is independent of any boundary conditions.

\subsection{Consistent reduction to 2D and the space of solutions}\label{subsec:dimreduce}

The starting point for the KK reduction of AdS$_3$ gravity is the ansatz
\be
\label{2dana}
ds^2_{\rm 3D}=ds^2_{\rm 2D}+e^{-2\j}(dw+A)^2,
\ee
where $\j$ and $A=A_\m dx^\m$ are respectively a scalar and a U(1) gauge field on the two dimensional base manifold with metric $g_{\m\n}$. The coordinate $w\sim w+2\pi R_w$ parameterizes the KK circle. Inserting this ansatz in the AdS$_3$ gravity action \eqref{3Daction} results in the Einstein-Maxwell-Dilaton theory \cite{Achucarro:1993fd,Castro:2014ima,Cvetic:2016eiv}
\be
\label{2Daction}
I_{\rm 2D}=\frac{1}{2\k^2_{\rm 2D}}\Big[\int d^2x \sqrt{-g}\;e^{-\j}\Big(R[g]+\frac{2}{\ell^2}-\frac14e^{-2\j}F_{\m\n}F^{\m\n}\Big)+2\int dt\sqrt{-\g}\;e^{-\j}K\Big]\,,
\ee
where $\k^2_{\rm 2D}=\k^2_{\rm 3D}/2\pi R_w={6\ell\over c R_w}$, $K$ is the trace of the extrinsic curvature of the induced metric, $\g$, on the (regulated) one dimensional boundary, and $F_{\m\n}=\pa_\m A_\n-\pa_\n A_\m$ is the field strength of $A_\m$. Crucially, the 2D theory \eqref{2Daction} is a consistent truncation of AdS$_3$ gravity \cite{Cvetic:2016eiv}. Namely, the equations of motion following from \eqref{2Daction} imply those obtained from \eqref{3Daction}, which means that any solution of the 2D theory uplifts to a solution of AdS$_3$ gravity. 

The field equations following from the 2D action \eqref{2Daction} can be solved exactly. Starting with the Maxwell equation
\be
\label{eoms-gauge}
\nabla_\mu(e^{-3\j}F^{\mu\n})=0\,,
\ee
its general solution can be expressed covariantly as
\be
\label{gaugecond}
F_{\m\n}=2Q e^{3\j}\e_{\m\n},\qquad F^{2}=-8Q^2e^{6\j}\,,
\ee
where the constant $Q$ is the 2D electric charge and $\e_{\m\n}$ is the Levi-Civita tensor in two dimensions. 

Using \eqref{gaugecond} and adopting (without loss of generality) the FG gauge 
\be
\label{2dana2}
ds^2_{\rm 2D}=d\r^2+\g_{tt}(\r,t)dt^2,\qquad A=A_t(\r,t) dt,
\ee
the field equations take the form
\bal
\label{2DeomFG}
&(\pa^2_{\r}-\ell^{-2})e^{-\j}+Q^2e^{3\j}=0\,,\NO\\
&\big(\pa^2_{\r}-\ell^{-2}-3Q^2e^{4\j}\big)\sqrt{-\g}=0\,,\NO\\
&\big(\square_t+K\pa_{\r}-\ell^{-2}\big)e^{-\j}+Q^2e^{3\j}=0\,,\NO\\
&\pa_{\r}\big(\pa_t e^{-\j}/\sqrt{-\g}\big)=0\,,\NO\\
&\pa_\r A_t=-2Qe^{3\j}\sqrt{-\g}\,,
\eal
where $\sqrt{-\g}=\sqrt{-\g_{tt}}$, $K=\pa_{\r}\log\sqrt{-\g}$ is the extrinsic curvature of $\g_{tt}$, $\square_{t}$ is the covariant Laplacian with respect to $\g_{tt}$, and in the last equation we have chosen the orientation of the two dimensional base such that $\e_{t\r}=\sqrt{-\g}$.

The general solution of the system of equations \eqref{2DeomFG} can be obtained analytically \cite{Cvetic:2016eiv}. The solution space consists of two distinct branches, depending on whether the dilaton, $\j$, is constant or not. However, running dilaton solutions that satisfy a certain extremality condition flow to constant dilaton solutions in the IR.  

\paragraph{Running dilaton solutions} Provided the dilaton, $\j$, is not a constant, the general solution of the system of equations \eqref{2DeomFG} takes the form
\begin{align}
\label{Rdilaton}
e^{-\j}=&\;\b(t)e^{\r/\ell}\sqrt{\Big(1+\frac{m-\b'^2(t)/\a^2(t)}{4\b^2(t)}\ell^2e^{-2\r/\ell}\Big)^2-\frac{Q^2\ell^2}{4\b^4(t)}e^{-4\r/\ell}}\,,\NO\\
\sqrt{-\g}=&\;\frac{\a(t)}{\b'(t)}\pa_te^{-\j}\,,\NO\\
A_t=&\;\mu(t)+\frac{\a(t)}{2\b'(t)}\pa_t\log\Big(\frac{4\ell^{-2}e^{2\r/\ell }\b^2(t)+m-\b'^2(t)/\a^2(t)-2Q/\ell}{4\ell^{-2}e^{2\r/\ell }\b^2(t)+m-\b'^2(t)/\a^2(t)+2Q/\ell }\Big)\,,
\end{align}
where primes denote a derivative with respect to $t$, $\a(t), \b(t)$ and $\mu(t)$ are arbitrary functions, and $m$ is an integration constant. Notice that the asymptotic behavior of the solution as $\r\to\infty$ is determined by the functions $\a(t), \b(t)$ and $\mu(t)$, namely
\be\label{asympt}
\g_{tt}\sim-\a^2 e^{2\r/\ell}+\co(1),\qquad e^{-\j}\sim\b e^{\r/\ell}+\co(1),\qquad A_t =\mu+\co(e^{-2\r/\ell})\,.
\ee
The asymptotic behavior of $e^{-\j}$ and $\sqrt{-\g}\sim \a\, e^{\r/\ell}$ require that both $\a$ and $\b$ are positive definite. As discussed in \cite{Cvetic:2016eiv}, for certain ranges of the parameters $m$ and $Q$, these solutions correspond extremal or non-extremal 2D black holes and can be uplifted to the BTZ black hole.

\paragraph{Constant dilaton solutions} The system \eqref{2DeomFG} admits a second branch of solutions where the dilaton is constant and determined by the electric charge \cite{Castro:2008ms,Castro:2014ima,Jensen:2016pah}. These solutions take the form
\be\label{cds}
e^{-2\j}=\ell |Q|,\quad
\sqrt{-\g}={\wt\a}(t) e^{2 \r/\ell} +\wt\b(t) e^{-2 \r/\ell}, \quad A_{t}=\wt\mu(t) -\ell Q e^{3\j}\big({\wt\a}(t) e^{2 \r/\ell}-\wt\b(t) e^{-2 \r/\ell}\big), 
\ee
where $\wt\a(t)>0$, $\wt\b(t)$ and $\wt\m(t)$ are again arbitrary functions. In this branch, the metric is locally AdS$_2$ with radius $\ell_{\rm 2D}=\ell/2$. In particular, in the absence of a horizon, the functions $\wt\a(t)$ and $\wt\b(t)$ correspond to 1D metrics on the two boundaries of global AdS$_2$. More generally, the constant dilaton solutions correspond to the Very Near Horizon region of near extremal black holes \cite{Strominger:1998yg} and can also be uplifted to the BTZ black hole.

\subsection{Extremality condition and interpolating flow}\label{sec:RGflow}

The fact that the running dilaton solutions are (in a suitable Weyl frame) asymptotically AdS$_2$ with radius $\ell$, while the constant dilaton ones are asymptotically AdS$_2$ with radius $\ell_{\rm 2D}=\ell/2$ suggests that there should exist a holographic RG flow that interpolates between the former in the UV and the latter in the IR. As we now review, this is indeed the case, provided the parameters of the two solution branches satisfy certain (near) extremality conditions.  

It is instructive to discuss this interpolating flow from two complementary viewpoints. Firstly, starting from the running dilaton solutions in the UV, we show that such solutions flow to the constant dilaton ones in the IR, provided the extremality conditions are fulfilled. In the second approach, we start from the constant dilaton solutions in the IR and consider linear perturbations around them, which take us into the running dilaton class of solutions. This procedure corresponds to conformal perturbation theory with an irrelevant deformation around the IR fixed point.

\paragraph{UV to IR} Starting with the running dilaton solutions \eqref{Rdilaton}, we are interested in extremal black holes which allow for an RG flow description from the UV at the asymptotic infinity $\rho\to\infty$ to the IR near the horizon. By examining the zeros of $\gamma_{tt}$ we observe that the extremality condition is $m-2|Q|/\ell=\b'^2/\a^2$, in which case the horizon is located at $\rho\to-\infty$. Requiring trivial holonomy at the horizon imposes the additional condition $\m=-{\rm sgn}(Q)\a/\b$. The solution \eqref{Rdilaton} then becomes
\be\label{RG-flow}
e^{-\j}=\sqrt{\ell |Q|+\b^2(t) e^{2\r/\ell}},\quad
\sqrt{-\g}=\frac{\a(t) \b(t)e^{2\r/\ell}}{\sqrt{\ell |Q|+\b^2(t)e^{2\r/\ell}}},\quad
A_t=-\frac{{\rm sgn}(Q)\a(t) \b(t)e^{2\r/\ell}}{\ell |Q|+\b^2(t)e^{2\r/\ell}}.
\ee
The asymptotic UV behavior of this solution as $\r\to+\infty$ is still given by \eqref{asympt}. However, since $\b^2$ enters always with a factor of $e^{2\r/\ell}$, the IR expansion as $\r\to -\infty$ is equivalent to the parameter limit $\b\to 0$, keeping $\a\b$ fixed, which also sets $m=2|Q|/\ell $ and $|\m|\to \infty$. 
As was shown explicitly in \cite{Cvetic:2016eiv}, 
the leading order terms in the near horizon expansion, which is equivalent to small $\beta $ expansion, correspond to a constant dilaton solution of the form \eqref{cds} with $\wt\a=\a\b/\sqrt{\ell |Q|}$, $\wt\b=0$, $\wt\m=0$.

In order to get a constant dilaton solution with non-zero $\wt\b$ and $\wt\m$ in the IR, we need to generalize this limit to near extremal running dilaton solutions. In particular, setting $\m=-{\rm sgn}(Q)\a/\b+\wt\m(t)$ and $m-\b'^2/\a^2=2|Q|/\ell+8\b^2\d(t)/\ell^2$ for some $\wt\m(t)$ and small $\d(t)\geq 0$, the limit $\b\to 0$ of the solutions \eqref{Rdilaton}, while keeping $\a\b$ fixed and  $0\leq \d(t) e^{-2\r/\ell} \approx \b^2(t)e^{2\r/\ell}$, takes the form
\begin{align}
\label{RGflow}
e^{-\j}=&\;\sqrt{\ell|Q|}+\frac{\b^2}{2\sqrt{\ell |Q|}}e^{2\r/\ell}+\sqrt{\ell|Q|}e^{-2\r/\ell}\d+\co\big(\b^4 e^{4\r/\ell}\big)\,,\NO\\
\sqrt{-\g}=&\;\frac{\a\b}{\sqrt{\ell |Q|}}e^{2\r/\ell}+\sqrt{\ell|Q|}e^{-2\r/\ell}\frac{\a\d'}{\b'}+\co\big(\b^2 e^{4\r/\ell}\big)\,,\NO\\
A_t=&\;\wt\m-\frac{\a\b}{\ell Q}e^{2\r/\ell}+ {\rm sgn}(Q)\frac{\a\d'}{\b'} e^{-2\r/\ell}+\co\big(\b^2 e^{4\r/\ell}\big)\,.
\end{align}
Up to $\co\big(\b^2 e^{4\r/\ell}\big)$ terms, this is a constant dilaton solution of the form \eqref{cds} with parameters
\be\label{UV-IR-relations}\boxed{
\wt\a=\frac{\a\b}{\sqrt{\ell |Q|}},\qquad \wt\b=\sqrt{\ell|Q|} \frac{\a\d'}{\b'},\qquad \wt\m=\m+{\rm sgn}(Q)\frac{\a}{\b},}
\ee
where
\be\label{extremality}
\d(t)=\frac{\ell^2}{8\b^2}\big(m-2|Q|/\ell-\b'^2/\a^2\big),
\ee
is small and positive. In particular, we identify $\wt\b$ or $\d$ with the near extremality parameter. It is worth mentioning that this RG flow can be uplifted to three dimensional gravity using the KK ansatz \eqref{2dana}-\eqref{2dana2}. In particular, the near extremal running dilaton solutions uplift to the near extremal BTZ black hole, with the limit $\r\to-\infty$ corresponding again to the near horizon limit.

It is important to realize that the relations \eqref{UV-IR-relations} are highly non-trivial and describe a renormalization group flow in the dual quantum mechanics. It is this flow that will enable us to model the CSYK model in the subsequent sections. Notice that \eqref{UV-IR-relations} relate the UV parameters of the running dilaton solutions to the IR variables of the constant dilaton ones. Such an explicit map is very rarely available in holographic dualities. Here it is made possible by the fact that the running dilaton solutions are known exactly. The map between UV and IR variables allows us to describe the RG flow in parameter space, i.e. directly in the dual quantum mechanics, without reference to the radial coordinate. As we will see below, this RG flow uplifts directly to the $\m^+\to 0$ limit of the DWCFT phase space of 3D gravity solutions.

\paragraph{IR to UV} The approximate solution \eqref{RGflow} can be alternatively obtained by perturbation theory around the constant dilaton solutions. Although such an approach is redundant when the running dilaton solutions are known, it has the advantage that it can be used in cases where the running dilaton solutions are not known. Such an example arises in the 2D dilaton theory obtained from the consistent KK reduction of five dimensional AdS gravity \cite{Castro:2018ffi}.

Starting with the constant dilaton solutions, we consider the linear perturbations 
\be 
\label{RGsolution}
e^{-\j}=e^{-\j_0}+(e^{-\j})^{(1)},\qquad  \sqrt{-\g}=(\sqrt{-\g})^{(0)}+(\sqrt{-\g})^{(1)},\qquad  A_{t}=A_{t}^{(0)}+A_{t}^{(1)},
\ee
where $e^{-\j_0}$, $(\sqrt{-\g})^{(0)}$ and $A_{t}^{(0)}$ denote the constant dilaton solution \eqref{cds}. Inserting this formal expansion in the system of equations \eqref{2DeomFG} leads to the following equations for the linear perturbations
\bal
&(\pa^2_{\r}-4\ell^{-2})(e^{-\j})^{(1)}=0\,,\NO\\
&(\pa^2_{\r}-4\ell^{-2})(\sqrt{-\g})^{(1)}+12Q^2e^{5\j_0}(e^{-\j})^{(1)} (\sqrt{-\g})^{(0)}=0\,,\NO\\
&\big(\square^{(0)}_t+K^{(0)}\pa_{\r}-4\ell^{-2}\big)(e^{-\j})^{(1)}=0\,,\NO\\
&\pa_{\r}\big(\pa_t (e^{-\j})^{(1)}/(\sqrt{-\g})^{(0)}\big)=0\,,\NO\\
&\pa_\r A_t^{(1)}=-2Q\big(e^{3\j_0}(\sqrt{-\g})^{(1)}-3e^{4\j_0}(\sqrt{-\g})^{(0)}(e^{-\j})^{(1)}\big)\,.
\eal
The general solution of this system of equations is parameterized by a function $\nu(t)$ and is given by
\bal\label{IRsolutions}
(e^{-\j})^{(1)}=&\;\n(t)e^{2\r/\ell}-\frac{\ell^2}{8\tilde\alpha}\Big(    \Big({\nu'\over\tilde\alpha}\Big)'+{8\tilde\beta\nu\over\ell^2} \Big)e^{-2\r/\ell},\NO\\
(\sqrt{-\g})^{(1)}=&\;-e^{\j_0}\Big((\sqrt{-\g})^{(0)}(e^{-\j})^{(1)}+\frac{\ell^2}{2}\Big({\n'\over \wt\a }\Big)'\Big),\NO\\
A_t^{(1)}=&\;\frac12 Qe^{4\j_0}\pa_\r\big((\sqrt{-\g})^{(0)}(e^{-\j})^{(1)}\big),
\eal
where $\n(t)$ satisfies the differential equation
\be\label{nueq}
\bigg(\Big({\nu'\over\tilde\alpha}\Big)^2-{2\nu\over\tilde\alpha}\Big({\nu'\over\tilde\alpha}\Big)'-{16\tilde\beta\nu^2\over\ell^2\tilde\alpha}\bigg)'=0\,.
\ee
Given the background specified by $\tilde\alpha,\,\tilde \beta,\,\tilde\mu$ and $Q$, the linearized perturbation $\nu$ can be obtained by solving the differential equation \eqref{nueq}. However, since $\nu$ describes the leading growth of the running dilaton, it is sometimes more convenient to treat $\nu(t)$ as arbitrary and use \eqref{nueq} to determine $\tilde \beta$ through the relation 
\be\label{betaeq}
\tilde \beta={\ell^2\over16\nu} 
\bigg( {\nu'^2\over\nu\tilde\alpha}-2\Big({\nu'\over\tilde\alpha}\Big)'-{\Delta_m\tilde\alpha\over\nu}\bigg)\,,
\ee 
with $\Delta_m$ an integration constant. Therefore, the IR fixed point and the perturbations around it can be described by the phase space parameters  $\tilde\alpha,\,\tilde\mu, Q,\,\tilde \beta$ together with the solution of \eqref{nueq}, or equivalently just $\tilde\alpha,\,\tilde\mu, Q,\Delta_m, \,\nu$. 

Comparing the solutions \eqref{IRsolutions} with the IR expansion of the running dilaton solutions in \eqref{RGflow}, we identify
\be\label{IR-perturbation-variables}
\n=\frac{\b^2}{2\sqrt{\ell|Q|}}, \qquad \Delta_m=m-2Q/\ell.
\ee 
This verifies that the perturbative analysis around the constant dilaton solutions reproduces the IR limit of the running dilaton ones. However, the identification between the IR and UV variables, i.e. the relations \eqref{IR-perturbation-variables}, would not be possible without knowledge of the full running dilaton solutions.

\paragraph{First order flow equations}

The RG flow associated with the running dilaton solutions can be further elucidated by the fact that they can be described in terms of first order gradient flow equations derived from a `superpotential'. The superpotential is an exact solution of the radial Hamilton-Jacobi equations\footnote{See \cite{Cvetic:2016eiv} for a derivation of the Hamilton-Jacobi equations from the radial Hamiltonian formulation of the 2D dilaton gravity theory \eqref{2Daction}.} 
\bal\label{HJeqn}
&-\frac{\k_{\rm 2D}^2}{{\sqrt{-\gamma}}}e^{\psi}\gamma_{tt} \left(2{\d\cs\over\d\gamma_{tt}}{\d\cs\over{\d\psi}}+e^{2\psi}\left({\d\cs\over\d A_t}\right)^2\right)-\frac{\sqrt{-\gamma}}{\k_{\rm 2D}^2}(\ell^{-2}-\square_t)e^{-\psi}=0,\NO\\
&-2D_t{\d\cs\over\d\gamma_{tt}}+{\d\cs\over\d\psi}{\pa^t\psi}=0,\NO\\
&-D_t\left({\d\cs\over\d A_t}\right)=0,
\eal 
and takes the form 
\be\label{exactsup1}
\cs[\g_{tt},\j,A_t]=\cu[\g_{tt},\j]-\frac{Q}{\k_{\rm 2D}^2}\int dt\, A_t,
\ee
where
\begin{align}
\label{exactsup2}
\boxed{
\begin{aligned}
\cu[\g_{tt},\j]=&\;\frac{1}{\k_{\rm 2D}^2}\int dt\,\sqrt{-\g}\Bigg[\sqrt{\ell^{-2}e^{-2\j}-\g^{tt}(\pa_t e^{-\j})^2+Q^2e^{2\j}-m}\\
&-\frac{\pa_t e^{-\j}}{\sqrt{-\g}}\log\bigg(\frac{\frac{\pa_t e^{-\j}}{\sqrt{-\g}}+\sqrt{\ell^{-2}e^{-2\j}-\g^{tt}(\pa_t e^{-\j})^2+Q^2e^{2\j}-m}}{\sqrt{\ell^{-2}e^{-2\j}+Q^2e^{2\j}-m}}\bigg)\Bigg]\,.
\end{aligned}
}
\end{align}
In particular, using the functional derivatives of $\cu$ 
\bal
\frac{\d\cu}{\d\j}=&\;-\frac{\sqrt{-\g}\, e^{-\j}\big(\ell^{-2}-Q^2e^{4\j}-\square_t\big)e^{-\j}}{\k_{\rm 2D}^2\sqrt{\ell^{-2}e^{-2\j}-\g^{tt}(\pa_t e^{-\j})^2+Q^2e^{2\j}-m}}\,,\NO\\
\g_{tt}\frac{\d\cu}{\d\g_{tt}}=&\;\frac{\sqrt{-\g}}{2\k_{\rm 2D}^2}\sqrt{\ell^{-2}e^{-2\j}-\g^{tt}(\pa_t e^{-\j})^2+Q^2e^{2\j}-m}\,,
\eal
one can show that the running dilaton solutions \eqref{Rdilaton} satisfy the flow equations 
\be
\pa_\r e^{-\j}=\frac{2\k_{\rm 2D}^2}{\sqrt{-\g}}\g_{tt}\frac{\d\cu}{\d\g_{tt}}\,,\qquad
\pa_\r\sqrt{-\g}=-\k_{\rm 2D}^2e^{\j}\frac{\d\cu}{\d\j}\,,\qquad
\pa_\r A_t=-2Qe^{3\j}\sqrt{-\g}\,.
\ee

The exact superpotential \eqref{exactsup1}-\eqref{exactsup2} fully describes the solution space of the 2D dilaton gravity \eqref{2Daction} and provides an alternative understanding of the RG flow discussed in this subsection. We will see later on that it is also related with the holographic effective action of the dual quantum mechanics. To that end, it is important to notice that the second line in \eqref{exactsup2} is independent of the radial coordinate. This can be proven as follows.

The radial derivative of the superpotential is given by
\be
\pa_\r \cu=\int dt \Big(-e^\j\pa_\r e^{-\j} \frac{\d\cu}{\d\j}+\frac{2}{\sqrt{-\g}}(\pa_\r\sqrt{-\g}) \g_{tt}\frac{\d\cu}{\d\g_{tt}}\Big)=\frac{2}{\k_{\rm 2D}^2}\int dt \sqrt{-\g}\,K \pa_\r e^{-\j}.
\ee    
Moreover, from the first and third equations in \eqref{2DeomFG} follows that
\be
\frac{1}{\k_{\rm 2D}^2}\pa_\r\int dt \sqrt{-\g}\, \pa_\r e^{-\j}=\frac{1}{\k_{\rm 2D}^2}\int dt \sqrt{-\g}\, (K\pa_\r +\pa_\r^2)e^{-\j}=\frac{1}{\k_{\rm 2D}^2}\int dt \sqrt{-\g}\, (2K\pa_\r +\square_t)e^{-\j}.
\ee
We conclude that 
\be
\pa_\r \cu=\frac{1}{\k_{\rm 2D}^2}\pa_\r\int dt \sqrt{-\g}\, \pa_\r e^{-\j}=\frac{1}{\k_{\rm 2D}^2}\pa_\r\int dt \sqrt{-\g}\,\sqrt{\ell^{-2}e^{-2\j}-\g^{tt}(\pa_t e^{-\j})^2+Q^2e^{2\j}-m}\,,
\ee
which shows that the second line in \eqref{exactsup2} is independent of the radial coordinate, as claimed. Its value, therefore, can be determined by inserting the asymptotic form of  the fields. We obtain
\be\label{sup-value}\boxed{
\cu=\frac{1}{\k^2_{\rm 2D}}\int dt\,\sqrt{-\g}\,\sqrt{\ell^{-2}e^{-2\j}-\g^{tt}(\pa_t e^{-\j})^2+Q^2e^{2\j}-m}-\frac{\ell}{\k^2_{\rm 2D}}\int dt\,\frac{(\pa_t\b)^2}{\a\b}.}
\ee

\subsection{3D uplift of 2D solutions}
\label{3D-uplift}

Since the 2D dilaton gravity \eqref{2Daction} is obtained by a consistent KK reduction of AdS$_3$ gravity, any solution of the 2D theory can be uplifted to three dimensions. In this subsection, we show that both the running and constant dilaton solutions can be embedded into the DWCFT phase space of AdS$_3$ gravity. Moreover, the interpolating flow between these solutions can be understood as the $\m^+\to 0$ limit within the DWCFT phase space, or equivalently, as a near horizon expansion of the near extremal BTZ black hole.  

\subsubsection{Embedding into the DWCFT phase space}

To make contact with the DWCFT phase space in section \ref{sec:phase-space}, we make the coordinate identification   
\be\label{coordinate-map}
t=x^+,\qquad w=x^-, 
\ee
while the FG radial coordinate, $\r$, is the same in three and two dimensions. With this coordinate identifications, the map between the 2D and 3D solutions is provided by the ansatz \eqref{2dana}. Namely,
\be\label{uplift}
h_{--}=e^{-2\j},\qquad h_{+-}=e^{-2\j}A_t,\qquad h_{++}=\g_{tt}+e^{-2\j}A_t^2.
\ee

Recall that the DWCFT phase space is described by the variables ${\m^\pm}(x^+)$ and $\F(x^+)$ that parameterize the boundary metric, as well as the functions $\cl_-(z^-)$ and $\cl_+(z^+)$ that enter only in the stress tensor. It follows that, if the solutions of 2D dilaton gravity uplift to the DWCFT phase space of AdS$_3$ gravity, there must exist a map between these variables and those parameterizing the 2D solutions. As we now review, such a map exits for both running and constant dilaton solutions \cite{Cvetic:2016eiv}.

\paragraph{Running dilaton solutions} The running dilaton solutions \eqref{Rdilaton} are parameterized by the functions $\a(x^+)$, $\b(x^+)$, and $\m(x^+)$, as well as as the constants $Q$ and $m$. Comparing boundary metrics provides a map between the DWCFT variables ${\m^\pm}(x^+)$, $\F(x^+)$ and the 2D variables $\a(x^+)$, $\b(x^+)$, and $\m(x^+)$. From the ansatz \eqref{2dana} follows that the boundary metric takes the form
\be\label{dilaton-metric}
ds_{(0)}^2=-\a(x^+)^2(dx^+)^2+\b(x^+)^2\big(dx^-+\m(x^+)dx^+\big)^2.
\ee
Comparing this with \eqref{BD-metricGEN} leads to the identifications
\be\label{map}
\a^2=-e^{2\F}\m^+\Big(\frac{1-\m^+\m^-}{2\m^+}\Big)^2,\qquad
\b^2=-e^{2\F}\m^+,\qquad
\m=\frac{1+\m^+\m^-}{2\m^+}.
\ee
Notice that the uplift of the running dilaton solutions always has $\m^+<0$. 

Inverting these relations determines the variables $\F$, ${\m^\pm}$ in terms of $\a$, $\b$ and $\m$, namely  
\be\label{Imap-prelim}
\frac{1}{\m^+}=\m\pm\frac{\a}{\b},\qquad \m^-=\m\mp\frac{\a}{\b},\qquad e^{2\F}=-\b^2\Big(\m\pm\frac{\a}{\b}\Big),
\ee
where the signs in the three expressions are correlated. Either choice of these signs leads to a consistent embedding of the running dilaton solutions into the DWCFT phase space. However, additional conditions arise from the limit $\m^+\to 0$. Since $\a$ and $\b$ are always positive and $\m^+$ is negative, the limit $\m^+\to 0$, while keeping $\m^-$ fixed, is consistent only if choosing the lower signs.
At the same time, the last relation in \eqref{UV-IR-relations} requires that the linear combination $\m+{\rm sgn}(Q)\a/\b$ be kept fixed in this limit. Clearly, these two conditions are not compatible unless $Q>0$. Similarly, had we identified the KK direction as $w=-x^-$, only negative $Q$ would be consistent with this limit. In the following, therefore, we require that $Q>0$ and set 
\be\label{Imap}
\frac{1}{\m^+}=\m-\frac{\a}{\b},\qquad \m^-=\m+\frac{\a}{\b},\qquad e^{2\F}=-\b^2\Big(\m-\frac{\a}{\b}\Big).
\ee
This map allows us to write the Ricci scalar \eqref{Ricci} in terms of the running dilaton variables, namely
\be
R[g\sub{0}]=\frac{2}{\a\b}\pa_{+}\Big(\frac{\b'}{\a}\Big),
\ee
while the function $\mathscr F$ defined in \eqref{anomaly} becomes 
\be\label{anomaly-ab}
\mathscr F(x^+)= -\frac{\b'}{\a}.
\ee

In order to complete the uplift of the running dilaton solutions we should also determine the variables $\cl_-(z^-)$ and $\cl_+(z^+)$ in the stress tensor in terms of the running dilaton solution parameters. Inserting \eqref{Rdilaton} in the relations \eqref{uplift} and comparing with the general solution of AdS$_3$ gravity in \eqref{3D-metric} determines that the stress tensor \eqref{DWCFT-stress-tensor} takes the form (cf. eq.~(4.21) in \cite{Cvetic:2016eiv}) 
\bal\label{Rdilaton-stress-tensor}
\ct_{--}=&\;\frac{c}{24\p}\Big(\frac{2\b}{\a}\pa_{+}\Big(\frac{\b'}{\a}\Big)-\frac{\b'^2}{\a^2}\Big)+\cl_-+\cl_+,\NO\\
\ct_{-+}=&\mu \ct_{--} +\frac{\a}{\b}\cl_--\frac{\a}{\b}\cl_+,\NO\\
\ct_{++}=&\mu \ct_{-+}-{c\over24\pi}\frac{\b'^2}{\b^2} +\frac{\a}{\b}\Big(\m+\frac{\a}{\b}\Big)\cl_--\frac{\a}{\b}\Big(\m-\frac{\a}{\b}\Big)\cl_+,
\eal
where $\cl_-$ and $\cl_+$ are constants determined in terms of the parameters $m$ and $Q$ via the relations
\be\label{2D-sol}
\cl_-=\frac{c}{48\p}(m+2|Q|/\ell),\qquad \cl_+=\frac{c}{48\p}(m-2|Q|/\ell).
\ee 

\paragraph{Constant dilaton solutions} The constant dilaton solutions \eqref{cds} can be similarly embedded in the DWCFT phase space. In fact, they uplift to the GWCFT space of solutions discussed in subsection \ref{subsec:RF-GWCFT-sol}, which corresponds to the $\m^+\to 0$ limit of the DWCFT phase space. Inserting the constant dilaton solutions \eqref{cds} in the relations \eqref{uplift} leads to the boundary metric
\be\label{cd3d}
ds^2_{(0)}=-2\,{\rm sgn}(Q)\sqrt{\ell|Q|}\,\wt\a\, dx^+\big(dx^-+\wt\m dx^+\big).
\ee
Comparing this with \eqref{BD-metricGEN}, we identify
\be
\label{cd2Ricciflat}
\mu^+=0,\qquad \mu_-=\wt\mu,\qquad e^{2\Phi}=2\,{\rm sgn}(Q)\sqrt{\ell |Q|}\,\wt\a.
\ee
As for the embedding of the $\m^+\to0$ limit of the running dilaton solutions, the uplift of the constant dilaton solutions (with the coordinate identification \eqref{coordinate-map}) requires that the charge $Q$ be positive. 

Matching the first subleading term in the asymptotic expansion the 3D metric obtained from the KK relations \eqref{uplift} with the general form in \eqref{3D-metric} determines that the stress tensor coincides with that of the GWCFT in \eqref{GWCFT-T}, namely (cf. eq.~(4.33)-(4.34) in \cite{Cvetic:2016eiv}) 
\be
\ct_{--}=\cl_-,\qquad 
\ct_{+-}=\wt\mu\,\cl_-,\qquad 
\ct_{++}={\wt\cl}(x^{+})+\wt\mu^{2}\cl_-,
\ee
with
\be\label{2D-sol-cd}
\cl_-=\frac{|Q|}{\k_{\rm 3D}^2},\qquad {\wt\cl}(x^{+})=-\frac{4}{\k_{\rm 3D}^2\ell}\wt\a\wt\b.
\ee

\paragraph{Interpolating flow within the DWCFT phase space} Given that both the running and constant dilaton solutions can be embedded in the DWCFT phase space, it should be possible to describe the entire interpolating flow discussed in subsection \ref{sec:RGflow} within this phase space. We showed earlier that this flow can be understood as the limit 
\be\label{IRlim}
\b\to 0,\quad \a\b\;\;\text{finite},\;\; \m+\frac{\a}{\b}\;\; \text{fixed}, 
\ee
of the running dilaton solutions. The relations \eqref{Imap} translate this to the limit
\be
\m^+\to 0^-,\quad \F\;\;\text{finite},\;\; \m^-\;\; \text{fixed},
\ee 
in the DWCFT phase space. Moreover, the quantities that remain finite coincide and the RG map \eqref{UV-IR-relations} correctly translates them to the parameters of the constant dilaton solutions. In particular, 
\be\label{3D-}
\m^-=\m+\frac{\a}{\b}=\wt\m,\qquad e^{2\F}\to 2\a\b=2\sqrt{\ell|Q|}\,\wt\a, 
\ee
in agreement with \eqref{cd2Ricciflat}. 

Besides the variables that parameterize the boundary metric, however, we need to confirm that the DWCFT stress tensor also interpolates between the running and constant dilaton solutions. In order to address this question, we must first examine the limit $\m^+\to0^-$ within the DWCFT phase space. The stress tensor remains well defined in this limit provided the function $\cl_+(z^+)$ satisfies (see \eqref{mp0} and related discussion in section \ref{sec:phase-space})
\be\label{limit}
-\lim_{\m^+\to0^-}\bar{\cO}_+=\lim_{\m^+\to0^-}\Big[(\m^+)^{-2}\Big(\cl_+(z^+)-\frac{c}{48\p}\big(\mathscr F^2-2\m^+\mathscr F'\big)\Big)\Big]={\wt\cl}(x^+),
\ee
where ${\wt\cl}(x^+)$ is the function appearing the GWCFT stress tensor \eqref{GWCFT-T}, while $\mathscr F$ is a local function of the variables ${\m^\pm}$ and $\F$ defined in \eqref{anomaly}. In order for \eqref{limit} to hold, the function $\cl_+(z^+)$ must be a constant that parametrically is $\co((\m^+)^2)$ as $\m^+\to0$. 

For the running dilaton solutions, this constant is is given in \eqref{2D-sol}, namely
\be
\cl_+=\frac{c}{48\p}\Delta_m.
\ee 
Using this result and the expression for $\mathscr F$ in \eqref{anomaly-ab}, we determine
\be\label{F-extremal}
-\bar{\cO}_+={(\m^+)}^{-2}\Big(\cl_+-\frac{c}{48\p}\big(\mathscr F^2-2\m^+\mathscr F'\big)\Big)=\frac{c}{12\p}\frac{\a^2}{\b^2}\Big[\Delta_m-\frac{\b'^2}{\a^2}+\frac{\b}{\a}\pa_t\Big(\frac{\b'}{\a}\Big)\Big].
\ee
The same result follows from the constant dilaton solution \eqref{2D-sol-cd}, using the UV-IR relation \eqref{UV-IR-relations}. This completes the proof that the $\m^+\to 0$ limit within the DWCFT phase space matches precisely the RG flow between running and constant dilatons solutions of 2D dilaton gravity.

\subsubsection{3D metric and comparison with the BTZ black hole} 

Although we have embedded the space of solutions of 2D dilaton gravity in the DWCFT phase space of 3D gravity, it is instructive to present the explicit form of the 3D metric corresponding to the uplift of generic 2D solutions and their near extremal limit. 

Inserting the boundary metric \eqref{dilaton-metric} and stress tensor \eqref{Rdilaton-stress-tensor} in the 3D metric \eqref{3D-metric} and using the nonlocal coordinates $z^\pm$ defined in \eqref{zs}, we find that the 3D metric obtained from the uplift of the running dilaton solutions can be written compactly as 
\bal
\label{3D-metric-uplift-RD}
ds^2_{\rm 3D}=&\;d\r^2+e^{2\r/\ell}\b^2\Big[1+e^{-2\r/\ell}\b^{-2}\k_{\rm 3D}^2\ell\Big(\Hat\cl_-dz^-(dz^+)^{-1}-\frac{c}{48\p}\frac{\b}{\a}\pa_+\Big(\frac{\b'}{\a}\Big)\Big)\Big]\times\NO\\
&\Big[1+e^{-2\r/\ell}\b^{-2}\k_{\rm 3D}^2\ell\Big(\Hat\cl_+dz^+(dz^-)^{-1}-\frac{c}{48\p}\frac{\b}{\a}\pa_+\Big(\frac{\b'}{\a}\Big)\Big)\Big]dz^-dz^+,
\eal
where 
\be
\Hat\cl_\pm\equiv \cl_\pm+\frac{c}{48\p}\Big(\frac{\b}{\a}\pa_+\Big(\frac{\b'}{\a}\Big)-\frac{\b'^2}{\a^2}\Big),
\ee
and we formally write $(dz^\pm)^{-1}$ to denote elements of the tangent bundle that act on the cotangent bundle through the canonical inner product as e.g.
\be
(dz^+)^{-1} dz^+\equiv \<(dz^+)^{-1},dz^+\>=1,\qquad (dz^+)^{-1} dz^-\equiv \<(dz^+)^{-1},dz^-\>=0.
\ee

Notice that for constant $\b$ the metric \eqref{3D-metric-uplift-RD} reduces to the Ba\~nados metric \cite{Banados:1998gg}. In that case, introducing the coordinates 
\bal
&z^\pm=R_w(\f\mp \bar t/\ell),\qquad \f\sim \f+2\p,\NO\\
&r^2=R_w^2\big(e^{\r/\ell}\b+e^{-\r/\ell}\b^{-1}\k_{\rm 3D}^2\ell\cl_-\big)\big(e^{\r/\ell}\b+e^{-\r/\ell}\b^{-1}\k_{\rm 3D}^2\ell\cl_+\big),
\eal
and setting
\be
2r_+r_-=2R_w^2\k_{\rm 3D}^2\ell(\cl_--\cl_+),\qquad r_+^2+r_-^2=2R_w^2\k_{\rm 3D}^2\ell(\cl_-+\cl_+),
\ee
brings the metric \eqref{3D-metric-uplift-RD} to the general BTZ metric \cite{Banados:1992wn} 
\be\label{BTZ}
ds^2_{\rm 3D}=\frac{\ell^2r^2}{(r^2-r_+^2)(r^2-r_-^2)}dr^2-\frac{(r^2-r_+^2)(r^2-r_-^2)}{\ell^2r^2}d\bar t^2+r^2\Big(d\f+\frac{r_+r_-}{\ell r^2}d\bar t\Big)^2.
\ee
The extremal limit corresponds to $\cl_+=0$ so that $r_+=r_-$.

The uplift of the constant dilaton solutions \eqref{cds} can be easily determined using the KK ansatz \eqref{2dana}, \eqref{2dana2}. The resulting 3D metric takes the form
\be
\label{3D-metric-uplift-CD}
ds^2_{\rm 3D}=d\r^2-2\sqrt{\ell |Q|}\big({\wt\a} e^{2 \r/\ell}-\wt\b e^{-2 \r/\ell}\big)dz^-dx^++\ell |Q|(dz^-)^2-4\wt\a \wt\b (dx^+)^2.
\ee
If $\wt\a$ and $\wt\b$ are constant, the change of coordinates
\bal
&z^-=R_w(\f+ \bar t/\ell),\qquad x^+=\bar t-\ell\f,\qquad \f\sim \f+2\p, \NO\\
&r^2=2\sqrt{\ell |Q|}R_w\ell\big({\wt\a} e^{2 \r/\ell}-\wt\b e^{-2 \r/\ell}\big)+\ell |Q|R_w^2-4\wt\a \wt\b\ell^2,
\eal
brings the metric \eqref{3D-metric-uplift-CD} to the BTZ form \eqref{BTZ} with
\be
2r_+r_-=2\big(\ell |Q|R_w^2+4\wt\a \wt\b \ell^2\big),\qquad r_+^2+r_-^2=2\big(\ell |Q|R_w^2-4\wt\a \wt\b\ell^2\big).
\ee
In the extremal limit $\wt\b=0$ and so $r_+=r_-$. In particular, the near extremal BTZ geometry corresponding to the uplift of the near extremal solutions \eqref{RGflow} to leading order takes the form \eqref{3D-metric-uplift-CD} with $\wt\a$, $\wt\b$ as given in \eqref{UV-IR-relations}-\eqref{extremality}.

\section{Effective action and the complex SYK model}
\label{sec:eff-action}

In section \ref{sec:Dimreduce} we showed that the solution space of 2D dilaton gravity is isomorphic to the DWCFT phase space of 3D gravity, with constant $\cl_-$ and $\cl_+$. We now evaluate the 2D on-shell action, including the KK reduction of the parity violating boundary term \eqref{Ipv}. By analyzing the corresponding local symmetries in two dimensions, we will demonstrate that, in the near extremal limit, the 2D dilaton gravity effective action with GCSS boundary conditions reproduces the low energy effective action of the complex SYK model. Remarkably, the emergence of the CSYK effective action in the near extremal limit is independent of the choice of the spectral function $\rg(\l)$ introduced in section \ref{sec:phase-space}, provided it satisfies the two conditions \eqref{RG-function} and \eqref{slope-condition}. Further evidence for this agreement will be provided by matching several thermodynamic quantities.

\subsection{GCSS on-shell action and variational principle in the near extremal limit}
\label{sec:GCSS-action-NE}

The 2D dilaton gravity action \eqref{2Daction} is renormalized and evaluated on-shell in appendix \ref{sec:on-shell-action-D} for the case of Dirichlet boundary conditions, i.e. without any additional boundary term besides the counterterm \eqref{counterterm}. The result is given in eq.~\eqref{2D-ren-action-2Dvariables-D}, which we reproduce here for convenience:
\be
I_{\rm 2D}^{\rm ren}=\frac{\ell}{2\k^2_{\rm 2D}}\int dt\,\Big(\frac{(\Delta_m+2Q/\ell)\a}{\b}+2\pa_t\Big(\frac{\pa_t\b}{\a}\Big)-\frac{(\pa_t\b)^2}{\a\b}+\frac{2Q\m}{\ell}-\frac{2Q}{\ell}A_t(\r_{\rm min},t)\Big).
\ee
We now include the contribution of the parity violating boundary term $I_{\rm pv}$, which imposes generalized Compère-Song-Strominger (GCSS) boundary conditions, as was shown in section \ref{sec:phase-space}. Moreover, we will focus on the near extremal limit and determine the form of the on-shell action near the bottom of the RG flow discussed in section \ref{sec:RGflow}. As we will show, it is this limit that coincides with the low energy effective dynamics of the CSYK model.

Dimensionally reducing the parity violating boundary term \eqref{Ipv}, the renormalized on-shell action for GCSS boundary conditions is given by
\be\label{ipv}
I_{\rm 2D}^{\rm GCSS}=I_{\rm 2D}^{\rm ren} - 2\p R_w\int dt\; \cj^-{\co}_{-}\,.
\ee
Using the definition of the spectrally flowed variables $\cj^-$, $\co_-$ in \eqref{flow}, as well as the embedding of the 2D solutions into the DWCFT phase space of 3D gravity specified by the relations 
\eqref{Imap} and \eqref{2D-sol} (recall that our choice of embedding requires that $Q>0$), we have
\be\label{JO2d}
\cj^{-}=\rg(\l)\Big(\m+\frac{\a}{\b}\Big),\qquad \co_{-}=-\frac{\ell}{8\p\k_{\rm 2D}^2R_w}\frac{1}{(\l\rg)'}\Big(\Delta_m-\frac{(\pa_t\b)^2}{\a^2}\Big),\qquad \l=\frac{\m+\a/\b}{\m-\a/\b}\,.
\ee
The renormalized on-shell action for GCSS boundary conditions, therefore, takes the form 
\bal\label{2D-ren-action-2Dvariables-GCSS}
I_{\rm 2D}^{\rm GCSS}=&\;\frac{\ell}{2\k^2_{\rm 2D}}\int dt\,\Big[\Big(\frac{\a}{\b}+\frac{\rg}{2(\l\rg)'}\Big(\m+\frac{\a}{\b}\Big)\Big)\Big(\Delta_m-\frac{(\pa_t\b)^2}{\a^2}\Big)+2\pa_t\Big(\frac{\pa_t\b}{\a}\Big)\NO\\
&\hskip5.cm+\frac{2Q}{\ell}\Big(1+\frac{\rg}{(\l\rg)'}\Big)\Big(\m+\frac{\a}{\b}\Big)-\frac{2Q}{\ell}A_t(\r_{\rm min},t)\Big].
\eal
where we have used the relations between the 3D and 2D gravitational constants with the Brown-Henneaux central charge, $c$, specified in eq.~\eqref{2D-3D-Newton}.

Recall that the near extremal limit discussed in section \ref{sec:RGflow} corresponds to $\b\to 0$, while $\a\b$ is kept fixed and $\Delta_m-(\pa_t\b)^2/\a^2=\co(\b^2)$. Assuming that the spectral flow function $\rg(\l)$ satisfies the two conditions \eqref{RG-function} and \eqref{slope-condition}, the leading non-trivial behavior of the renormalized on-shell action \eqref{2D-ren-action-2Dvariables-GCSS} in this limit is 
\be\label{2D-ren-action-2Dvariables-GCSS-beta}\boxed{
I_{\rm 2D}^{\rm GCSS}=\frac{1}{2\k^2_{\rm 2D}}\int dt\,\Big(2Q\big(2\wt\m+\frac{\wt\m^2\n}{\wt\a}-A_t(\r_{\rm min},t)\big)-{8\over \ell}\nu\tilde\beta+\ell\pa_t\Big(\frac{\pa_t\n}{\wt\a}\Big)\Big),}
\ee
where we have used the UV-IR relation \eqref{UV-IR-relations} and \eqref{IR-perturbation-variables}. Note that $\nu$ and $\tilde\beta$ in the on-shell action \eqref{2D-ren-action-2Dvariables-GCSS-beta} are not independent. 
Starting from the background \eqref{cds} parameterized by $\tilde \beta$, the leading coefficient of the running dilaton $\nu$ can be determined by solving the differential equation \eqref{nueq}. Alternatively, we can use \eqref{betaeq} to eliminate $\tilde\beta$, and write the on-shell action in terms of $\nu$ as 
\be\label{2D-ren-action-2Dvariables-GCSS-NE}\boxed{
I_{\rm 2D}^{\rm GCSS}=\frac{1}{2\k^2_{\rm 2D}}\int dt\,\Big(2Q\big(2\wt\m-A_t(\r_{\rm min},t)\big)+\frac{\ell\Delta_m\wt\a}{2\n}+2\ell\pa_t\Big(\frac{\pa_t\n}{\wt\a}\Big)-\frac{\ell(\pa_t\n)^2}{2\n\wt\a}+\frac{2Q\wt\m^2\n}{\wt\a}\Big).}
\ee


\paragraph{Variational principle} The variational principle can be obtained by adding the variation of the new boundary term \eqref{Ipv} to the Dirichlet variational principle \eqref{var2D}, namely  
\be\label{new-principle-2d}
\delta I^{\rm GCSS}_{\rm 2D}=\delta I^{\rm ren}_{\rm 2D}+\delta I_{\rm pv}\equiv \int dt \, \Psi^{\rm 2D}_{\rm GCSS}.
\ee
Focusing on the RG flow solution \eqref{RGsolution} describing the near extremal limit and linearized perturbations, the phase space is parameterized by the arbitrary functions $\tilde \alpha,\,\tilde \mu, \,\nu,\, $ and constants $Q,\, \Delta_m$. 
Using the 3D-2D map \eqref{Imap}, \eqref{2D-sol},  \eqref{JO2d}, together with the UV-IR relations \eqref{UV-IR-relations} and \eqref{IR-perturbation-variables}, the symplectic potential for the new variational principle can be written in terms of the IR variables as  
\be\label{symplectic2d} 
\Psi^{\rm 2D}_{\rm GCSS}=\mathcal O_\alpha\delta\tilde\alpha+\mathcal O_\nu\delta\nu-\mu_{Q}\delta \mathcal O_Q, 
\ee
where 
\bea\label{newIRvariables}
&&\mathcal O_\alpha=-\frac{\ell}{2\k_{\rm 2D}^2} \Big(\frac{\Delta_m}{2\nu}-\frac{(\nu')^2}{2\tilde\alpha^2\nu}-\frac{2Q\nu\tilde\mu^2}{\ell\tilde\alpha^2}\Big),\nonumber
\\&&\mathcal O_\nu=-\frac{\ell}{2\k_{\rm 2D}^2}\Big(\frac{-\Delta_m\tilde\alpha^3+2\tilde\alpha'\nu\nu'+\tilde\alpha(\nu'^{2}-2\nu\nu'')}{2\tilde\alpha^2\nu^2}+\frac{2Q\tilde\mu^2}{\ell\tilde\alpha}\Big)=-{\tilde\alpha\over \nu}\mathcal O_\alpha+{{\ell}\over{2}\nu}\pa_t\Big({\nu'\over\kappa_{\rm 2D}^2\tilde \alpha}\Big), \nonumber
\\&&\mu_{Q}=\tilde\mu\Big(1-\frac{\nu\tilde\mu}{\tilde\alpha}\Big),\nonumber
\\&&\mathcal O_Q=-{Q\over\k_{\rm 2D}^2}\Big(1+{2\nu \tilde\mu\over\tilde \alpha}\Big),
\eea
and we have kept terms up to linearized order in $\nu$. 

The variational principle \eqref{symplectic2d} generalizes that in eq.~(6.28) of \cite{Cvetic:2016eiv} that was obtained from CSS (as opposed to GCSS) boundary conditions in three dimensions, and which was shown to correspond to Neumann boundary conditions on the 2D gauge field $A_t$. The variational principle  \eqref{symplectic2d} includes the leading non-trivial dependence of all symplectic variables on the deformation parameter $\n$, while in \cite{Cvetic:2016eiv} $\n$ appeared only in the variation $\d\n$. In fact, the general form of the new boundary term in \eqref{Ipv} allows us to obtain a consistent variational principle, given in \eqref{new-principle-2d}, at arbitrary values of $\n$.

A natural question is how one could motivate and understand the new boundary term \eqref{Ipv} within the 2D dilaton theory. The technical answer to this question is that the boundary term \eqref{Ipv} determines the class of consistent boundary conditions (parameterized by the spectral function $\rg(\l)$) for the running dilaton solutions that flow to Neumann boundary conditions for the gauge field at the IR fixed point. Notice that these do not amount simply to Neumann boundary conditions for the gauge field in the running dilaton solutions, because the asymptotic behavior of the gauge field in the running and constant dilaton solutions is qualitatively different.

This difference makes the extrapolation of Neumann boundary conditions for the gauge field away from the IR fixed point a non-trivial problem. To leading order in the deformation $\n$ away from the IR fixed point, one could determine the correct boundary term using the variational principle obtained from holographic renormalization near the IR fixed point, which was developed in \cite{Cvetic:2016eiv,Castro:2018ffi}. However, the fact that we have control of the UV completion in this case allows us to adopt a more powerful approach that determines the form of the boundary term throughout the RG flow. 

In particular, we start from the renormalized variational principle \eqref{var2D} of the UV theory, which in 2D is formulated on the phase space of running dilaton solutions. As for 3D gravity, consistent boundary conditions are classified by finite boundary terms that generate canonical transformations on this phase space \cite{Papadimitriou:2007sj}. The procedure of determining the appropriate boundary term within the phase space of running dilaton solutions of 2D dilaton gravity mirrors the analysis of section \ref{sec:phase-space} for the DWCFT phase space of 3D gravity. Using the map between 3D and 2D variables given in \eqref{Imap} and \eqref{2D-sol}, one can follow the procedure in section \ref{sec:phase-space} entirely within the phase space of 2D dilaton gravity, arriving directly at the boundary term \eqref{Ipv}.   

A few remarks may help further elucidate the significance of the new boundary conditions of 2D dilaton gravity imposed by \eqref{Ipv}. Firstly, one may wonder if it is possible to express the boundary term \eqref{Ipv} in terms of the fields of 2D dilaton gravity. It is possible to express any boundary term in terms of the canonical variables of a radial Hamiltonian formulation of the bulk dynamics \cite{Cvetic:2016eiv,Castro:2018ffi}. However, these variables correspond to unrenormalized couplings and observables and do not have well defined transformations under the RG flow. Boundary conditions can be studied more efficiently using renormalized canonical variables, which correspond to symplectic conjugate pairs on the phase space of solutions. This is the approach we adopted here. Nevertheless, one may express the phase space variables, as well the boundary term \eqref{Ipv}, in terms of bulk fields and their conjugate momenta using the explicit form of the running dilaton solutions in \eqref{Rdilaton}.    

Secondly, we should emphasize that the full non-linear boundary term \eqref{Ipv} is consistent only for the specific 2D dilaton gravity theory obtained from 3D gravity, since it is intimately connected with the non-relativistic deformation of 2D CFTs discussed in section \ref{sec:phase-space}. However, its perturbative form around the IR fixed point and the associated variational principle \eqref{symplectic2d}-\eqref{newIRvariables} are applicable to any 2D dilaton gravity with a U(1) gauge field, such as those arising from the consistent KK reduction of higher dimensional black holes \cite{Castro:2018ffi,Moitra:2018jqs}. 

The universality of the perturbative result \eqref{symplectic2d}-\eqref{newIRvariables} is reflected in the fact that it does not depend on the choice of spectral function $\rg(\l)$ parameterizing the full boundary term \eqref{Ipv}, as long as it satisfies the two conditions \eqref{RG-function} and \eqref{slope-condition}. It is this universal perturbative result near extremality that can be compared with e.g. \cite{Hartman:2008dq} and the more recent work \cite{Gonzalez:2018enk,Afshar:2019axx,Godet:2020xpk}. Contrary to our setup, these earlier works adopt exclusively a 2D perspective to the dynamics of near extremal black holes, without reference to a possible UV completion. The fact that they obtain the same symmetry breaking pattern as that derived from the boundary term \eqref{Ipv} in the next section indicates a possible connection between the perturbative form of the boundary conditions used here and in these earlier works. We further comment on some of these earlier works in section \ref{sec:comparison}.

\subsection{Symmetries and conserved charges}\label{Sec:symmetry2D}

The symmetries of the 2D theory can be determined from the symplectic potential which can be obtained from the variational principle \eqref{symplectic2d}. Under 2D PBH and residual gauge transformations
\bea\label{PBH2D}
\xi=\tilde\varepsilon(t)\pa_t+\sigma(t)\pa_\rho,\qquad \delta A_t= \pa_t \big(\tilde\varphi(t)+\tilde\varepsilon(t) A_t\big),
\eea
where $\tilde\varepsilon(t)$ denotes time reparameterizations, $\sigma(t)$ parameterizes Weyl transformations, and $\tilde\varphi(t)$ is an arbitrary gauge parameter.
The parameters of the 2D RG flow solutions \eqref{RGsolution} transform as 
\be\label{1dPBH}
\delta Q=0,\qquad 
\delta\tilde\alpha=\pa_t(\tilde\varepsilon\tilde\alpha)+2\tilde\alpha\sigma/\ell,\qquad 
\delta\nu=\tilde\varepsilon\pa_t\nu+2\nu\sigma/\ell,\qquad 
\delta\tilde\mu=\pa_t(\tilde\varepsilon\tilde\mu+\tilde\varphi).
\ee 
Moreover, evaluated on such transformations the symplectic potential \eqref{symplectic2d} becomes 
\bal\label{psigcss2d0}
\Psi^{\rm 2D}_{\rm GCSS}=&\;
-\frac{\ell}{\k_{\rm 2D}^2}\pa_t\Big[\Big({\ell\Delta_m\over 4}-\frac{\nu'^{2}}{4\tilde\alpha^2}-\frac{ Q\nu^2\tilde\mu^2}{\tilde\alpha^2}\Big)\tilde\varepsilon-\frac{2\nu Q\tilde\mu}{\ell\tilde\alpha} \varphi\Big] \NO\\
&+\sigma\pa_t\Big(\frac{\nu'}{\k_{\rm 2D}^2\tilde\alpha}\Big) -\varphi\pa_t \Big(\frac{2\nu Q\tilde\mu}{\k_{\rm 2D}^2\tilde\alpha}\Big)+\mathcal{O}(\nu^2).
\eal
As discussed in section \ref{sec:phase-space}, local symmetries are symmetries that leave the action invariant, which means that the second line of \eqref{psigcss2d0} vanishes given that the Ward identities are satisfied. The symplectic potential \eqref{symplectic2d} implies that the sources are now $\tilde\alpha,\nu$ and ${\mathcal O}_Q$.
The global symmetries can be determined by imposing the boundary condition
\be\label{glsymondition}
\delta\tilde \alpha=\delta\nu=\delta{\mathcal O}_Q=0, 
\ee
which we will discuss separately for $\nu=0$ and for small but non-zero $\nu$.

\paragraph{The extremal limit}
At the extremal limit, $\nu=0$, time reparameterizations, Weyl and gauge transformations are all local symmetries and leave the action invariant, as the corresponding Ward identities are all satisfied. 
Imposing the boundary condition \eqref{glsymondition} we get the global symmetries
\bea\label{1dbckv0}
\xi= \tilde\varepsilon(t)\pa_t-{\ell\over\tilde \alpha}\pa_t\big(\tilde\varepsilon(t) \tilde \alpha\big)\pa_\rho, \qquad \tilde\varphi(t),
\eea
which form a Witt-Kac-Moody algebra with all the conserved charges vanishing. This is consistent with the fact that there are no finite energy excitations in AdS$_2$ with constant dilaton.

\paragraph{Away from extremality}
For small but non-vanishing $\nu$,  the Ward identities can be read from the second line of the symplectic potential \eqref{psigcss2d0}.  For general time-dependent solutions, time reparameterizations are still a local symmetry, while the the Ward identities for the Weyl symmetry and U(1) guage symmetry are both anomalous. We are interested in the charged black hole solutions \eqref{RGsolution} with linear dilaton specified by
\be\label{stationary}
\tilde\alpha=\tilde\alpha_0, \qquad \tilde\mu=\tilde\mu_0,\qquad\nu=\nu_0.
\ee
On such backgrounds, the Ward identities are satisfied and the local symmetries are still parameterized by $\tilde \varepsilon,\, \tilde\varphi,$ and $\sigma$.

The global symmetries are two $U(1)s$
\be\label{1dbckv}
\tilde\varepsilon=b_1\frac{\nu}{\tilde\alpha},\qquad \sigma=-\frac{b_1\ell\nu'}{2\tilde\alpha},\qquad\tilde\varphi=b_2-b_1\frac{\tilde\mu\nu}{\tilde\alpha}.
\ee
To calculate the corresponding charges we use the Noether method and promote $b_1$ and $b_2$ to arbitrary functions of time. Substituting the variation \eqref{1dbckv} into the symplectic potential \eqref{symplectic2d} we get
\bal\label{psigcss2d}
\Psi^{\rm 2D}_{\rm GCSS}=&\;
-\pa_t\bigg(\Big[\frac{\ell}{\k_{\rm 2D}^2}\Big({\ell\Delta_m\over 4}-\frac{\nu'^{2}}{4\tilde\alpha^2}+\frac{ Q\nu^2\tilde\mu^2}{\tilde\alpha^2}\Big)\Big]b_1-\frac{2\nu Q\tilde\mu}{\k_{\rm 2D}^2\tilde\alpha} b_2\bigg)\NO\\
&+\frac{b_1\ell}{\k_{\rm 2D}^2}\pa_t\Big({\ell\Delta_m\over 4}-\frac{\nu'^{2}}{4\tilde\alpha^2}+\frac{ Q\nu^2\tilde\mu^2}{\tilde\alpha^2}\Big)-b_2\pa_t \Big(\frac{2\nu Q\tilde\mu}{\k_{\rm 2D}^2\tilde\alpha}\Big).
\eal

The second line in \eqref{psigcss2d} vanishes on such black hole solutions, and we can read the conserved charges from the first line
\be\label{charges2D}
Q[b_1]=\frac{b_1\ell}{\k_{\rm 2D}^2} \Big(\frac{\ell\Delta_m}{4}+\frac{Q\nu^2\tilde\mu^2}{\ell\tilde\alpha^2}\Big),\qquad Q[b_2]=-b_2\frac{2\nu Q\tilde\mu}{\k_{\rm 2D}^2\tilde\alpha}.
\ee
In order words, the linearized perturbation breaks the Witt-Kac-Moody symmtry to two U(1)s.
This result is consistent with our analysis in section \ref{sec:phase-space}.
In particular, the global symmetry \eqref{1dbckv} for $\nu\neq0$ is consistent with result from 3D analysis at finite $\mu^+$, while \eqref{1dbckv} at $\nu=0$ is consistent with the $\mu_+\to0$ limit. 

\subsection{2D black hole thermodynamics with GCSS boundary conditions}
\label{sec:2dthermo}
The black hole solution specified by \eqref{stationary} describes a near extremal black hole
and can be expanded as a constant dilaton solution with perturbations in the form of \eqref{RGsolution}.
The constant dilaton background is 
\bal\label{IRs0} 
e^{-\psi_0}=&\;\sqrt{\ell Q},\nonumber\\
(\sqrt{-\gamma})^{(0)}=&\;\tilde\alpha_0e^{2\rho/\ell}+\tilde\beta_0e^{-2\rho/\ell},\nonumber\\
A_t^{(0)}=&\;\tilde\mu_0-\frac{1}{\sqrt{\ell Q}}\big(\tilde\alpha_0 e^{2\rho/\ell}-\tilde\beta_0e^{-2\rho/\ell}\big).
\eal
The linearized order solution drives the RG flow and can be written as
\bal\label{IRs1}
(e^{-\psi})^{(1)}=&\;\nu_0\Big(e^{2\rho/\ell}-\frac{\tilde\beta_0}{\tilde\alpha_0}e^{-2\rho/\ell}\Big),\nonumber\\
(\sqrt{-\gamma})^{(1)}=&\;-e^{\psi_0}(\sqrt{-\gamma})^{(0)}(e^{-\psi})^{(1)},\nonumber\\
A_t^{(1)}=&\;\frac{2\nu_0\tilde\alpha_0}{\ell Q}\Big(e^{4\rho/\ell}+\frac{\tilde\beta_0^2}{\tilde\alpha_0^2}e^{-4\rho/\ell}\Big).
\eal
The extremal limit corresponds to $\nu_0\to 0$ with all other variables kept fixed. In the perturbative analysis hereafter, we will keep terms up to the linearized order in $\nu_0$.

Now we calculate the thermodynamic variables for the constant dilaton black hole solution \eqref{IRs0}  together with the linearized perturbation \eqref{IRs1}. The  horizon of the above 2D black hole  can be determined by solving the equation $\gamma_{tt}=0$, and is given by 
\be e^{2\rho_h\over\ell}= \sqrt{-\tilde\beta_0\over\tilde\alpha_0}.
\ee 
To calculate the temperature and 2D mass, we need to first determine the normalization of the horizon generator 
\be 
\xi_h=b_1 {\nu_0\over\tilde \alpha_0}\pa_t.
\ee 
A natural normalization is to choose $\xi_h=\pa_t$ and hence $b_1={\tilde \alpha_0\over\nu_0}.$ \footnote{See \cite{Castro:2018ffi} for a discussion of this normalization factor.}
The temperature is the surface gravity associated to the horizon generator  
\bea\label{T2d}
&&T_{H}=\frac{1}{2\pi}\pa_\rho\sqrt{-\gamma}\Big|_{\rho_h}={2\over\pi\ell}\Big(\sqrt{-\tilde\alpha_0\tilde\beta_0}+{2\tilde\beta_0\nu_0\over\sqrt{\ell Q}}\Big)\,.
\eea 
The 2D mass $M_{\rm 2D}$ is the conserved charge associated with the  global symmetry parameterized by $b_1$ 
\be\label{mass2D} 
M_{\rm 2D}\equiv Q[b_1={\tilde \alpha_0\over\nu_0}]={\nu_0\over\k_{\rm 2D}^2}\Big(-4\tilde\beta_0+\frac{\ell Q\tilde\mu_0^2}{\tilde\alpha_0}\Big),
\ee
where we have used that $\Delta_m=-{16\tilde\beta_0\nu_0^2\over\ell^2 \tilde\alpha_0}$ from \eqref{betaeq}.
The entropy is determined by the value of the dilaton at the horizon and takes the form
\be\label{s2d}  
S_{\rm 2D}=\frac{2\pi}{\k_{\rm 2D}^2}e^{-\psi(\rho_h)}=
\frac{2\pi}{\k_{\rm 2D}^2}\bigg(\sqrt{\ell Q}+2\nu_0\sqrt{-\tilde\beta_0\over\tilde\alpha_0}\bigg).
\ee
Finally, from \eqref{charges2D},  we see that the global U(1) charge from the Maxwell field is given by
\be\label{charge2D}
\mathcal Q_{\rm 2D}= Q[b_2=1]=-\frac{2\nu_0 Q\tilde\mu_0}{\k_{\rm 2D}^2\tilde\alpha_0}.
\ee

Notice that the temperature and entropy coincide with the corresponding quantities for the near extremal limit of the BTZ black hole \eqref{3D-metric-uplift-CD} (see eq.~(2.13) in \cite{Cvetic:2016eiv}), but the new boundary conditions modify the mass and electric charge of the 2D black hole.
However, the deviation of the mass from extremality is not affected by our choice of boundary conditions and, as we will see shortly, reproduces the well known ``mass gap'' for the BTZ black hole \cite{Preskill:1991tb,Almheiri:2016fws}. Moreover, the modified mass and electric charge still satisfy the expected thermodynamic relations.

\subsection{Near extremal effective action and the CSYK model}
\label{sec:effective-action}

In this subsection we derive the effective action near the IR fixed point described by the constant dilaton solution \eqref{IRs0}, and show that it agrees with the effective action of the SYK model with complex fermions \cite{Davison:2016ngz,Chaturvedi:2018uov}. 

\subsection*{Effective action of 2D EMD gravity}
Let us consider performing finite 2D PBH transformations which are obtained by integrating the infinitesimal transformations \eqref{PBH2D}. Note that $\sigma$ and $\tilde\varphi$ can be directly integrated and therefore we keep the same notation for finite boundary Weyl and gauge transformations. We use $f(t)$ to denote finite boundary time reparameterizations so that
\be
f(t)=t+\tilde\varepsilon+\mathcal O(\varepsilon^2).
\ee
Under the finite PBH and gauge transformations
\be
t\to f(t)+\mathcal O(e^{-2\rho/\ell}),\qquad  e^{2\rho/\ell}\to e^{2\rho/\ell}(e^{2\sigma(t)/\ell}+\mathcal O(e^{-2\rho/\ell})),\qquad A_tdt\to \pa_t(\tilde\varphi+A_tf)dt\,,
\ee
the phase space parameters transform as 
\be
\tilde\alpha\to\tilde\alpha f'e^{2\sigma/\ell},\qquad \nu\to\nu e^{2\sigma/\ell},\qquad \tilde\mu\to\tilde\mu f'+\tilde\varphi'.
\ee
The phase space of 2D EMD theory can then be classified by distinct orbits under these transformations. Any point in the phase space can be expressed in terms of the symmetry parameters $f,\sigma,\tilde\varphi$, together with a reference solution. As discussed earlier in section \ref{Sec:symmetry2D}, time reparameterization is a local symmetry, and leaves the local part of the action invariant. It is also straightforward to check that the global part of the action \eqref{Iglobal} is also invariant. 
Therefore, we can use this freedom to choose $f$ such that $\tilde\alpha$ is fixed , i.e.
\be\label{param2}
e^{-2\sigma/\ell}=f'.
\ee In the following, we will use \eqref{param2} to equivalently eliminate $\sigma$, while keeping $f$ arbitrary. 
We start with the stationary constant dilaton background \eqref{cds} with 
$\tilde\alpha=\alpha_0, \, \tilde\beta=\beta_0,\,\tilde\mu_0$, 
and a PBH transformation with \eqref{param2} will generate a new constant dilaton background with 
\be\label{background}
\tilde\alpha=\tilde\alpha_0,\qquad\tilde\beta=(f')^2\tilde\beta_0+\frac{\ell^2}{8\tilde\alpha_0}\{f,t\},\qquad \tilde\mu=\tilde\varphi'+\tilde\mu_0f'.
\ee
Evaluating the total action \eqref{2D-ren-action-2Dvariables-GCSS-beta} under the reparameterizations \eqref{background}  for the background gives
\bea\label{sch-act}
I^{\rm GCSS}_{\rm 2D}&=&-\frac{\ell}{2\kappa_{\rm 2D}^2}\int dt\Big[\frac{\nu}{\tilde\alpha_0}\Big(\{f,t\}+{8\over\ell^2}\tilde\alpha_0\tilde\beta_0(f')^2-\frac{2Q}{\ell}[\pa_t(\tilde\varphi+\tilde\mu_0f)]^2\Big)
\\&&-\pa_t\Big(\frac{\pa_t\nu}{\tilde\alpha_0}\Big)-\frac{4Q\pa_t(\tilde\varphi+\tilde\mu_0f)}{\ell}+\frac{2Q}{\ell}A(\rho_{min},t)\Big]\,.\nonumber
\eea
The leading running dilaton coefficient $\nu(t)$ in \eqref{sch-act} is not arbitrary but is determined by the equation of motion 
\eqref{nueq}.

Now let us view \eqref{sch-act} as an effective action for the fields $f(t)$ and $\tilde \varphi$, with $\nu$ an external coupling. Varying $f(t)$ and $\tilde \varphi$ we obtain the following equations of motion
\be\label{nueq2}
\Big[\frac{1}{f'}\Big(\frac{(\nu f')'}{f'}\Big)'+{16\over\ell^2}\tilde\alpha_0\tilde\beta_0(\nu f')\Big]'=0,\qquad
\pa_t( \nu \pa_t(\tilde \varphi+\tilde\mu_0 f))=0.
\ee
It is straightforward to check that  the left hand side of the bulk equation \eqref{nueq} evaluated on the background \eqref{background} is just \eqref{nueq2} multiplied  by $-2\nu f'$. 
Therefore, for non-zero perturbation $\nu$ and non-singular transformation function $f$, \eqref{nueq} and \eqref{nueq2} are equivalent.
The general solution of \eqref{nueq2} is given by 
\bea\label{sl2}
&&\tanh \pi T f(u)={a\tanh\pi T u+b\over c\tanh\pi T u+d},   \qquad \tilde\varphi=-\tilde\mu_0 (f(u)-u)+\tilde\varphi_0,\NO\\
&& ad-bc=1, \qquad T={2\over\pi\ell}\sqrt{-\tilde\alpha_0\tilde\beta_0},\label{T-from-action}
\eea
where $u$ is the coordinate reparameterization 
\be
du=\nu_0dt/\nu(t),
\ee
and $\nu_0$ is chosen so that the period of the imaginary time is preserved, namely $u\sim u+{i\over T}, \, t\sim t+{i\over T}$. The solution \eqref{sl2} can be specified by four independent parameters, and can be generated by an $SL(2,\mathbb R)\times U(1)$ transformation. 

Using the property of the Schwarzian derivative
\be
(\pa_t u)^2\{f,u\}=\{f,t\}-\{u,t\},
\ee
we can rewrite the action \eqref{sch-act} in the $u$ coordinate as 
\be\label{sch-act-nu0}
I^{\rm GCSS}=I^{\rm GCSS}_0+I^{\rm GCSS}_{\rm eff},
\ee
where
\bal
I^{\rm GCSS}_0=&\;\frac{Q}{\kappa_{\rm 2D}^2}\int du \big(2\pa_u(\tilde\varphi+\tilde\mu_0f)-A(\rho_{min},u)\big),\NO\\
I^{\rm GCSS}_{\rm eff}=&\;-\frac{\nu_0\ell}{2\tilde\alpha_0\kappa_{\rm 2D}^2}\int du\Big(\{f,u\}-2\pi^2 T^2\pa_u f^2-\frac{2Q}{\ell}\big(\pa_u(\tilde\varphi+\tilde\mu_0f)\big)^2\Big),\label{sch-act-eff}
\eal
and we have omitted terms for $\nu$ which are decoupled from the $f(t)$ and $\tilde\varphi$ degrees of freedom, and separated the dynamical part $I^{\rm GCSS}_{\rm eff}$ from the background part $I_0^{\rm GCSS}$. We have shown that the solution of \eqref{sch-act-nu0} is \eqref{sl2}, which reproduces the linearized perturbation from the bulk analysis \eqref{nueq}. It is therefore consistent to treat \eqref{sch-act-nu0} as an off-shell action.

\paragraph{From 2D EMD gravity to complex SYK}
The SYK model with $N$ complex fermions \cite{Davison:2016ngz} has an emergent conformal symmetry in the low temperature and large-$N$ limit. In \cite{Chaturvedi:2018uov} it was observed that the low temperature dynamics of the complex SYK model can be described by the symmetry broken phase of a WCFT. We now show that 2D EMD gravity with generalized CSS boundary conditions has the same low energy effective action.

To set up the connection, we need to perform a Wick rotation of the 2D EMD theory using
\be\label{wick2}
 u\to i\tau, \qquad \tilde\mu_0\to-i\bar\mu\,,\qquad f\to i\bar f,\qquad \tilde\varphi\to i\Lambda,
\ee
which put the theory on the Euclidean circle
\be \tau\sim \tau+{1\over T}.\ee
The Euclidean effective action \eqref{sch-act-eff} then takes the following form
\begin{align}
\begin{aligned}\label{Ieff}
I_{\rm eff}^{\rm GCSS}=&-\frac{\nu_0\ell}{2\tilde\alpha_0\kappa_{\rm 2D}^2}\int d\tau \Big(\{\tan(\pi T_{}\bar f,\tau\}+\frac{2Q}{\ell}(\Lambda'-i\bar\mu\bar f')^2\Big),
\end{aligned}
\end{align}
and agrees with the effective action of the complex SYK model in the IR
\be\label{sch-SYK}
I_{\rm eff}^{\rm GCSS}=I_{\rm eff}^{\rm CSYK}=-\frac{N\gamma}{4\pi^2}\int d\tau\Big(\{\tan(\pi T\bar f,\tau\}-\frac{2\pi^2K}{\gamma}(\Lambda'-i\bar\mu\bar f')^2\Big)\,,
\ee
provided we identify 
\be
\frac{\nu_0\ell}{2\tilde\alpha_0\kappa_{\rm 2D}^2}=\frac{N\gamma}{4\pi^2}\,,\qquad {Q\over\ell}=-{\pi^2 K\over \gamma}\,, \ee
where $N$ is the number of fermions,  $T$  is the temperature, $K$ is the compressibility and $\g$ is the specific heat. 
That is, the effective action of 2D EMD theory with our new boundary term agrees with that of the complex SYK models, with the dictionary \eqref{sch-SYK}.

\paragraph{From 3D gravity to complex SYK}
Recall that the central charge $c$ and level $k$ obtained from AdS$_3$ are given by 
\be
\frac{c}{12}=\frac{\ell}{2R_w\kappa_{\rm 2D}^2}\,,\qquad k=-\frac{2cQ}{3\ell}\,.
\ee
It follows that we can further relate AdS$_3$ gravity with generalized CSS boundary conditions with the complex SYK theory by identifying 
\be
\frac{c}{k}=\frac{3\gamma}{2\pi^2K}\,,\qquad \frac{R_w\nu_0c}{12\tilde\alpha_0}=\frac{N\gamma}{4\pi^2}\,.
\ee
The first relation is exactly the result obtained in \cite{Chaturvedi:2018uov}, where it was observed that complex SYK models can be described as the symmetry broken phase of WCFT.

\paragraph{Thermodynamics at the saddle point}

We are particularly interested in the saddle point \be
\bar f=\tau,\qquad \Lambda=0\,,
\ee which corresponds to the stationary RG flow solution with $\tilde \alpha_0,\,\tilde\beta_0,\tilde\mu_0 ,\,\nu_0$,  namely the background \eqref{IRs0} with linearized perturbation \eqref{IRs1}.  In the following we show that the on-shell action can correctly reproduce all the thermodynamic quantities that  we directly calculated from the bulk analysis in section \ref{sec:2dthermo}. 

First of all, the temperature arising from the periodicity \eqref{T-from-action} is just the black hole temperature derived using the smoothness condition at the horizon, 
\be 
T=T_H\,.
\ee
It is natural to choose $\rho_{min}$ in \eqref{sch-act-nu0} to be the horizon value $\rho_h$, and impose the trivial holonomy condition so that $A(\rho_h)=0$. 
Evaluating then the saddle point value of the Euclidean effective action \eqref{sch-act-nu0} gives
\be
I_E^{\rm GCSS}= I_0^{\rm GCSS}+I^{\rm GCSS}_{\rm eff},
\ee
where
\be
I_0^{\rm GCSS}=-\frac{2\pi\sqrt{\ell Q}}{\k_{\rm 2D}^2},\qquad 
I^{\rm GCSS}_{\rm eff}=-\frac{\nu_0\ell }{\tilde\alpha_0\k_{\rm 2D}^2 T}\Big(\pi^2T^2-\frac{Q\bar\mu^2}{\ell}\Big).
\ee
The on-shell action is a function of the temperature and chemical potential and can be identified with the grand canonical potential as 
\be
\Omega=\Omega_0+\Omega_{\rm eff}\equiv T(I_0^{\rm GCSS}+I^{\rm GCSS}_{\rm eff}).
\ee
From the grand canonical potential we obtain the thermal entropy, charge, and energy as 
\bea\label{S1d}
S&\equiv& -\Big(\frac{\pa\Omega}{\pa T_{\rm 2D}}\Big)_{\bar \mu}=S_0+S_1, \\
 S_0&\equiv& -\Big(\frac{\pa\Omega_{0}}{\pa T_{\rm 2D}}\Big)_{\bar \mu}=\frac{2\pi\sqrt{\ell Q}}{\k_{\rm 2D}^2},
\qquad S_1\equiv -\Big(\frac{\pa\Omega_{\rm eff}}{\pa T_{\rm 2D}}\Big)_{\bar \mu}={\pi^2 \ell\over 3}c_{\rm eff} T,
\\\mathcal Q&\equiv&-\Big(\frac{\pa\Omega}{\pa \bar\mu}\Big)_{T}={1\over2}k_{\rm eff }\bar \mu\ell,
\\E&\equiv&\Omega+T S+\bar\mu\mathcal Q={\pi^2\ell\over6}c_{\rm eff} T_{}^2+{{\mathcal Q}^2\over \ell k_{\rm eff}},\label{E1d}
\eea
where we have introduced the effective central charge $c_{\rm eff}$ and level $k_{\rm eff}$ as 
\bea\label{ceff}
c_{\rm eff} = \frac{6\nu_0}{\tilde\alpha_0\k^2_{\rm 2D}}, \qquad k_{\rm eff}=-{4Q\nu_0\over \ell\tilde\alpha_0\k_{\rm 2D}^2}.\label{k2d}
\eea
We have also split the total entropy into a zero temperature part, $S_0$, and a deviation part, $S_1$. The entropy increase $S_1$ from zero temperature can alternatively be written in the microcanonical ensemble as 
\be 
S_1=2\pi \sqrt{{c_{\rm eff}\over6}\Big(\ell E-{\mathcal{Q}^2\over  k_{\rm eff} }\Big) },
\ee 
which formally looks like the entropy formula for the right moving part of WCFT \cite{Chaturvedi:2018uov}.
It is also interesting to note that the ratio between the effective central charge and  effective level in 2D is the same as those in 3D gravity with CSS boundary conditions, i.e. 
\be 
{c_{\rm eff}\over k_{\rm eff}}={c\over k},
\ee 
which is consistent with the dimensional reduction. 

Finally, as a consistency check, one can verify that the thermodynamic quantities  \eqref{S1d}-\eqref{E1d} agree with the Euclidean version of the bulk results \eqref{mass2D}-\eqref{charge2D}, with 
\be 
S=S_{\rm 2D},\qquad \mathcal{Q}=\mathcal{Q}_{\rm 2D}, \qquad E=M_{\rm 2D}.
\ee
Therefore, the boundary effective action \eqref{sch-act-nu0} can indeed reproduce the thermodynamics of near extremal black holes with running dilaton. Moreover, from \eqref{E1d} follows that, in agreement with the first law of thermodynamics, the deviations of the mass and entropy away from extremality satisfy
\be
E-{{\mathcal Q}^2\over \ell k_{\rm eff}}=M_{\rm gap}^{-1} T^2,\qquad S_1= 2M_{\rm gap}^{-1} T,
\ee
where the ``mass gap'' $M_{\rm gap}$ \cite{Preskill:1991tb,Almheiri:2016fws,Castro:2018ffi,Ghosh:2019rcj} is given by
\be
M_{\rm gap}^{-1}={\pi^2\ell\over6}c_{\rm eff}=\frac{\pi^2\ell\nu_0}{\tilde\alpha_0\k^2_{\rm 2D}}.
\ee
This result depends on the normalization of the timelike Killing vector, but seems to be a universal near horizon property of near extremal black holes. In particular, it is not affected by our choice of boundary conditions and coincides with the mass gap of the near extremal Kerr-AdS$_5$ black hole \cite{Castro:2018ffi}. With a suitable choice for the normalization of the timelike Killing vector, this mass gap agrees with that obtained from the far-from-the-horizon asymptotic region of near extremal black holes, but this normalization is not universal since it depends on the specific UV completion.   

\subsubsection{Comments on related models}
\label{sec:comparison}

Finally, let us briefly compare our results with other literature on two-dimensional models of dilaton gravity with a Maxwell field. Our bulk action and the Hartman-Strominger model can both be put into the following general form, 
\bea\label{generalbulk}
I_{\rm bulk}=\frac{1}{2\k^2_{\rm 2D}}\int d^2x \sqrt{-g}\; e^{-\j}\Big(R[g]+{1\over\ell^2}V[\psi]-\frac{\ell^2} {4} Z[\psi]F_{\m\n}F^{\m\n}\Big)\,.
\eea
In particular, for \cite{Hartman:2008dq}, $V[\psi]=8,\quad Z[\psi]=e^{\psi}.$ In our model, $V[\psi]=2,\quad Z[\psi]=\frac {e^{-2\psi}}{\ell^2}.$ 
The phase space of \cite{Hartman:2008dq} always contains a locally AdS$_2$ metric, with constant dilaton as well as linear dilaton. 
\cite{Hartman:2008dq,Castro:2014ima} discussed the unbroken Virasoro-Kac-Moody symmetry on the constant dilaton background of both models. The currents generating these symmetries have two anomalies, the Virasoro anomaly $c_{\rm 2d}$ and the Kac-Moody anomaly $k_{\rm 2d}$\footnote{The ratio \eqref{ratio} refers to Appendix B of \cite{Castro:2014ima}, which puts the currents of \cite{Hartman:2008dq} into the standard Virasoro-Kac-Moody form.}. 
It is interesting to observe that the two anomalies in both models are related by 
\be
{c_{\rm 2d}\over k_{\rm 2d}}=-{3\over2}{e^{-\psi_0}\over Z[\psi_0] }\,.\label{ratio}
\ee
In this paper, we further consider linear dilaton perturbations around the constant dilaton background, and after adding the boundary term \eqref{ipv}, the  Virasoro-Kac-Moody algebra is both explicitly and spontaneously broken.  We note the coefficients in front of the Schwarzian term and broken $U(1)$ term  \eqref{Ieff} inherit the ratio \eqref{ratio}, namely, 
\be 
{c_{\rm eff}\over k_{\rm eff}}= {c_{\rm 2d}\over k_{\rm 2d}}
\ee
where $c_{\rm eff}$ and $k_{\rm eff}$ are given by \eqref{ceff}.
The Schwarzian term is a universal term appearing in many two-dimensional dilaton-gravity models, with \cite{Cvetic:2016eiv} and without \cite{Maldacena:2016upp} Maxwell fields. On the other hand, the breaking of the local $U(1)$ symmetry is due to the new boundary term \eqref{ipv}. 
We expect a term similar to \eqref{ipv} can be added to a more general class of models \eqref{generalbulk}, so that the effective action is also similar to that of the complex SYK model  \eqref{Ieff}, with coefficients compatible with \eqref{ratio}.

Another bulk theory featuring the effective action of the complex SYK model has also been proposed in \cite{Gaikwad:2018dfc}, which comes from the dimensional reduction of three-dimensional gravity coupled with Chern-Simons $U(1)$ theory. The effective action also has the structure of \eqref{Ieff}, but the broken $U(1)$ term is purely from the additional Chern-Simons $U(1)$ action. In fact, the analog of $k_{\rm eff}$ is just the Chern-Simons level, and is independent of the value of the constant. Another important difference is that the analysis of \cite{Gaikwad:2018dfc} is directly on the linear dilaton background, instead of a perturbation above constant dilaton background as in the present paper.

\section{Discussion}\label{sec:discussion}

In this paper we determined a non-conformal generalization of the CSS boundary conditions of AdS$_3$ Einstein gravity that can be holographically identified with a local irrelevant deformation of WCFT. Using a consistent KK reduction to two dimensions and evaluating the renormalized on-shell action on near extremal solutions subject to this new boundary condition, we have uncovered specific relations connecting AdS$_3$ Einstein gravity, 2D dilaton gravity with a U(1) gauge field, deformed WCFTs, and complex SYK models. These relations are pictorially summarized in fig. \ref{relations}. In particular, our new boundary condition reproduces the symmetry breaking pattern exhibited at low energies by the complex SYK models. 

Besides the concrete application to the complex SYK models we focused on here, the framework we developed for classifying AdS$_3$ and AdS$_2$ boundary conditions maybe useful in the context of a number of recent developments. For example, it would be interesting to classify consistent boundary conditions in the presence of a gravitational Chern-Simons term \cite{Castro:2019vog}. Moreover, the local WCFT deformation we have identified corresponds to a non-relativistic 2D RG flow that could potentially shed light on the existence of a non-relativistic c-theorem. Another interesting application of our framework would be in the context of non-local integrable deformations of WCFTs. Finally, recent approaches to AdS$_2$ \cite{Saad:2019lba,Witten:2020wvy,Johnson:2019eik} and AdS$_3$ \cite{Cotler:2018zff,Cotler:2020ugk,Maxfield:2020ale,Cotler:2020lxj} quantum gravity typically focus on specific boundary boundary conditions. It would be very interesting apply our techniques to identify and explore different boundary conditions, as was recently done for AdS$_2$ in \cite{Goel:2020yxl}.

\section*{Acknowledgments}
We are grateful to Luis Apolo, Alejandra Castro, Yingfei Gu, and Yuan Zhong for helpful discussions. The work of PC, WS and BY was supported by the National Thousand-Young-Talents Program of China and NFSC Grant No. 11735001. The work of IP is supported by a KIAS Individual Grant (PG064402) at the Korea Institute for Advanced Study. We thank the Tsinghua Sanya International Mathematics Forum for hospitality during the workshop and research-in-team program on “Black holes, Quantum Chaos, and Solvable Quantum Systems”, where the collaboration got started. 

\appendix

\renewcommand{\theequation}{\Alph{section}.\arabic{equation}}

\setcounter{section}{0}

\section{Local symmetry transformations of spectrally flowed variables}
\label{sec:transformations}

In this appendix we provide the transformations of the spectrally flowed canonical variables \eqref{flow} under generic PBH diffeomorphisms, as well as those under the DWCFT phase space symmetries.

\paragraph{General PBH transformations} From the action of PBH transformations \eqref{pbh-trans} on the boundary metric and stress tensor follows that the spectrally flowed canonical variables in \eqref{flow} transform as 
\bal\label{PBH-flow}
\d\cj^+=&\;-{\m^+}(\m^+\pa_+-\pa_-)\big(\x_o^-+(\m^+)^{-1}\x_o^+\big),\NO\\
\rule{0pt}{.7cm}\d\cj^-=&\;(\rg+\l\rg') (\pa_+-{{\m^-}}\pa_-)(\x_o^-+{{\m^-}}\x_o^+)-\mu^-\l\rg'(\m^+\pa_+-\pa_-)\big(\x_o^-+(\m^+)^{-1}\x_o^+\big),\NO\\
\rule{0pt}{.7cm}\d{{\co}_{-}}=&\;\x^+_o\pa_{+}\co_{-}+\x^-_o\pa_-\co_{-}+\Big(2(\pa_{-}\x^-_o+\m^-\pa_{-}\x^+_o)-\frac{(1-\l)(\l\rg)''}{(\l\rg)'}(\m^-\pa_{-}\x^+_o+\m^+\pa_{+}\x^-_o)\Big)\co_{-}\NO\\
&+\frac{c} {12\pi\ell(1-\l)(\l \rg)'}\bigg(\mathscr F\m^+\pa_{+}\s-\Big(\mathscr F-\frac{2\m^+\pa_{+}\l}{(1-\l)^2}\Big)\pa_{-}\s+\frac{\m^+(3+\l)}{(1-\l)}\pa_{+}\pa_{-}\s-\frac{1+\l}{1-\l}\pa^2_{-}\s\bigg)\NO\\
&+\frac{c} {12\pi(\l \rg)'}\bigg(\frac{\mu^+\big((\m^+)^2\pa_{+}\m^-+\pa_{+}\m^+\big)}{(1-\l)^3}\pa_{-}\pa_{+}+\frac{(\m^+)^2}{(1-\l)^2}\pa^2_{+}\pa_{-}\bigg) \big((1+\l)(\m^+)^{-1}\x_o^++2\x_o^-\big)\NO\\
&+\frac{c} {12\pi(\l \rg)'}\Big(\frac{\m^+\mathscr F}{1-\l}\pa_+\pa_- \big((1+\l)(\m^+)^{-1}\x_o^++\x_o^-\big)+\frac{(3+\l)\pa_{+}\mathscr F}{2(1-\l)}\pa_{-}\x_o^+\Big),\NO\\
\rule{0pt}{.7cm}\d{\co}_{+}=&\;\x^+_o\pa_{+}\co_{+}+\x^-_o\pa_{-}\co_{+}+2(\m^+\pa_{+}\x^-+\pa_{+}\x^+_o)\co_{+}\NO\\
&+\frac{\m^-(1-\l)\big((\m^+)^2 (2 {\rg'}^2 - \rg \rg'')\pa_{-}\x_o^{+} - \rg (\l \rg)''\pa_{+}\x_o^{-}\big)}{(\l \rg)'}\co_{-}-\frac{c}{12\pi\ell}\frac{\pa^2_{+}\s}{(1-\l)}\NO\\
&+\frac{c}{12\pi\ell}\frac{\m^{-}(3+\l)\rg}{(1-\l)^2(\l\rg)'}\pa_{+}\pa_{-}\s-\frac{c}{12\pi\ell}\frac{(\m^{-})^2 (2\rg-(1-\l)\rg')}{(1-\l)^2(\l\rg)'}\pa^{2}_ {-}\s\NO\\
&+\frac{c}{12\pi\ell}\frac{\mathscr F (1-\l)(\rg+\l(1-\l)\rg')-(\l\rg)' \big((\m^+)^2\pa_{+}\m^{-}+\pa_{+}\m^{+}\big)}{\m^+(1-\l)^2(\l\rg)'}\pa_{+}\s \NO\\
&-\frac{c}{12\pi\ell}\frac{\m^{-}\mathscr F\rg(1-\l)^2-\big((1+\l)\rg+\l(1-\l)\rg'\big)\pa_{+}\l}{\m^+(1-\l)^3(\l\rg)'}\pa_{-}\s\NO\\
&+\frac{c}{{12}\pi}\frac{\rg}{(1-\l)^2(\l\rg)'}\Big(\l\pa^2_{+}\pa_{-}+\frac{\m_-\big((\m^+)^2\pa_ {+}\m^{-} + \pa_{+}\m^{+}\big)}{(1-\l)}\pa_{+}\pa_{-}\Big)\big((1+\l)(\m^+)^{-1}\x_o^++2\x_o^-\big)\NO\\
&+\frac{c}{12\pi}\frac{\m^{-}\rg\mathscr F}{(1-\l)(\l\rg)'} \pa_{+}\pa_{-}\big((1+\l)(\m^+)^{-1}\x_o^++\x_o^-\big)-\frac{c}{24\pi}\frac{(\m^{-})^2(3+\l)\rg'\mathscr F'}{(1-\l)(\l\rg)'}\pa_{-}\x^{+}_o\NO\\
&+\frac{c}{24\p}\frac{\mathscr F'}{(1-\l)}\big(2\l(1+\l)(\m_+)^{-2}\pa_{-}\x_o^{+}+(1-\l)\pa_{+}\x_o^{-}\big).
\eal
Moreover, the composite quantity $\l={{\m^+}}{{\m^-}}$ transforms as
\bal\label{PBH-flow-composite}
\d\l=&\;\x^+\pa_+\l+(1-\l)\big(\m^+\pa_+\x^-+\m^-\pa_-\x^+\big).
\eal

\paragraph{Residual DWCFT local symmetries} The DWCFT phase space symmetry transformations \eqref{resPBH-coords} and \eqref{resPBH-G} imply that the canonical variables \eqref{flow} transform as
\bal\label{resPBH-flow}
\d\cj^-=&\;(\l\rg)'\pa_+(\vf+{{\m^-}}{{\ve}})-\l^2\rg'\pa_+((\m^+)^{-1}{{\ve}}),\NO\\
\rule{0pt}{.5cm}\d\cj^+=&\;-({\m^+})^2\pa_+((\m^+)^{-1}{{\ve}}),\NO\\
\rule{0pt}{.5cm}\d{{\co}_{-}}=&\;(\vf+{{\m^-}}{{\ve}}+{\e^-})\pa_-{{\co}_{-}}+\Big(2(\e^-)'-\frac{(\l\rg)''}{(\l\rg)'}{{\m^+}}\big(\pa_+(\vf+{{\m^-}}{{\ve}})-\l\pa_+((\m^+)^{-1}\ve)\big)\Big){{\co}_{-}}\NO\\
&+\frac{c}{24\p}\frac{{{\m^+}}\mathscr F(\pa_+\om-\pa_+\mathscr F(\vf+{{\m^-}}{{\ve}}-(\m^+)^{-1}{{\ve}}))}{(\l\rg)'(1-\l)}+\frac{c}{24\p}\frac{1}{(\l\rg)'}\big((\e^-)'''-\mathscr F^2(\e^-)'\big),\NO\\
\rule{0pt}{.5cm}\d{\co}_{+}=&\;\big((\m^+)^{-1}{{\ve}}+{\e^+}\big)\pa_-{\co}_{+}+2{{\m^+}}\pa_+\big((\m^+)^{-1}{{\ve}}+{\e^+}\big){\co}_{+}+\frac{c}{24\p}\pa_+\mathscr F\pa_+\big((\m^+)^{-1}{{\ve}}\big)\NO\\
&+{(\m^-)}^2\rg'\Big(\big((\m^+)^{-1}{{\ve}}+{\e^+}\big)\pa_-{{\co}_{-}}+2(\e^+)'{{\co}_{-}}-\d{{\co}_{-}}\Big)\NO\\
&-{{\m^-}}(\l\rg)''\Big(\pa_+(\vf+{{\m^-}}{{\ve}})-\l\pa_+\big((\m^+)^{-1}{{\ve}}\big)\Big){{\co}_{-}}+\frac{c}{24\p}(\m^+)^{-2}\times\\
&\times\bigg((\e^+)'''-(\mathscr F^2-2{{\m^+}}\pa_+\mathscr F)(\e^+)'+(\mathscr F-{{\m^+}}\pa_+)\Big(\frac{{{\m^+}}\big(\pa_+\om-\pa_+\mathscr F(\vf+{{\m^-}}{{\ve}}-(\m^+)^{-1}{{\ve}})\big)}{1-\l}\Big)\bigg).\NO
\eal
Moreover, the transformations of the composite quantities $\l={{\m^+}}{{\m^-}}$ and $\cj^-{{\co}_{-}}$ take the form 
\bal\label{resPBH-flow-composite}
\d\l=&\;{{\m^+}}\big(\pa_+(\vf+{{\m^-}}{{\ve}})-\l\pa_+((\m^+)^{-1}{{\ve}})\big),\NO\\
\rule{0pt}{.4cm}\d(\cj^-{{\co}_{-}})=&\;(\vf+{{\m^-}}{{\ve}}+{\e^-})\cj^-\pa_-{{\co}_{-}}+2(\e^-)'\cj^-{{\co}_{-}} +\frac{c}{24\p}\frac{\cj^-}{(\l\rg)'}\big((\e^-)'''-\mathscr F^2(\e^-)'\big)\NO\\
&+\bigg(\pa_\l\Big(\frac{\l\rg}{(\l\rg)'}\Big)\pa_+(\vf+{{\m^-}}{{\ve}})-\l^2\pa_\l\Big(\frac{\rg}{(\l\rg)'}\Big)\pa_+((\m^+)^{-1}{{\ve}})\bigg)(\l\rg)'{{\co}_{-}}\NO\\
&+\frac{c}{24\p}\frac{\l\rg\mathscr F(\pa_+\om-\pa_+\mathscr F(\vf+{{\m^-}}{{\ve}}-(\m^+)^{-1}{{\ve}}))}{(\l\rg)'(1-\l)}.
\eal

\section{Evaluation of the Dirichlet on-shell action}
\label{sec:on-shell-action-D}

The 2D dilaton gravity action \eqref{2Daction} can be evaluated on-shell on general running dilaton solutions in three distinct but complementary ways. Since all three approaches have been used in the literature, we take this opportunity to demonstrate their equivalence. In this appendix we evaluate on-shell the Dirichlet action \eqref{2Daction}. The on-shell action in the presence of the 2D version of the parity violating term \eqref{Ipv} is discussed in section \ref{sec:GCSS-action-NE}.

\paragraph{Integrating the variational principle} The first way to evaluate the renormalized on-shell action is to integrate the variational principle in two dimensions, as was done in \cite{Cvetic:2016eiv}. Since the solution space of 2D dilaton gravity may be embedded into the DWCFT phase space of 3D gravity, the variational principle and the on-shell action in two dimensions can be obtained by dimensional reduction. In particular, the renormalized Dirichlet variational principle for the 2D action \eqref{2Daction} follows from \eqref{var-DWCFT}-\eqref{Ilocal} and takes the form
\be\label{var2D}
\d I_{\rm 2D}^{\rm ren}=2\p R_w\int dt\;\big(-\cL_{-}\d\m^{-}+\cL_{+}\d(\m^+)^{-1}\big)+2\p R_w \frac{c}{24\p} \d \int dt\;\frac{1-\m^+\m^-}{2\m^+}\mathscr F^2.
\ee   

Since $\cL_{-}$ and $\cL_{+}$ are constants on the solution space of 2D dilaton gravity (see eq.~\eqref{2D-sol}), this variational principle can be integrated with respect to $\m^-$ and $(\m^+)^{-1}$ to obtain the corresponding on-shell action, namely  
\be\label{2D-ren-action-3Dvariables}
I_{\rm 2D}^{\rm ren}=2\p R_w\int dt\;\Big(-\cL_{-}\m^-+\cL_{+}(\m^+)^{-1}+\frac{c}{24\p}\frac{1-\m^+\m^-}{2\m^+}\mathscr F^2\Big)+I\sbtx{global}\,,
\ee
where $I\sbtx{global}$ denotes an integration constant in the functional integration. Using the identifications \eqref{Imap}, \eqref{anomaly-ab} and \eqref{2D-sol} between the 3D and 2D variables, as well as the relations 
\be\label{2D-3D-Newton}
\frac{1}{2\k^2_{\rm 2D}}=\frac{2\p R_w}{2\k^2_{\rm 3D}}=2\p R_w\frac{c}{24\p \ell}\,,
\ee
the on-shell action \eqref{2D-ren-action-3Dvariables} takes the form \cite{Cvetic:2016eiv}
\be\label{2D-ren-action-2Dvariables}
I_{\rm 2D}^{\rm ren}=-\frac{\ell}{2\k^2_{\rm 2D}}\int dt\;\Big(\frac{m\a}{\b}+\frac{2|Q|\m}{\ell}+\frac{(\pa_t\b)^2}{\a\b}\Big)+I\sbtx{global}\,.
\ee

\paragraph{Exact superpotential} An alternative derivation of the on-shell action \eqref{2D-ren-action-2Dvariables} that leads to an explicit expression for $I\sbtx{global}$ can be obtained from the exact superpotential \eqref{exactsup1}-\eqref{exactsup2}. In terms of this superpotential, the regularized on-shell action \eqref{2Daction} can be expressed as \cite{Castro:2018ffi}
\bal\label{2DactionHJ}
I_{\rm 2D}=&\;\cs\big|_{\r_{\rm min}}^{\r_o}+\frac{1}{2\k^2_{\rm 2D}}\int dt\,\sqrt{-\g}\, e^{-\j}2K\Big|_{\r_{\rm min}}\NO\\
=&\;\cu\big|_{\r_{\rm min}}^{\r_o}-\frac{Q}{\k_{\rm 2D}^2}\int dt\, A_t\big|_{\r_{\rm min}}^{\r_o}+\frac{1}{2\k^2_{\rm 2D}}\int dt\,\sqrt{-\g}\, e^{-\j}2K\Big|_{\r_{\rm min}},
\eal
where $\r_o$ is a regulating surface and $\r_{\rm min}$ is the yet unspecified lower limit of the radial integration. 

In order to evaluate this expression we observe that, for large radial cutoff $\r_o$, the superpotential \eqref{exactsup1} admits the covariant asymptotic expansion
\be
\cu[\g_{tt},\j]\big|_{\r_o}=\frac{1}{\k_{\rm 2D}^2}\int dt\,\sqrt{-\g}\Big(\ell^{-1}e^{-\j}-\frac{\ell}{2}e^{\j}\Big(\frac{\pa_t e^{-\j}}{\sqrt{-\g}}\Big)^2-\frac{m\ell}{2}e^{\j}+\co(e^{2\j})\Big)_{\r_o}\,.
\ee
Using the leading asymptotic behavior of the running dilaton solutions in eq.~\eqref{asympt}, this implies that adding the following boundary counterterm to the regularized action in \eqref{2DactionHJ}
\be\label{counterterm}
-\frac{1}{\k_{\rm 2D}^2}\int dt\,\sqrt{-\g}\,\ell^{-1}e^{-\j}\big|_{\r_o},
\ee
and sending the regulator $\r_o\to\infty$ results in the renormalized on-shell action \eqref{2D-ren-action-2Dvariables} with 
\bal\label{Iglobal-prelim}
I_{\rm global}=&\,-\cu\big|_{\r_{\rm min}}+\frac{Q}{\k_{\rm 2D}^2}\int dt\, A_t\big|_{\r_{\rm min}}+\frac{1}{2\k^2_{\rm 2D}}\int dt\,\sqrt{-\g}\, e^{-\j}2K\Big|_{\r_{\rm min}}\\
=&\,\frac{\ell}{\k^2_{\rm 2D}}\int dt\,\frac{(\pa_t\b)^2}{\a\b}-\frac{Q}{\k_{\rm 2D}^2}\int dt\, A_t\big|_{\r_{\rm min}}-\frac{1}{\k^2_{\rm 2D}}\int dt\,\Big(\sqrt{-\g}\,\pa_\r e^{-\j}-e^{-\j}\pa_\r\sqrt{-\g}-2Q A_t\Big)_{\r_{\rm min}}\,.\NO
\eal
Notice that the second line follows from the relation $K=\pa_\r\log \sqrt{-\g}$ and the identity \eqref{sup-value}, which allows us to write the superpotential in the form
\be\label{supID}
\cu=\frac{1}{\k^2_{\rm 2D}}\int dt\,\sqrt{-\g}\,\pa_\r e^{-\j}-\frac{\ell}{\k^2_{\rm 2D}}\int dt\,\frac{(\pa_t\b)^2}{\a\b}.
\ee

Finally, the 2D field equations \eqref{2DeomFG} imply that the quantity in the last parenthesis of eq.~\eqref{Iglobal-prelim} is independent of the radial coordinate, namely
\be\label{identity}
\sqrt{-\g}\,\pa_\r e^{-\j}-e^{-\j}\pa_\r\sqrt{-\g}-2Q A_t=-\ell\Big(\frac{m\a}{\b}-\frac{(\pa_t\b)^2}{\a\b}+\pa_t\Big(\frac{\pa_t\b}{\a}\Big)+\frac{2Q\m}{\ell}\Big).
\ee
It follows that the general form of $I\sbtx{global}$ is 
\be\label{Iglobal}
I_{\rm global}=\frac{\ell}{\k^2_{\rm 2D}}\int dt\,\Big(\frac{m\a}{\b}+\pa_t\Big(\frac{\pa_t\b}{\a}\Big)+\frac{2Q\m}{\ell}\Big)-\frac{Q}{\k_{\rm 2D}^2}\int dt\, A_t(\r_{\rm min},t)\,.
\ee
Combining this with \eqref{2D-ren-action-2Dvariables}, the full expression for the renormalized Dirichlet on-shell action is
\be\label{2D-ren-action-2Dvariables-D}\boxed{
I_{\rm 2D}^{\rm ren}=\frac{\ell}{2\k^2_{\rm 2D}}\int dt\,\Big(\frac{m\a}{\b}+2\pa_t\Big(\frac{\pa_t\b}{\a}\Big)-\frac{(\pa_t\b)^2}{\a\b}+\frac{2Q\m}{\ell}-\frac{2Q}{\ell}A_t(\r_{\rm min},t)\Big).}
\ee
From the analysis in section \ref{sec:effective-action} one sees that $I\sbtx{global}$  does not contribute to the Schwarzian term in the effective action.

\paragraph{Direct integration} The renormalized on-shell value  \eqref{2D-ren-action-2Dvariables-D} of the 2D action may also be obtained by direct integration. Observe that, in the FG gauge \eqref{2dana2}, the bulk Ricci scalar and the extrinsic curvature of the constant radial slices can be expressed, respectively, as
\be
\sqrt{-g}\,R[g]=-2\pa_\r\big(\sqrt{-\g}\,K\big),\qquad K=\pa_\r\log \sqrt{-\g}\,.
\ee
Using these identities, the difference between the first two equations of motion in \eqref{2DeomFG} implies that the bulk Lagrangian can be written as a total derivative in infinitely many ways, namely
\be\label{2DLagrangian}
\sqrt{-g}\;e^{-\j}\Big(R[g]+\frac{2}{\ell^2}+2Q^2e^{4\j}\Big)=\pa_\r\Big(a(t)\big(\sqrt{-\g}\,\pa_\r e^{-\j}-e^{-\j}\pa_\r\sqrt{-\g}\big)+\big(1-a(t)\big)2QA_t\Big),
\ee
where we have added zero to the right hand side of this relation in the form
\be
0=\pa_\r\big((\sqrt{-\g}\,\pa_\r e^{-\j}-e^{-\j}\pa_\r\sqrt{-\g})-2QA_t\big),
\ee
multiplied by an arbitrary function $a(t)$ of $t$.

From the identity \eqref{2DLagrangian} follows that the action \eqref{2Daction} can be expressed on-shell as
\bal
I_{\rm 2D}=&\;\frac{1}{2\k^2_{\rm 2D}}\int dt\bigg(\int_{\r_{\rm min}}^{\r_o} d\r \sqrt{-g}\;e^{-\j}\Big(R[g]+\frac{2}{\ell^2}+2Q^2e^{4\j}\Big)+2\sqrt{-\g}\;e^{-\j}K\big|_{\r_o}\bigg)\NO\\
=&\;\frac{1}{2\k^2_{\rm 2D}}\int dt\,\Big(a(t)\big(\sqrt{-\g}\,\pa_\r e^{-\j}-e^{-\j}\pa_\r\sqrt{-\g}-2Q A_t\big)+2e^{-\j}\pa_\r\sqrt{-\g}+2QA_t\Big)_{\r_o}\NO\\
&-\frac{1}{2\k^2_{\rm 2D}}\int dt\,\Big(a(t)\big(\sqrt{-\g}\,\pa_\r e^{-\j}-e^{-\j}\pa_\r\sqrt{-\g}-2Q A_t\big)+2QA_t\Big)_{\r_{\rm min}}\,.
\eal
Notice that for $a(t)=2$ this coincides with \eqref{2DactionHJ}. However, since the terms multiplying $a(t)$ are independent of the radial coordinate by virtue of \eqref{identity}, $a(t)$ cancels between $\r_o$ and $\r_{\rm min}$, giving
\be
I_{\rm 2D}=\frac{1}{2\k^2_{\rm 2D}}\int dt\,\big(2e^{-\j}\pa_\r\sqrt{-\g}+2QA_t\big)_{\r_o}\NO\\
-\frac{1}{2\k^2_{\rm 2D}}\int dt\,2QA_t\big|_{\r_{\rm min}}\,.
\ee
Notice that this expression reduces to the Gibbons-Hawking term when $Q=0$, reproducing the known result for Jackiw-Teitelboim dilaton gravity \cite{Maldacena:2016upp}. Adding the boundary counterterm \eqref{counterterm}, the limit $\r_o\to\infty$ of the renormalized on-shell action can be evaluated using the asymptotic expansions 
\be
e^{-\j}\pa_\r\sqrt{-\g}=\ell^{-1}\sqrt{-\g}\, e^{-\j}+\frac{\ell}{2}\Big(\frac{m\a}{\b}-\frac{(\pa_t\b)^2}{\a\b}+2\pa_t\Big(\frac{\pa_t\b}{\a}\Big)\Big)+\co(e^{-2\r/\ell}),\quad A_t =\mu+\co(e^{-2\r/\ell})\,,
\ee 
and reproduces the result in \eqref{2D-ren-action-2Dvariables-D}.



\bibliographystyle{JHEP}
\bibliography{refs}

\end{document}